\documentclass[preprint,5pt]{elsarticle}

\usepackage{amssymb}
\usepackage{graphics}
\usepackage{subfigure}
\usepackage{amsmath}
\usepackage{threeparttable}
\usepackage{caption}
\usepackage{hyperref}[colorlinks=true]
\begin{document}

\begin{frontmatter}
\journal{\ }


\title{China's Easily Overlooked Monetary Transmission Mechanism: Monetary Reservoir}


\author[inst1]{Lai Xinglin and Xiao Shuguang$^*$\footnote{Corresponding author,Email:\href{mailto:shuguangxiao@126.com}{shuguangxiao@126.com}} }

\affiliation[inst1]{organization={Economics Department of Economics},%Department and Organization
            addressline={The Guangdong university of Financial and Economics}, 
            city={Guangzhou},
            postcode={510320}, 
            state={Guangdong},
            country={China}}
\author[inst2]{Peng Jiamin}

\affiliation[inst2]{organization={School of Social Sciences, Faculty of Humanities},%Department and Organization
            addressline={The University of Manchester}, 
            city={Oxford Rd},
            postcode={M139PL}, 
            state={},
            country={UK}}
\begin{abstract}
%
%传统的货币传导机制通常以证券资本市场作为货币蓄水池，然而由于中国独特的财税金融体系，房地产部门多年来事实上成为了中国“视而未见”的非传统货币蓄水池。首先，本文构建包含财政投融资和金融加速器的动态一般均衡模型，利用中国住房市场数据以及中央银行对金融机构再贷款利率数据进行参数估计，揭示了货币蓄水池-财政融资投资的货币传导机制规律：一种资产只要满足杠杆交易制度、余额承诺支付以及存在政府效用三个条件即可称之为贷款品，并由地方政府进行投融资形成财政赤字最终走向货币化。本文将这一机制称为货币蓄水池。货币蓄水池机制将推高贷款品的溢价乃至形成资产泡沫并且对货币政策的有效性产生明显的影响。地方政府通过财政融资影响金融中介资产负债表的方式提高贷款品部门和自身的杠杆率以获取短期增长，但这通过挤出人力资本和技术积累的方式损害了长期增长的根基。
The traditional monetary transmission mechanism usually views the equity  markets as the monetary reservoir that absorbs over-issued money, but due to China's unique fiscal and financial system, the real estate sector has become an "invisible" non-traditional monetary reservoir in China for many years. First, using data from Chinese housing market and central bank for parameter estimation, we constructs a dynamic general equilibrium model that includes fiscal expansion and financial accelerator to reveal the mechanism of monetary reservoir. An asset can be called a loan product, which worked as financed asset for local fiscal expansion, as long as it satisfies the following three conditions: leveraged trading system, balance commitment payment, and the existence of the utility of local governments. This paper refers to this mechanism as the monetary reservoir that will push up the premium of loan product, form asset bubbles and has a significant impact on the effectiveness of monetary policy. Local governments leverage the sector of the loan product to obtain short-term growth by influencing the balance sheets of financial intermediaries through fiscal financing, expenditure and also investment, but this mechanism undermines the foundations of long-term growth by crowding out human capital and technological accumulation.
\end{abstract}




\end{frontmatter}


\section{Introduction}
%自2008年金融危机以来，中国的住房市场和美国的股票市场 都进入了显著高于于消费品及工业品价格水平的繁荣时期，由于这些资产相对其他资产的价格和市值高速膨胀特征与中美两国宽松货币政策周期重合，“中房美股”是“货币蓄水池”的说法不胫而走。图1说明了中房美股两个市场相对于PPI和CPI的累积溢价涨幅水平，2009年以来美国联邦储备局启动量化宽松政策，中国为维持出口价格优势对增发美元通过人民币超发进行冲销导致输入型通胀
\indent Since the financial crisis in 2008, both the Chinese housing market and the American stock market have experienced a boom period that was significantly higher than consumer price index (CPI) and producer price index (PPI). Since the rapid expansion of the price and market value of these assets relative to other assets coincides with the loose monetary policy of China and the United States, the notion that "Chinese housing and US stocks" are "monetary reservoirs" has become widespread. Premium price lever is that the nth-year price of product X divides the base-year price, and also divides CPI or PPI cumulative level which is the multiplier of each CPI or PPI from base year to nth year. Figure 1 below illustrates that cumulative premium increase trend of two markets relative to the PPI and CPI, with Federal Reserve's initiation of quantitative easing policy since 2009 and China's offsetting of additional US dollar issuance through RMB over-issuance in order to maintain its export volume.

%基于此背景“中房美股”的变动趋势都呈现一定的共性以及各自的特点。中美两国的共性便是房市和股市都呈现对CPI和PPI明显更高的价格涨幅过程，这一点在中国尤为突出，在不到十年间中国房价水平对消费品价格水平的溢价累积到45倍以上，对工业品价格水平的溢价水平也累积到5倍左右。在美国该现象同样存在，但是溢价累积水平远低于中国。另一方面，两者同样有显著的差异。首先，中国的货币蓄水池资产价格增幅远远高于美国，其次，在中国房地产部门成为货币蓄水池而在美国则是股票市场承担该“功能”。事实上，美国房地产部门在2008年房屋抵押支持证券泡沫破灭之前也吸纳了大部分的超发货币和投机资金，美国股票市场成为货币蓄水池资产的重要原因是美国政府对房屋证券化的严厉监管措施。
\indent The trends in both the US and China have certain commonalities as well as individual characteristics. The commonality between them is that both the housing and stock markets have experienced significantly higher price increases compared to the CPI and PPI, particularly in China, where the housing price premium to CPI has accumulated to over 45 times within less than a decade, and the price premium to PPI has also accumulated to around 5 times. 
This phenomenon also happened in the US, but the level of premium accumulation is much lower than the one in China. On the other hand, there are also remarkable differences between these markets. Firstly, the increase in asset prices is much higher in China than in the US, and secondly, in China the real estate sector acts as a monetary reservoir, whereas in the US it is the stock market that performs this function. Indeed, the US real estate sector also absorbed much of the excess money and speculative capital before the bursting of the mortgage-backed securities bubble in 2008, and the US equity market became a monetary reservoir asset because of the US government's heavy-handed regulatory measures on housing securitisation.\newline
%货币蓄水池资产现象的主要特征为货币蓄水池部门的逆周期繁荣。认为宽松货币政策是货币蓄水池的主要依据是中央银行货币宽松周期与“货币蓄水池”资产价格上涨的过程基本重合，然而这至多说明“货币蓄水池”现象与中央银行的宽松货币政策有相关关系，但是并不能说明以下问题：①为何在同等货币超发条件下，“货币蓄水池”资产呈现出一枝独秀、溢价涨幅明显的特征？ ②总量上“货币蓄水池”的新增市值是否全部来源于中央银行的货币宽松？第一个问题的意义在于“货币蓄水池”资产如果存在天然吸引超发货币的特征，一方面它有利于缓解通胀现象并且发挥其增长效应服务实体经济，但另一方面它可能导致严重的资产价格泡沫并最终引发严重的系统性金融危机。
\indent The main feature of the monetary reservoir asset phenomenon is a counter-cyclical boom. The main basis of the view that loose monetary policy regards as a monetary is that 
the central bank's monetary easing cycle largely coincides with the rise in the prices of monetary reservoir assets, however, this point just confirms a correlation between the monetary reservoir phenomenon and the expansionary monetary policy of the central bank. But it does not explain the following questions: (i) Why did the monetary reservoir assets only stand out and show the significant premium trend under the same conditions of monetary excess? (ii) Did all the newly increased market value of the monetary reservoir in aggregate come from monetary easing policy? The significance of the first question lies in the fact that if monetary reservoir assets have the characteristic of naturally attracting excess money, on the one hand they can help alleviate inflation and serve the real economy with their financing effect, but on the other hand they can lead to serious asset price bubbles and eventually to serious systemic financial crises.\newline
\indent To answer these problems, the traditional theory of the top-down monetary transmission mechanism still needs to be supplemented. Relevant research normally ignores the possibility of local governments expanding credit and issuing more money, therefore there is the biased understanding in local governments' financing and investment behaviors. In fact, local governments have certain pricing rights over production factors and tax guarantees, so they finance and invest through financial intermediaries, forming a monetary transmission mechanism of monetary issuance - increasing credit - driving economic growth or over-issuing money to cultivate growth. This mechanism has been "invisible" to the academic community, i.e. only partial but not overall, only ranked secondly but not firstly in importance.\newline
%本文的边际贡献在于：①构建了包含公共品部门以及金融加速器的动态一般均衡模型，对中国地方政府的高债务问题和房地产部门的逆周期繁荣现象给出详细合理的解释；②分析货币蓄水池除了财政赤字货币化之外对货币政策有效性的影响；③通过货币蓄水池，本文对中国进入新世纪以来的以财政投资和外资共同拉动经济发展的增长模式提供更微观的解释，并且对高杠杆和高资产泡沫的问题提出对应的政策建议。
\indent The marginal contributions of this paper are: (i) the construction of a dynamic general equilibrium model that includes the public goods sector and the financial accelerator to provide a detailed and rational explanation for the high debt problem of local governments and the counter-cyclical boom in the real estate sector in China; (ii) the analysis of the impact of the monetary reservoir on the effectiveness of monetary policy in addition to the monetisation of the fiscal deficit; and (iii) through the monetary reservoir, this paper provides a more microscopic explanation for China's growth model since the beginning of the new century, which has been driven by a combination of fiscal investment and foreign investment, and proposes corresponding policy recommendations for the problems of high leverage and high asset bubbles.\newline
%本文接下来的内容安排如下：第二部分为文献综述；第三部分从传统的货币传导机制理论开始，结合中国的独特的财税金融体系，对地方政府主导的货币蓄水池机制进行理论分析；第四部分基于Bernanke et al.(1998)的金融加速器模型构建一个包含地方政府、贷款合同以及公共品部门的动态一般均衡(DSGE)模型，第四部分根据我国住房市场以及相关宏观经济数据对上述DSGE模型进行参数校准及贝叶斯模拟矩估计；第五部分通过分析具有贷款交易制度的部门(贷款品)的市值和单位商品价格相对于非贷款品部门的溢价因素；第六部分通过脉冲响应函数和方差分解的分析方法研究货币蓄水池机制对货币政策的调控效果的影响；第七部分从增长速度、资源配置、增长结构和财政投资的作用等方面分析货币蓄水池在短期对宏观经济系统的作用；第八部分本文在DSGE模型中嵌入了Romer(1987\citep{romer1987growth}, 1990\citep{romer1990endogenous})的内生增长模型，研究货币蓄水池对研发劳动的挤出在长期的经济效应。
\indent The following parts of this paper are organized as follows: the section 2 is a literature review; the section 3 starts from the traditional theory of monetary transmission mechanism and theoretically analyzes the local government-led monetary reservoir mechanism in the context of China's unique fiscal and financial system; the section 4 constructs a dynamic general equilibrium (DSGE) model based on Bernanke et al.'s (1998) financial accelerator model that includes local government, loan contracts, and The fourth part constructs a dynamic general equilibrium (DSGE) model based on Bernanke et al.'s (1998) financial accelerator model that includes local governments, loan contracts, and the public goods sector; the section 5 calibrates the parameters of the above DSGE model and estimates Bayesian simulation moments based on China's housing market and relevant macroeconomic data; the section 6 analyzes the market value of the sector with a loan transaction system (loan goods) and the premium factors of unit commodity prices relative to the non-loan goods sector; the section 7 investigates the monetary reservoir mechanism through the impulse response function and variance decomposition analysis to study the impact of the monetary reservoir mechanism on the efficiency of monetary policy; the section 8 analyzes the role of monetary reservoirs on the macroeconomic system in the short run in terms of growth rate, resource allocation, growth structure and the role of fiscal investment; the section 9 of this paper embeds the DSGE model with Romer (1987\citep{romer1987growth}, 1990\citep{romer1990endogenous}) endogenous growth model in the DSGE model to study the economic effects of the crowding out of R\&D labour by the monetary reservoir in the long run.
\section{Literature Review}
%舆论多将房地产和股票市场的繁荣归因于中央银行的货币超发，认为房地产市场和股票市场作为货币蓄水池吸纳过量的超发货币。这种观点本质上仍是以中央银行为核心和自上而下的视角，即便许多学者对土地财政或地方债务等问题有所研究，但是学者们对中国地方财政与货币传导机制两者仍旧是各自独立地分析。
\indent Public opinions has mostly attributed the boom in the real estate and stock markets to the over-issue of money by central bank, arguing that the real estate and stock markets acted absorb overflow money as monetary reservoirs. This is still essentially a central bank-centred and top-to-bottom perspective. Even though many scholars have studied issues such as land finance or local debt, they still analyse local finance and the monetary transmission mechanism in China independently.\newline
%理解中央银行外的货币增发需要厘清货币增发主体及动机两个核心问题。货币增发本质是信用扩张，地方政府的信用扩张转化为货币增发需要具备一定条件，判断标准在于财政融资投资是否转为政府赤字，如果信用扩张中的财政融资投资带来高于债务的收益，那么不应称之为货币增发，反之多余政府赤字的积累通过金融中介转化为货币，这种转化并不限于中央银行的兑付。在1994年分税制改革后，中央政府得到更多财权的同时地方政府承担更多事权，财政资金缺口迫使地方政府寻找财政融资，此外，在很长的时期内中国地方政府的官员评比中GDP是最主要的绩效指标，地方政府由于官员绩效评比、财政体制等因素通过财政补贴返还、配套设施建设、税收优惠、土地折价等方式推动地方经济增长或财政收入最大化(杨帆,2010;陈超,2011;黄少安,2012;朱英姿,2013;王媛,2016;邵朝对,2016)，实证方面的证据如宫汝凯(2015)、王弟海(2015)等，地方政府通过信用扩张和金融杠杆推动房地产部门投资，然而这也导致了地方政府的增发货币速度较快，其中大部分增发货币流入货币蓄水池。地方政府信用扩张的核心支持资产是土地财政、影子银行以及隐形担保等(梅冬州,2021；付敏杰,2017；梁琪,2019；吴盼文,2013；Wu et al.,2016)。基于土地财政的财政融资实质是在垄断性土地市场条件下地方政府的土地资源资本化(杨帅和温铁军,2010；邵新建,2012)，地方政府通过银行项目贷款、城投债以及资本市场融资等方式间接地形成地方债务。地方政府的信用扩张虽然是以土地资本作为基础资产，但其快速增长依赖于政府税收的背书，地方政府为持续融资通过多种方式限制土地供应推动地价上升，进一步增加金融部门投向房地产部门的贷款总额，金融中介配合地方政府的信用扩张主要有两方面原因：一是以地方融资平台为主的金融中介的股权多数为地方政府所有或者在中国银行体系中金融中介配合地方政府官员的政策意愿可能有更大的经营便利，总而言之，金融中介与地方政府存在共同利益；二是金融中介以及社会资本“相信”中央及地方政府的担保。
\indent To understand the monetary over-issuance out of central bank, it is necessary to clarify the two core issues. The essence of currency additional issuance is credit expansion. There are some certain conditions for transforming the credit expansion into the additional monetary issuance. Specifically, it depends on whether the fiscal financing investment has turned into a government deficit. If fiscal financing and investment of the credit expansion brings higher return rather than debt, then it should not be called the monetary over-issuance. On the contrary, the accumulation of surplus government deficit is converted into money through financial inter-mediation, and this conversion is not limited to the payments of central bank. After the reform of the tax-sharing system in 1994, the central government gained more financial power while local governments took larger fiscal powers. The gap in financial funds forced local governments to seek financial financing. In addition, GDP was the main performance indicator in the evaluation of local government officials in China for a long period of time. The officials promoted local economic growth or maximizes fiscal revenue through the return of financial subsidies, the construction of supporting facilities, tax incentives, land discounts, etc.[Huang (2012)\citep{Huang2012Renttax};Shao (2016)\citep{Shao2016HousingPrice}], empirical evidence such as Wang(2015)\citep{Wang2015Thesupply}. Local governments promote real estate sector investment through credit expansion and financial leverage. However, this also leads to rapid issuance of currency by local governments, most of which flow into currency reservoirs.The core support assets for local government credit expansion are land finance, shadow banks and invisible guarantees [Mei(2021)\citep{Mei2020External};Fu(2017)\citep{Fu2017Evo}; Liang(2019)\citep{Liang2019Local}; Wu(2013)\citep{Wu2013TheImpact}; Wu et al. (2016)\citep{wu2016evaluating}. The essence of fiscal financing based on land finance is the capitalization of land resources by local governments under the condition of monopoly land market. Local governments indirectly form local debts through bank project loans, city investment bonds and capital market financing. Although the credit expansion of local governments is based on land capital as the underlying asset, its rapid growth relies on the endorsement of government tax revenue. restricts the supply of land in various ways to promote land prices for continuous financing, further increasing the financial sector's loans to the real estate sector. In total, there are two main reasons for financial intermediaries to cooperate with local governments in their credit expansion: In total, there are two main reasons for financial intermediaries to cooperate with local governments in their credit expansion: First, the equity of financial intermediaries based on local financing platforms is mostly owned by local governments, or the policy willingness of financial intermediaries to cooperate with local government officials in the Chinese banking system may be possible. There is greater operational convenience in this situation. All in all, financial intermediaries and local governments have common interests; the second is that financial intermediaries and social capital "trust" the guarantees of the central and local governments
\newline
%为详细区分两种不同的机制，我们首先对被广泛接受的货币传导机制进行分析。货币传导机制相关的实证和理论模型历经多年仍有较多争论，总的来说，货币传导机制分为三种观点：一是凯恩斯主义观点，认为货币政策通过操作短期利率影响长期利率从而影响实际投资和长期增长(Bernanke & Blinder et al.,1992)；二是货币主义观点，认为货币政策通过短期利率调整资产相对收益产生财富效应和通过Tobin Q值推动投资；三是80年代以来的信贷观点，认为货币政策首先是对资产价格发生作用，影响企业和银行的资产负债状况，最终通过社会总信贷水平影响产出。三种观点所讨论的传导渠道主要是四个中介变量：利率、汇率、资产价格以及信贷，且无一例外都是研究从中央银行到商业银行再到企业的传导路径。
\indent In order to distinguish in detail between the two different mechanisms, we first analyse the widely accepted monetary transmission mechanism. The empirical and theoretical models related to the monetary transmission mechanism have remained relatively controversial over the years and, in general, there are three views on the monetary transmission mechanism: Firstly, the Keynesian view, which argues that monetary policy affects long-term interest rates through the manipulation of short-term interest rates, thereby influencing real investment and long-term growth [Bernanke et al.(1992)\citep{bernanke1990federal}]. The second is the monetarist view, which sees monetary policy as generating wealth effects through short-term interest rate adjustments in relative asset returns and driving investment through Tobin Q values. Thirdly, the credit view, which argues that monetary policy acts first on asset prices, affecting the asset and liability positions of firms and banks, and ultimately affecting output through the level of aggregate social credit. The transmission channels discussed in the three perspectives are mainly four mediating variables: interest rates, exchange rates, asset prices and credit. Without exception they all study the transmission path from the central bank to commercial banks and finally to enterprises. \newline
%按照上述分析，对传统观点主要应该作两点补充：一方面区别于中央银行是货币传导机制唯一中心的传统视角，地方政府的信用扩张在一定条件下也存在货币发行-流通的传导机制。地方政府财政投资收益低将导致扩张信用转化为增发货币，地方政府高债务和低增长的局面将出现，为了偿还债务地方政府只能继续扩张信用借新还旧，加剧货币蓄水池资产和其他资产的信贷不平衡，形成非中央银行的货币传导机制。另一方面需要考虑财政投资的效益，例如公共品投资对产出效率的作用，例如在资源配置失衡条件下增发货币反而加重金融资源配置效率降低的矛盾，而源于财政融资的公共品投资则提升产出效率。
\indent According to the above analysis, there are two main additions to the traditional view: on the one hand, in contrast to the traditional view that the central bank is the only centre of the monetary transmission mechanism, the credit expansion of local governments also has a monetary issuance-circulation transmission mechanism under certain conditions . On the other hand, there is a need to consider the benefits of fiscal investment, such as the role of public goods investment on output efficiency, for example, the contradiction that increased monetary issuance under conditions of resource allocation imbalance instead aggravates the reduced efficiency of financial resource allocation, while public goods investment originating from fiscal financing enhances output efficiency [Chen(2016)\citep{chen2016hunman}], Liu(2021)\citep{Liu2021ThePuzzle}]. \newline
%以往文献对房地产部门与经济增长关系的研究主要分为两条脉络：资产价格渠道和金融体系渠道。从上述两条主要脉络看房地产投资挤出了消费、投资和降低金融效率是认可程度较高的结论，该结论至少说明两点：一、房地产部门为代表的货币蓄水池占用了较多的金融资金即信贷偏好；二、房地产部门的过度繁荣通过挤出消费和创新投资以及降低金融效率削弱长期增长的潜力。
\indent The previous literature on the relationship between the real estate sector and economic growth has been divided into two main splits of research, which are the asset price channel [Chaney(2012)\citep{chaney2012collateral} and Atif(2014)\citep{mian2014house}, Yang(2014)\citep{Yang2014DualRole}, Li(2014)\citep{Li2014RealAssets}, Xiao(2014)\citep{Xiao2014NonlinearInfluence}, Song(2021)\citep{Song2020DoExcessive}, Bernstein(2021)\citep{bernstein2021household}] and the financial system channel.[Luo(2015)\citep{Luo2015Credit}]. The conclusion that real estate investment crowds out consumption, investment and reduces financial efficiency is the more widely accepted conclusion in the two main splits mentioned above, which suggests at least two facts: first, that the monetary reservoir represented by the real estate sector takes up more financial funds, i.e. credit preference; and second, that the excessive boom in the real estate sector weakens the potential for long-term growth by crowding out consumption and innovative investment and also reducing financial efficiency.\newline
\indent 
%信贷偏好的结果是金融中介的资金更多地流向房地产部门、杠杆率增加并推高该部门市值，其中金融资金包括增发货币，这为货币蓄水池资产现象的逆周期繁荣现象提供一种解释，一些学者也提出了相关的实证证据。Landvoig et al.t(2011)用圣地亚哥的住房微观数据从实证发现对家庭尤其贫穷家庭的住房信贷对住房价格上涨有显著的影响。Fang et al.(2016)通过对中国住房市场的数据调查发现各地房地产市场繁荣都与信贷扩张有密切联系。Cong et al. (2019)使用19家中国上市银行对制造业企业贷款的数据发现自2008年之后政府隐性担保和经济刺激计划提高了金融中介对国有公司的贷款比例。朱太辉(2018)认为政府隐性担保和抵押属性与其他因素相互促进导致银行资金更多流向基础设施和房地产行业。魏玮(2017)发现杠杆与房价的关系具有负向、免疫和正向三种反应机制，赵胜民(2011)则发现金融信贷对房地产价格影响有限，对股票市场影响更大。需要区分的是，金融信贷与资产杠杆并不存在直接因果的关系，金融信贷一般采用金融中介贷款量作为代理变量，但是贷款量增加并不一定提高资产杠杆，而是由交易制度如抵(质)押、首付或担保决定资产杠杆，以贷款数量解释房地产部门的繁荣并不足够，因此解释房地产部门的货币蓄水池现象本质是以杠杆解释信用扩张和经济周期。国外相似的结论包括Justiniano.et.al.(2013), Mian et al.(2016)。Justiniano(2013)通过动态一般均衡模型研究美国次贷危机的信贷宽松与信贷杠杆的关系，认为杠杆周期主要由影响房价的因素所决定，而借贷人和放贷人的双向反应抵消大部分杠杆增减的宏观效应。Mian(2016)发现家庭部门债务对GDP比例的上升伴随着消费的繁荣。Schularick(2012)则认为信贷增长是金融危机的有力预测工具。Iacoviello(2010)研究美国1980~2010住房市场波动的来源，认为来自于货币政策冲击的部分少于20%，但对房地产市场周期影响显著。因此，信贷偏好与货币蓄水池资产的繁荣应该是互为因果的关系，杠杆性导致最初的信贷偏好，信贷偏好推动信贷规模的增加，产生繁荣周期，进而又增加信贷偏好。
The result of credit preference is an increased flow of financially intermediated funds to the real estate sector, increased leverage and a push up in the market value of the monetary reservoir, where financial funds include increased money issuance, providing an explanation for the countercyclical boom of the monetary reservoir, for which some scholars have also presented empirical evidence [Landvoig et al.(2011)\citep{landvoigt2015housing}, cong et al.\citep{cong2019credit}, Wei(2017)\citep{Wei2017DoesHigher}]. Financial credit generally uses the volume of loans from financial intermediaries as a proxy variable, but an increase in the volume of loans does not necessarily increase asset leverage, but rather the transaction regime such as down payment or guarantee determines asset leverage, which explains the boom in the real estate sector. The essence of this phenomenon is to explain credit expansion and economic cycles in terms of leverage. Similar findings from other countries include [Justiniano.et.al. (2015)\citep{justiniano2015household}, Mian et al.(2016)\citep{mian2014house},schularick (2012)\citep{schularick2012credit},  [Iacoviello(2010)\citep{iacoviello2010housing}]. Thus, credit preference and monetary reservoir booms should be causally related, with leverage leading to initial credit appetite, which drives an increase in credit size, generating boom cycles, which in turn increase credit appetite.
\begin{figure}[htbp]  
\centering
\subfigure[1]{\begin{minipage}{6cm}

\includegraphics[width=6cm]  {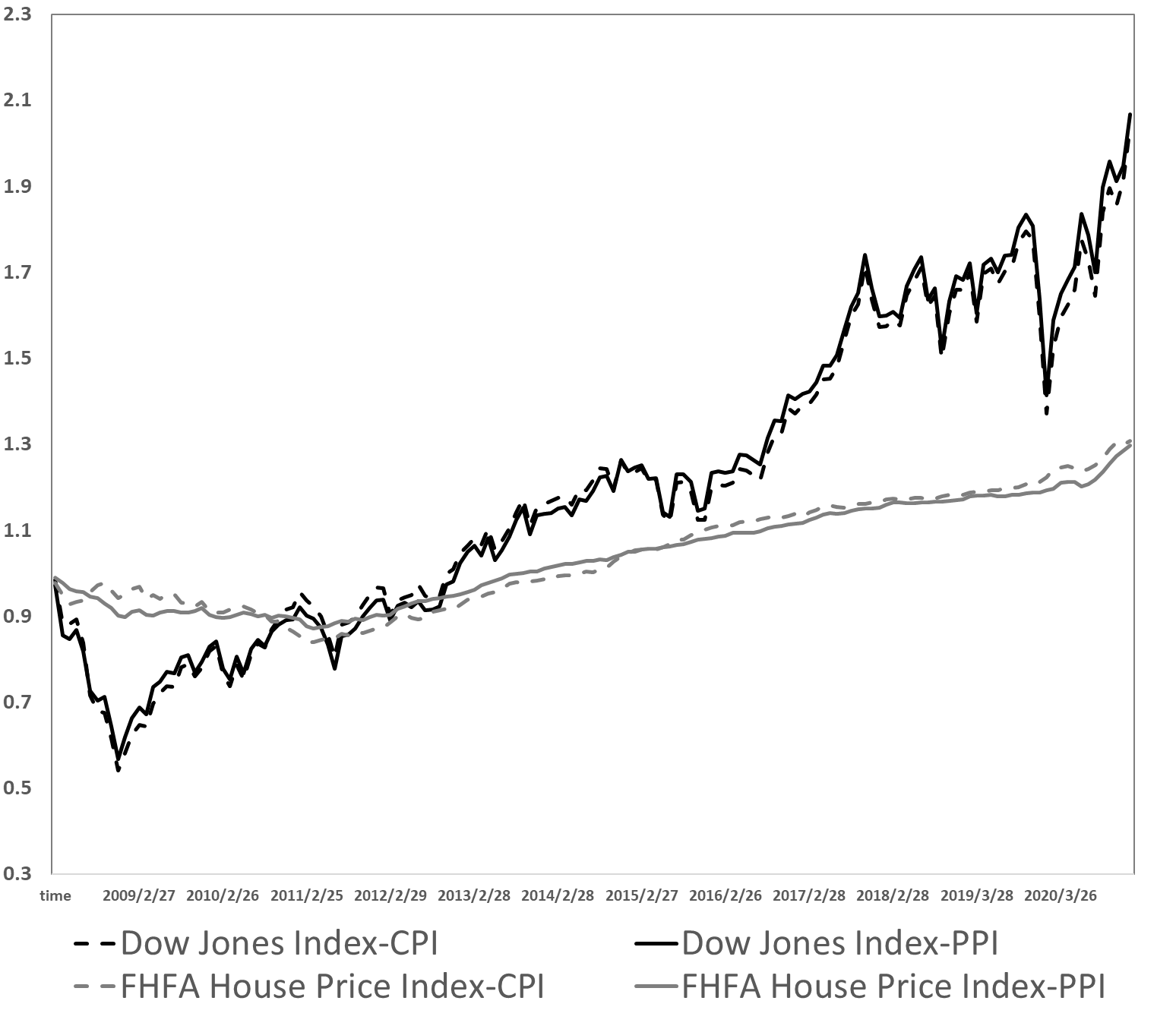}   
\end{minipage}
}

\subfigure[2]{\begin{minipage}{6cm}
  \includegraphics[width=6cm]  {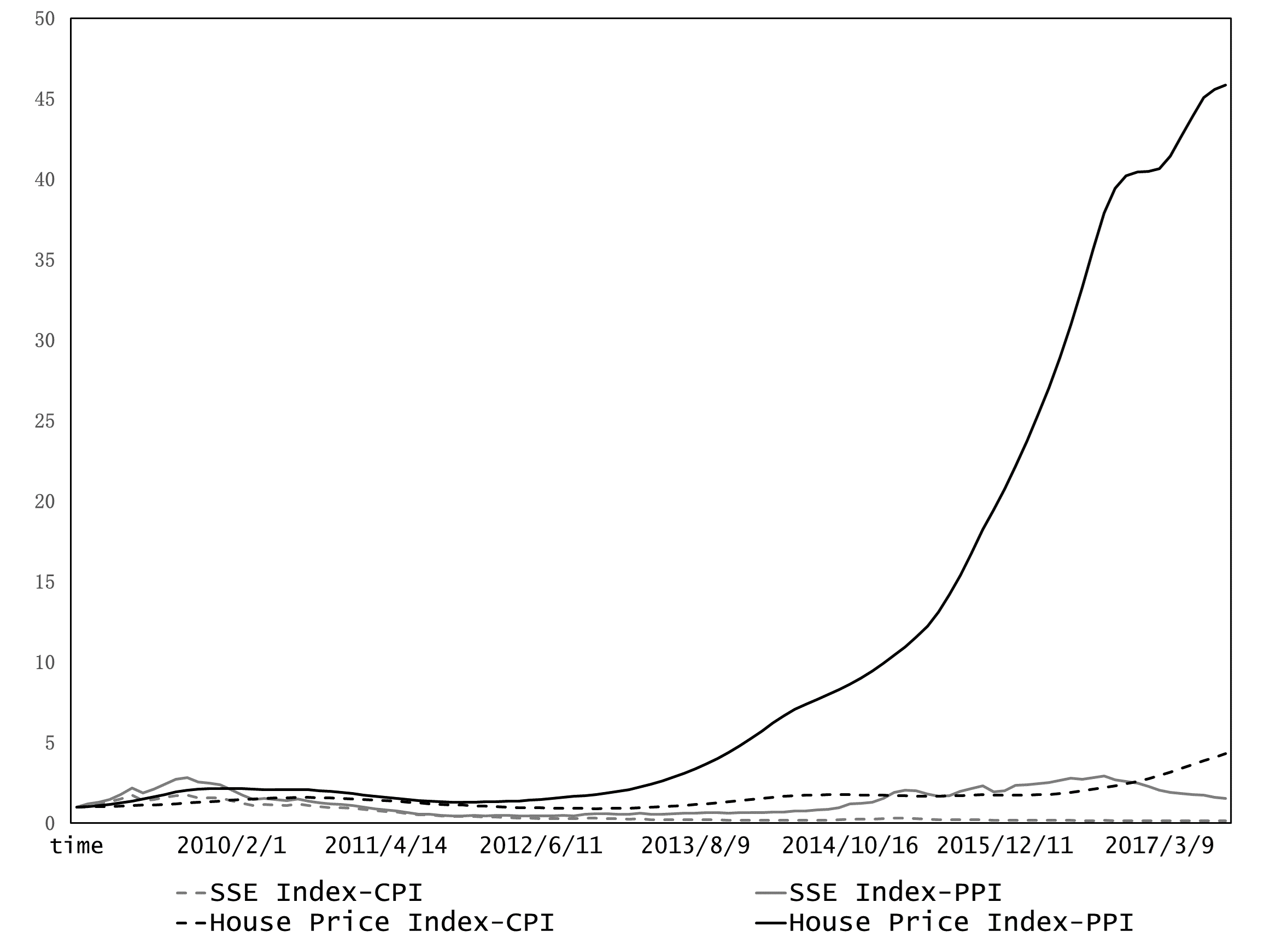}    
\end{minipage}
}
\caption{Comparative chart of price increases in the Chinese housing and US equity markets}

\end{figure}
\section{Theoretical analysis}
\indent There are two main aspects of leverage: the trading system and the refinancing mechanism. The leveraged trading system consists of a down payment ratio or margin system, while the refinancing mechanism is indirectly leveraged. Assuming that an asset has to be purchased in full, financial intermediary provides a partial "rebate" of the amount paid, and the purchaser of the asset can continue to repay the loan with a loan, the actual effect of refinancing is basically the same as that of a leveraged trading system. The interest rate is the central factor in a leveraged trading regime.In addition, due to information asymmetry and financial friction, financial intermediaries form an interest rate premium, that is, the financial accelerator mechanism [Bernanke et al.,(1999\citep{bernanke1990federal})]] This paper introduces the mortgage of housing assets on the basis of the financial accelerator. 
\begin{figure}[htb]   \center{\includegraphics[width=12cm]  {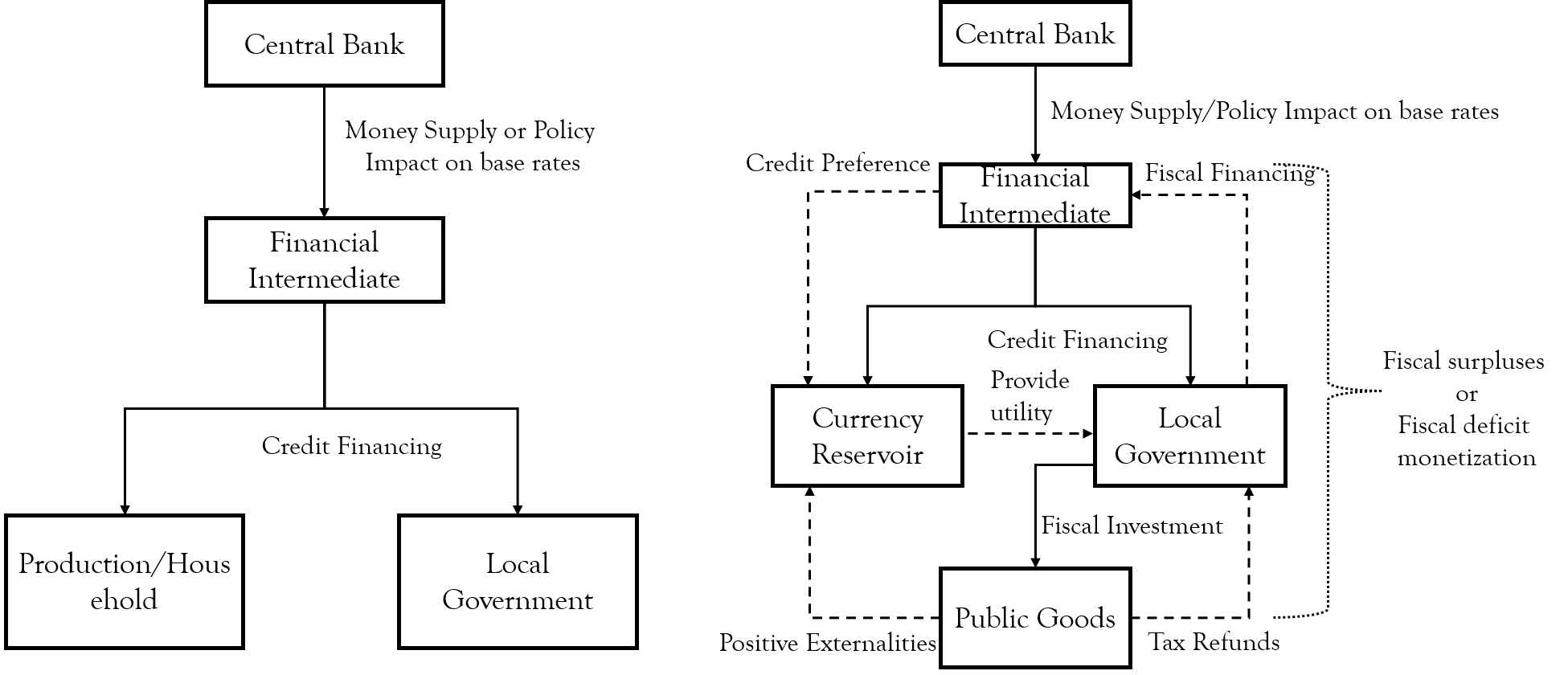}}   \caption{\label{1} Two channels of monetary transmission}   \end{figure}\newline
\indent 
%图2 左右两幅图分别描述传统货币传导机制和货币蓄水池-财政融资投资的货币传导机制。传统观点认为由金融中介为生产部门、家庭部门和各级政府提供资金融通服务，中央银行通过多种政策变量对金融中介的资金成本和可贷资金等产生影响，因此传统传导机制是单方向且自上而下的信贷供给过程；货币蓄水池-财政融资投资货币传导机制与传统机制的最大不同在于考虑了地方政府同样可以财政赤字货币化，随着财政投资的边际收益递减，财政赤字增加将促使政府债务货币化，形成与传统路径截然不同的货币传导机制。
In figure 2 the left and right panels depict the traditional monetary transmission mechanism and the monetary reservoir-fiscal financing and investment monetary transmission mechanism respectively. The traditional view is that financial intermediaries provide financing services to the productive sector, the household sector and the government at all levels, and that the central bank influences the cost of capital and loanable funds of financial intermediaries through various policy variables. Therefore, the traditional transmission mechanism is a unidirectional and top-to-bottom money supply process. The biggest difference between these two mechanism is that the second one takes into account that local governments can also monetize fiscal deficits. As the marginal returns of fiscal investment decrease, fiscal deficits increase, which will promote the monetization of government debt and form the currency transmission mechanism that is completely different from the traditional path.
 \newline
 %中央银行外货币增发的观点可能会受到如此辩驳：如果中国房地产部门存在中央银行体系外的货币增发，为何房价与货币供应量存在显著的相关性而非呈现独立变动甚至逆周期的特征？不可否认的是，货币蓄水池资产的价格涨落与中央银行的货币宽松政策有很强的相关性，但是两者的运行周期相关性并不强。为了度量运行周期，本文对增长率序列进行傅里叶变换将原始信号转换为频率及其对应振幅，根据振幅的大小对相位进行顺序排列(由于对称性只考虑绝对值大于0的部分)，最后通过计算两变量同等频率对应的秩(振幅大小排序)相关系数，结果列示于表1中。表1结果表明，中美两国的房市和股市都与货币供应量和价格型政策变量有显著的相关性，其中中国房市溢价增幅与M2货币供应量具有很强的周期相关性，而住房价格与M2货币供应量的周期相关性并不明显(负周期相关性也意味着弱周期相关)。作为对比，美国股市指数与市场利率具有与CPI/PPI更强的周期相关性，美国联邦储备局自2008年以来量化宽松及财政赤字货币化的货币超发推动消费品和工业品价格水平的上升，因此同时具有价格序列和运行周期的强相关性(Iacoviello,2010)，货币供应量作为中国主要的货币政策调控变量并不对各部门的价格水平运行周期产生明显的影响，这起码说明仅从中央银行角度理解货币增发并不全面。
\indent The argument for over-issuance out of central bank might be argued as follows: If there is the over-issuance existing in the china's real estate sector instead of the central bank, why is there a significant correlation between housing prices and money supply instead of independent changes or even counter-cyclical characteristics? It is undeniable that there is a strong correlation between the price fluctuations of currency reservoir assets and the central bank's monetary easing policy, but the correlation between the two operating cycles is not strong. In order to measure the operating period, this paper performs Fourier transform on the growth rate sequence to convert the original signal into frequency and its corresponding amplitude, and arranges the phases in order according to the magnitude of the amplitude (due to symmetry, only the part whose absolute value is greater than 0 is considered), and finally By calculating the rank (amplitude order) correlation coefficient corresponding to the same frequency of the two variables, the results are shown in Table 1. The results in Table 1 show that both the housing and stock markets in China and the US are significantly correlated with money supply and price-based policy variables. Among them, the growth of China's housing market premium has a strong cyclical correlation with M2 money supply, while housing prices have a strong cyclical correlation. The cyclical correlation with M2 money supply is not obvious (negative cyclical correlation also means weak cyclical correlation). In contrast, the U.S. stock market index and market interest rates have a stronger cyclical correlation with CPI/PPI. Since 2008, the U.S. Federal Reserve’s loose monetary pocily and monetary over-issue of fiscal deficit monetization have driven the price level of consumer and industrial products to rise. Therefore, there is a strong correlation between the price series and the operating cycle at the same time [Iacoviello(2010)\citep{iacoviello2010housing}]. As China's main monetary policy control variable, money supply does not have a significant impact on the operating cycle of price levels in various sectors, which at least shows that it is not comprehensive to understand the additional issuance of money only from the perspective of the central bank.
\begin{table}[]
\resizebox{\textwidth}{!}{
\begin{threeparttable}
\begin{tabular}{ccccccccc}
\hline
  & & CPI     & PPI   & SSE index    & $P_h$ &  $\frac{P_h}{P_{s,c}}$ & $\frac{P_h}{PPI}$ & $\frac{P_h}{CPI}$   \\  
  \hline 
$M_2$& Corr.& -0.19 & -0.02 & 0.016& -0.134& -0.052& -0.071& -0.009 \\$M_2$  & Rank Corr. & -0.10 & -0.34   & -0.098 & -0.310 & 0.229& 0.074 & 0.130  \\

\hline
  & & CPI     & PPI   & SSE index    & $P_s$ &  $\frac{P_s}{P_h}$ & $\frac{P_s}{PPI}$ & $\frac{P_s}{CPI}$   \\  
  \hline 
Interest Rate& Corr.& 0.03 & 0.137 & -0.091& -0.440& -0.097& -0.108& -0.086 \\Interest Rate & Rank corr. & 0.043 & 0.04  & 0.044 & 0.075 & -0.01& -0.092 & -0.096   \\
\hline
\end{tabular}
\begin{tablenotes}
\item[1]$P_h$ denotes the house price index made by data in over 30 cities in China.$P_{s,c}$ is the SSE index.
\item[2]$P_s$ denotes the Dow Jones industrial index. $P_h$ is the house price index in U.S.
\end{tablenotes}
\end{threeparttable}
}
\caption{Comparison of  stock market indices and house price levels in China and U.S.}
\end{table}
\begin{figure}[htb]   
\centering
\includegraphics[width=10cm]  {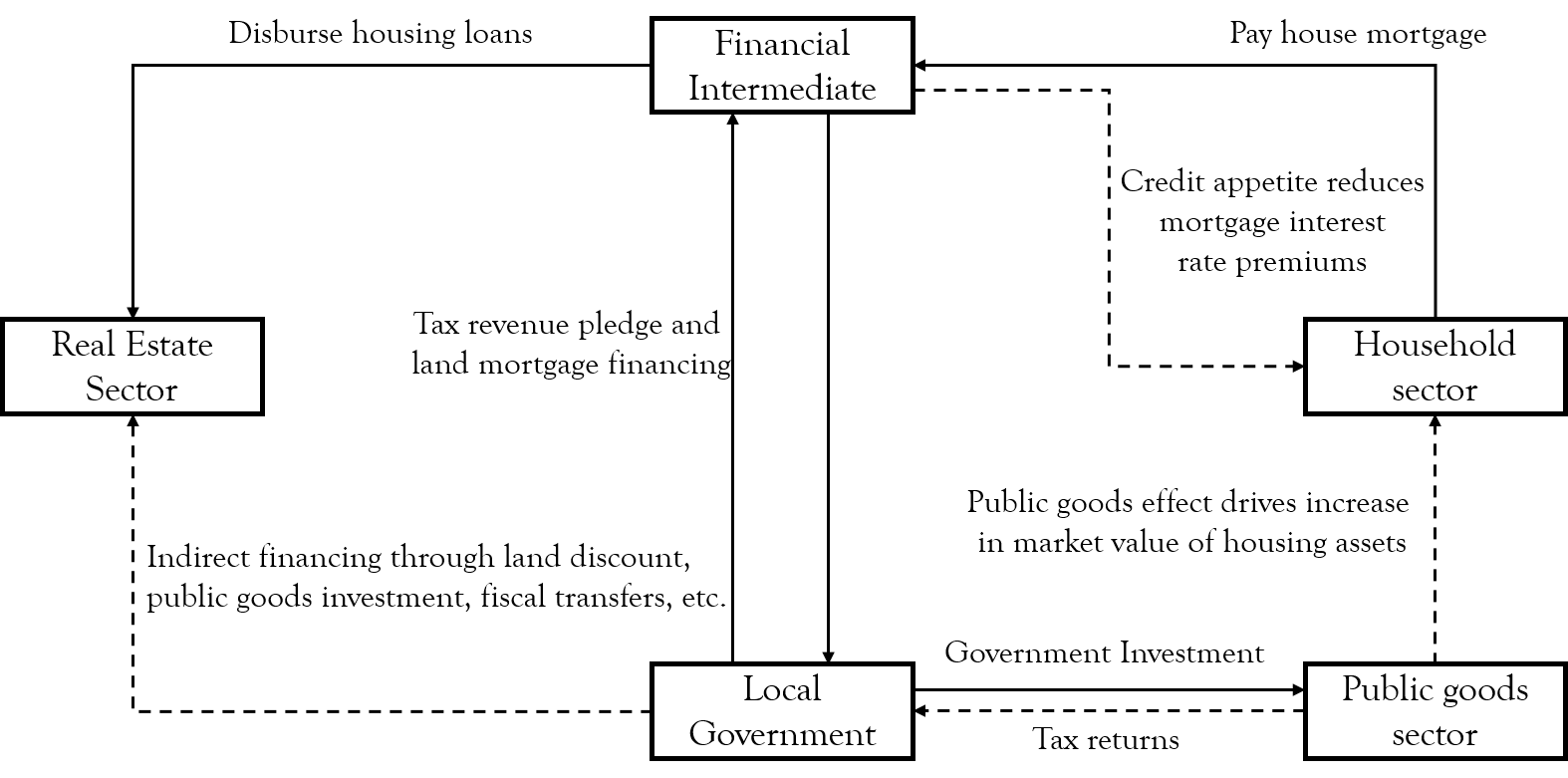}   \caption{\label{1}  Diagram of the process of increasing money in the real estate money reservoir}   \end{figure}
\newline
%图3描述了金融中介、房地产部门、家庭部门、地方政府以及公共品部门的关系，图中直线部分表示明显可见的资金流关系，即房产买卖关系，而虚线表示隐蔽的资金流关系，表示地方政府信用扩张。地方政府并没有参与房产买卖，但是地方政府通过两种途径影响房地产市场，一方面地方政府通过土地转让金折价、加大配套设施建设(例如引进商场、修建公共交通系统)等方式保证房产功能的实现和增值，另一方面地方政府通过财政融资向金融中介筹措资金投向公共品部门，公共品部门的正外部性同时提高了所有生产部门的生产效率，占公共品投资较大份额的基础设施的最大受益方仍是家庭部门和房地产部门(相比房地产部门，非房地产部门市值受基础设施正外部性的影响并不直接)。此外，房地产部门具有的杠杆性也意味着高价值，更有利于短期提高官员绩效，地方政府和金融中介在共同利益下通过财政融资为房地产部门提供更多贷款，日渐增加的房地产部门杠杆造就房地产的繁荣周期，但是过多金融资源向房地产部门的倾斜导致金融部门资源配置效率低下，信贷偏好在长期必然导致生产效率的边际递减。由于地方政府事权与财权的不平衡以及重短期成绩而忽视长期效益的官员绩效考核制度，地方政府形成路径依赖并形成债务扩张的循环，地方政府通过财政融资形成大量政府债务影响金融中介的资产负债表并在信贷偏好作用下信贷主要进入房地产部门，地方政府和房地产部门的杠杆率同时提高，长期的财政赤字推动债务货币化造成货币增发，构成所谓货币蓄水池机制。
\indent Figure 3 depicts the relationship among financial intermediaries, the real estate sector, the household sector, local governments and the public goods sector. The straight line in the diagram indicates the obvious visible financial flow relationship, i.e. the buying and selling of property, while the dotted line indicates the hidden financial flow relationship, indicating the expansion of local government credit. Local governments are not involved in the sale of property, but they influence the real estate market in two ways: On the one hand, local governments guarantee realization and value-added of real estate functions by discounting land transfer fees and increasing the construction of supporting facilities (such as building shopping malls and building public transportation systems). On the other hand, local governments raise funds from financial intermediaries through financial financing to invest in the public goods sector. The biggest beneficiaries are still the household sector and the real estate sector (compared to the real estate sector, the market value of the non-real estate sector is not directly affected by the positive externalities of infrastructure). In addition, the leverage of the real estate sector also implies high value and is more conducive to improving the performance of officials in the short term. Local governments and financial intermediaries provide more loans to the real estate sector through fiscal financing in their mutual interest, and the increasing leverage of the real estate sector contributes to the boom cycle of real estate. In the long run, this leads to a marginal decrease in productivity. As a result of the imbalance between local government authority and financial authority and the performance appraisal system of officials that emphasises short-term achievements at the expense of long-term benefits, local governments form a path dependency and a cycle of debt expansion, with local governments forming a large amount of government debt through fiscal financing that affects the balance sheets of financial intermediaries and, under the effect of credit preference, credit mainly goes to the real estate sector, with the leverage of local governments and the real estate sector increasing simultaneously. Long-term fiscal deficits drive debt monetisation resulting in increased monetary issuance, constituting a so-called monetary reservoir mechanism.
\section{Model Setup}
%模型假设贷款品部门和非贷款品部门劳动要素占总产出的比例相同，即ϕ_h+ψ_h=ϕ_f+ψ_f，附录A的证明说明该假设只是为了方便结果展示，并不影响核心结论。模型中贷款品指的是交易制度允许部分支付，购买者可以用少数的成本购买高价值的商品。在本文中，贷款品的典型例子是房地产部门。
\indent The model assumes that the labor factor accounts for the same proportion of total output in the loan good and non-loan good sectors, i.e., $\phi_h + \psi_h = \phi_f + \psi_f$. The proof in Appendix A illustrates that this assumption is made only to facilitate the presentation of the results and does not affect the core conclusions. Loan goods in the model refer to transactions where the system allows partial payments and purchasers can buy commodities with a small cost. In this paper, a typical example of a loan good is the real estate sector.
\subsection{Mortgage contracts}
%借贷个体每期主动根据基准利率选择最佳贷款期限使得自身杠杆率最高从而保证每期还本付息金额最少。贷款合同被设计为：首付比例为θ_t，金融中介每期向债务人家庭收取每期等额的房贷款项，贷款合同总额为：
\indent The individual borrower actively chooses the best loan term for each period based on the prime rate to maximize his or her leverage and thus ensure the least amount of debt service per period. The loan contract is designed as follows: the down payment ratio is $\theta_t$, the financial intermediary charges the debtor’s family an equal amount of the mortgage in each installment. The total loan contract is
\begin{equation}
    (1-\theta_{t})P_{t}*y_t*(1+i_{t})^T=T*D
\end{equation}
%其中D表示每期等额偿还的房贷金额，杠杆率被定义为每期贷款总额与每期等额偿付数额的比值，定义杠杆率为ω_t，i_t表示贷款利率：
\indent where D denotes the amount of mortgage repaid in equal installments per period, the leverage ratio is defined as the ratio of the total loan amount per period to the amount repaid in equal installments per period, and the leverage ratio is defined as $\omega_t$ and $i_t$ denotes the loan interest rate.
\begin{equation}
    \omega_{t}=\frac{T}{(1-\theta_t)*(1+i_{t})^T}=\frac{P_{t}y_t}{D}
\end{equation}
%债务人首先选择最优的期限T，一阶条件为T^*≈1/i_t ，假设债务人选择在给定贷款总额P_h y_t条件下选择期限T最小，杠杆率最大并将其代入原式，可以求得ω_t约等于下式：
\indent The debtor first selects the optimal term T with the first-order condition that $T^{*}\approx \frac{1}{i_t}$. Assuming that the debtor chooses the minimum term T and the maximum leverage ratio under the condition of the given total loan amount $P_h y_t$, and ubstitutes it into the original equation. We can find $\omega_t$ is approximately equal to the following equation.
\begin{equation}
    \omega_{t} = \frac{\frac{1}{i_t}}{(1-\theta_t)*(1+i_t)^{\frac{1}{i_t}}}\approx \frac{1}{(1-\theta_t)*e*i_t}
\end{equation}
\subsection{Financial intermediaries}
%金融中介通过发放贷款获得贷款利息。贷款利率由下式表示:
\indent Financial intermediaries earn interest on loans by lending. The interest rate on loans is represented by the following equation. 
\begin{equation}
    i_{t}=f_h(\omega_{h,t})R_t,f_h(x<1)=1,f^{'}_{h}>0,f^{''}_h(.)<0
\end{equation}
%其中R_t为基准利率，ω_t表示杠杆率，f_h (ω_t )=ω_t^(ξ_t ),ξ_t>0，表示金融中介的贷款利率溢价随贷款杠杆单调递增。金融中介决定家庭住房再融资比例μ_t。μ_t与中央银行的货币供应量、经济繁荣程度、监管政策等因素相关，本文将其看做数量型货币政策外生冲击的代理变量。
\indent where $R_t$ is the base interest rate and $\omega_{t}$ denotes leverage rate. $f_h(\omega_{t})=\omega^{\xi_{t}}_{t}$, where ${\xi_{t}}>0$ denotes that the financial intermediary's mortgage interest rate premium increases monotonically with the mortgage leverage. Financial intermediaries determine the proportion of leveraged commodities refinancing $\mu_t$ and the coefficient of the mortgage interest rate premium $\xi_(t)$. $\mu_t$ is correlated with the central bank's money supply, economic prosperity and regulatory policy, and is regarded as a proxy variable for exogenous shocks to quantitative monetary policy and risk preference of financial intermediaries in this paper.\newline
%结合公式3，公式4可以重新表示为
\indent Combining Equation 3, Equation 4 can be re-expressed as :
\begin{equation}
    i_t=R^{\frac{1}{1+\xi_t}}_t[\frac{1}{(1-\theta_t)*e}]^{\frac{\xi_t}{1+\xi_t}}
\end{equation}
%将公式3和公式5结合，将首付比例为θ_t的杠杆率表示为：
\indent Combining Equation 3 and Equation 5, the leverage ratio for a down payment ratio of $\theta_t$ is expressed as
\begin{equation}
    \omega_t=(1-\theta_t)^{\frac{\xi_t}{1+\xi_t}-1}e^{-1-\frac{\xi_t}{1-\xi_t}R^{-\frac{1}{1+\xi_t}}_t}
\end{equation}
\subsection{Representative households}
%本文考虑无限期生存的家庭，每期选择购买消费品C_t、贷款品和非贷款品的消费量Y_(h,t)和Y_(f,t)，提供劳动N_t并持有货币M_t和借贷B_t：
\indent In this paper, we consider households surviving indefinitely, choosing in each period to purchase consumption of consumption goods $C_t$, loan and non-loan goods $Y_(h,t)$ and $Y_(f,t)$, providing labor $N_t$ and holding money $M_t$ and borrowing $B_t$.
\begin{equation}
Max:E\sum^\infty_{t=0}{\frac{C^{1-\sigma}_t}{1-\sigma}+\chi*(M_t-B_t)+j_h*log[Y_{h,t}]+j_f*log[Y_{f,t}]-\kappa*N_t}
\end{equation}
%其中E表示预期，M_t为居民持有的货币余额，σ是家庭的风险厌恶系数，β为贴现因子，j为住房偏好，κ表示家庭对劳动的厌恶程度。代表性家庭分别向非贷款品f和贷款品h两个部门提供劳动n_(f,t)和n_(h,t)，代表性家庭的劳动总供给约束设为n_t，两种劳动之间对代表性家庭带来的效用是等同的：
\indent where E denotes expectations, $M_t$ is the money balance held by residents, $\sigma$ is the household's risk aversion coefficient, $\beta$ is the discount factor, $j_h$and $j_f$ is the preference of leveraged and non-leveraged commodities respectively. $\kappa$ denotes the household's aversion to labor. The representative household supplies labor $n_{f,t}$ and $n_{h,t}$ to two sectors, non-loan good f and loan good h, respectively, and the total labor supply constraint for the representative household is set to $n_t$. The utility to the representative household is equivalent between the two types of labor.
\begin{equation}
    N_t=n_{f,t}+n_{h,t}
\end{equation}
%家庭部门的预算约束为：
\indent The budget constraints for the household sector are.
\begin{equation}
\begin{aligned}
    &M_t+C_t+\frac{E(\sum^{+\infty}_{k=t+1}\frac{P_{h,t}Y_{h,t}}{(1+R_k)^k})}{\omega_{h,t}}+P_{f,t}Y_{f,t}+\theta_t P_{h,t}Y_{h,t}+i_{t-1}B_{t-1}\\
    &=(1+R_{t-1})M_{t-1}+B_t+W_{f,t}n_{f,t}+W_{h,t}n_{h,t}
\end{aligned}
\end{equation}
%其中E(∑_(k=t+1)^(+∞)▒(P_(h,t) Y_(h,t))/ρ^k )/ω_(h,t) 表示按照贷款合同还本付息的支出折现值，P_(f,t)和P_(h,t)分别表示非贷款品和贷款品产品的价格。代表性家庭还需要偿还上一期的借款利息i_(t-1) B_(t-1)和支出贷款品的首付额θ_t P_(h,t) Y_(h,t)。同时代表性家庭得到劳动收入W_(f,t) n_(f,t)+W_(h,t) n_(h,t)。
\indent where $\frac{E(\sum^{+\infty}_{k=t+1}\frac{P_{h,t}Y_{h,t}}{(1+R_k)^k})}{\omega_{h,t}}$ denotes the discounted value of the expenditure on debt service according to the loan contract, and $P_{f,t}$ and $P_{h,t}$ denote the prices of the non-loan and loan products, respectively. The representative household also needs to repay the interest on the previous period's borrowing $i_{t-1} B_{t-1}$and the down payment on the expenditure loan good $\theta_t\ P_{h,t} Y_{h,t}$. Meanwhile the representative household receives labor income $W_{f,t} n_{f,t} + W_{h,t} n_{h,t}$.
%本文考虑了将μ_t比例的累积贷款品作为贷款额度的借贷约束，其中μ_t表示贷款品的再融资比例：
\indent In this paper, we consider the borrowing constraint of using $\mu_t$ proportion of cumulative loan items as the loan amount, where $\mu_t$ denotes the refinancing proportion of loan items.
\begin{equation}
    B_{k,t}<=M_t+\mu_t*\sum^t_{k=0}{P_{h,k}(Y_{h,k})}
\end{equation}
%家庭决定消费、两个部门的劳动、货币和借贷、以及两大部门产品的一阶条件：
\indent Households determine consumption, labor in both sectors, money and borrowing, and first-order conditions for the products of both sectors.
\begin{equation}
    C^{-\sigma}_t=\lambda_t
\end{equation}
\begin{equation}
    -\kappa*\frac{1}{n_{h,t}}+\lambda_t W_{h,t}+\mu_t=0,-\kappa*\frac{1}{n_{f,t}}+\lambda_t W_{f,t}+\mu_t=0
\end{equation}
\begin{equation}
    \chi-\lambda_t+\beta\ E(\lambda_{t+1})[1+R_t]=0
\end{equation}
\begin{equation}
-\chi+\gamma_t+\lambda_t-\beta\ i_t\ E(\lambda-{t+1})
\end{equation}
\begin{equation}
\frac{j_h}{Y_{h,t}}-E(\sum^{+\infty}_{k=t}{\frac{\beta^k\lambda_k}{(1+R_k)^k}})\frac{P_{h,t}}{\omega_{h,t}}-\theta_t\ P_{h,t}+\gamma_t\mu_t\ P_{h,t}=0
\end{equation}
\begin{equation}
\frac{j_f}{Y_{f,t}}-\ P_{f,t}=0
\end{equation}
%其中λ_t，γ_t和u_t分别是预算约束、信贷约束和劳动供给约束的拉格朗日乘子。整理贷款品部门和非贷款品部门得到关于两大部门的需求曲线：
\indent where $ \lambda_t, \gamma_t and \mu_t$ are Lagrange multipliers for the budget constraint, the credit constraint and the labor supply constraint, respectively. Collapsing the loan goods sector and the non-loan goods sector yields the demand curves with respect to the two major sectors.
\begin{equation}
    P_{f,t}Y_{f,t}=j_f
\end{equation}
\begin{equation}
    P_{h,t}Y_{h,t}=\frac{j_h}{\theta_t+E(\sum^{+\infty}_{k=t}{\frac{\beta^k\lambda_k}{(1+R_k)^k}})-\gamma_t\mu_t}
\end{equation}
%在满足家庭部门所有的一阶条件下，由于P_(h,t) Y_(h,t) 正比于ω_(h,t)本文假设θ_t=γ_t μ_t以化简公式方便读者理解。
\indent Under all first-order conditions satisfying the household sector, since $P_{h,t} Y_{h,t}$ is proportional to $\omega_{h,t}$ this paper assumes that $\theta_t = \gamma_t \mu_t$ to simplify the formula for the reader's understanding.
\begin{equation}
    P_{h,t}(Y_{h,t}=\frac{j_h}{E(\sum^{+\infty}_{k=t}{\frac{\beta^k\lambda_k}{(1+R_k)^k}})}
\end{equation}
\subsection{Leveraged commodities production sector}
%贷款品企业生产除了需要资本以外，每期还雇佣劳动n_(h,t)和购买土地L_(h,t)，同时贷款品部门和非贷款品部门的生产函数都受益于公共品部门的正外部性Φ_t。Φ_t表示无需增加企业增加要素投入也能增加的产出。对应的生产函数如下：
\indent In addition to the capital required to produce the loan good firm, it also employs labor $n_{h,t}$ and purchases land $L_{h,t}$ in each period, while the production functions of both the loan good and non-loan good sectors benefit from the positive externality $\Phi_t$ of the public good sector. $\Phi_t$ denotes the output that can be increased without increasing the firm's additional factor inputs. The corresponding production functions are as follows.
\begin{equation}
    Y_{h,t}=\Phi_t K^{\rho_h}_{h,t}L^{\psi_h}_{h,t}n^{1-\rho_h-\psi_h}_{h,t}
\end{equation}
%ϕ_h,ψ_h∈(0,1)，其中ψ_h是土地的要素报酬在总要素中的比重， ψ_h越大说明土地在生产函数中的作用越大； ϕ_h是资本回报在总的要素报酬中的份额。
\indent $\psi_h$ is the share of rewards of land in total factors, and a larger $\psi_h$ indicates a larger role for land in the production function, while $\phi_h$ is the share of returns to capital in total factor rewards. The real estate sector satisfies the following conditions in its profit maximisation.
%定义利润函数如下：
\indent Define the profit function as follows.
\begin{equation}
    \prod_{h,t}=P_{h,t}Y_{h,t}-W_{h,t}n_{h,t}-PL\ L_{h,t}-i^P_{h,t}K_{h,t}
\end{equation}
%W_(h,t)是单位劳动的工资，PL_(h,t)是土地价格，由地方政府外生给定，i_(h,t)^p是对贷款品生产商的资本借贷利率，选择三种生产要素使得利润函数最大化
\indent $W_{h,t}$ is the wage per unit of labor, $PL_{h,t}$ is the price of land, exogenously given by the local government, and $i_{h,t}^p$ is the interest rate on capital borrowed from producers of loan goods, choosing three factors of production such that the profit function is maximized.
\begin{equation}
    W_{h,t}=\frac{P_{h,t}*(1-\phi_h-\psi_h)}{n_{h,t}*Y_{h,t}},PL_{h,t}=\frac{P_{h,t}*\psi_h Y_{h,t}}{L_{h,t}},PK_{h,t}=\frac{P_{h,t}*\phi_h*Y_{h,t}}{n_{h,t}}
\end{equation}
%根据金融中介的设定，资本价格等于借贷利率，杠杆率定义为总资产P_(h,t) Y_(h,t)与借入资本K_(h,t)的比值。$i^P_{h,t}=R_t(\omega^P_{h,t})^{\xi_t}=R_t(\frac{P_{h,t}Y_{h,t}}{K_{h,t}})^{\xi_t}=R_t(\frac{i^P_{h,t}}{\phi_h})^{\xi_t}$
\indent According to the financial intermediation setting, the price of capital is equal to the borrowing rate and the leverage ratio is defined as the ratio of total assets $P_{h,t} Y_{h,t}$ to borrowed capital $K_{h,t}$. Because $i^P_{h,t}=R_t(\omega^P_{h,t})^{\xi_t}=R_t(\frac{P_{h,t}Y_{h,t}}{K_{h,t}})^{\xi_t}=R_t(\frac{i^P_{h,t}}{\phi_h})^{\xi_t}$
\begin{equation}
    i^P_{h,t}=[\frac{R_t}{\phi^{\xi_t}_h}]^{\frac{1}{1-\xi_t}}
\end{equation}
\subsection{Non-Leveraged Commodities sector}
%非贷款品部门的一阶条件与贷款品部门的不同仅仅在于土地、劳动和资本各生产要素对产出的贡献程度不同：
\indent First-order conditions in the non-real estate sector differ from those in the real estate sector only in the extent to which each factor of production - land, labour and capital - contributes to production.
\begin{equation}
\scriptstyle
    W_{f,t}=\frac{P_{f,t}*(1-\phi_f-\psi_f)}{n_{f,t}*Y_{f,t}},PL_{f,t}=\frac{P_{f,t}*\psi_f Y_{f,t}}{L_{f,t}},PK_{f,t}=\frac{P_{f,t}*\phi_f*Y_{f,t}}{n_{f,t}},i^P_{f,t}=[\frac{R_t}{\phi^{\xi_t}_f}]^{\frac{1}{1-\xi_t}}
\end{equation}
\subsection{Public goods sector}
%本文定义G_t表示政府直接投资进行公共品建设。公共品包括基础设施、科技和人力资本积累等因素，TF_t表示对家庭部门的财政转移支付，对于公共品部门及财政转移满足以下条件：
\indent In this paper, we define $G_t$ to denote direct government investment for public goods construction. Public goods include factors such as infrastructure, science and technology, and human capital accumulation, and $TF_t$ denotes fiscal transfers to the household sector, for which the public goods sector and fiscal transfers satisfy the following conditions.
\begin{equation}
    \Phi_t=exp(\lambda^P_t[G_t-\delta_k\sum^{T-1}_0{G_i}])
\end{equation}
%其中Φ表示公共品为生产部门带来的正外部性，使生产部门的生产效率提高，Φ由G_t 、资本折旧率δ_k决定。公式中参数λ_t^P小于0时，公共品部门对生产部门生产效率的作用系数小于1，公共品部门反而导致同等要素投入的产出水平下降。由于总需求曲线不变，产出水平下降引起价格的上涨。
\indent where $\Phi_t$ denotes the positive externality that the public good brings to the production sector, causing the production efficiency of the production sector to rise, and $\Phi_t$ is determined by $G_t$ , the capital depreciation rate $\delta_k$.When the parameter $\lambda_t^P$ in the formula is less than 0, the coefficient of action of the public goods sector on the production efficiency of the production sector is less than 1. The public goods sector instead leads to a decrease in the level of output with the same factor inputs. As the aggregate demand curve remains unchanged, the fall in the level of output causes an increase in prices.
\subsection{Local government}
%本文将地方政府的目标决策函数分为两部分。地方政府的目标是当地经济总量和税收收入最大化。两种政策目标的权重分别为W_G和1-W_G。地方政府能够决定公共品投资G_t和土地供给L_t影响经济增长。本文将地方政府目标函数定为：
\indent In this paper, the objective decision function of the local government is divided into two parts. The objective of the local government is to maximize the total local economy and tax revenue. The weights of the two policy objectives are $W_G$ and $1-W_G$. Local governments are able to decide public goods investment $G_t$ and land supply $L_t$ to influence economic growth. In this paper, the local government objective function is set as.
\begin{equation}
\begin{aligned}
Max:E_t \sum^{+\infty}_{t}\beta^t_{G}[W_G(P_{h,t}\Phi_t K^{\rho_h}_{h,t}&L^{\psi_h}_{h,t}n^{1-\rho_h-\psi_h}_{h,t}+P_{f,t}\Phi_t K^{\rho_f}_{f,t}L^{\psi_f}_{f,t}n^{1-\rho_f-\psi_f}_{f,t})\\
&+(1-W_G)(PL_{h,t}\ L_{h,t}+PL_{f,t}\ L_{f,t})]
\end{aligned}
\end{equation}
%假设地方政府也面临一个零首付的贷款合同进行财政融资，财政杠杆为ω_(G,t)。地方政府用土地出让收入作为质押向金融中介借入资金，并预期在未来ω_(G,t)期每期等额还本付息{PL_(h,i) L_(h,i)+PL_(f,t) L_(f,t) }。
\indent Assume that the local government also faces a zero down payment loan contract for fiscal financing with fiscal leverage of $\omega_{G,t}$. The local government borrows funds from the financial intermediary using the land concession revenue as a pledge and expects to repay the debt in each period for  $\omega_{G,t}$ times at the equal cost of $PL_{h,i} L_{h,i}+PL_{f,t} L_[f,t]$.
\begin{equation}
    G_t+[PL_{h,t}\ L_{h,t}+PL_{f,t}\ L_{f,t}]\sum^{\omega_{G,t}}_{p=t}{\frac{1}{(1+i_t)^p}}<=[PL_{h,t}\ L_{h,t}+PL_{f,t}\ L_{f,t}]*\omega_{G,t}
\end{equation}
%其中
\indent where
\begin{equation}
   \omega_{G,t}=e^{-1-\frac{\xi_t}{1-\xi_t}R^{-\frac{1}{1+\xi_t}}_t},i_{G,t}=R^{\frac{1}{1+\xi_t}}_t\ e^{-\frac{\xi_t}{1+\xi_t}}
\end{equation}
%subsection:资本累积方程
\subsection{Capital accumulation equation}
\indent Each period the firm depreciates the remaining capital goods $(1 - \delta_K)$ at a depreciation rate $\delta_k$ to realise them, and the capital goods producer can likewise produce the final capital goods in conjunction with the newly increased investment $I_t$, with the final capital goods being produced in the next period. The capital accumulation equation is as follows.
\begin{equation}
    K_{t+1}=(1-\delta_k)K_t+I_{t+1}
\end{equation}
\subsection{Central bank}
\indent In most DSGE models, the central bank sets monetary policy according to the Taylor monetary rule.
\begin{equation}
    \frac{R_t}{\bar{R}}=[\frac{R_{t-1}}{\bar{R}}]^{\rho_R}[\frac{GDP_{t}}{\bar{GDP}}]^{\rho_Y}
\end{equation}
\subsection{Market clearing conditions}
%模型假定市场处于均衡状态时各个市场都出清，符合凯恩斯产出恒等式。劳动和土地供给是外生给定的。
\indent The model assumes that markets are in equilibrium when each market is cleared, consistent with the Keynesian output constancy equation. Labor and land supply are exogenously given.
\begin{equation}
    Y_{f,t}+Y_{h,t}=C_t+I_t+G_t
\end{equation}
\begin{equation}
    n_{f,t}+n_{h,t}=n_{f,t}
\end{equation}
\begin{equation}
    L_{h,t}+L_{f,t}=L_t
\end{equation}
\section{Calibration and Estimation of Parameters in Models}
%参数β和β_G分别表示居民和地方政府的主观贴现率，根据经典DSGE模型，设置为0.98；资本品取折旧率〖δ﷩k〗约为0.01；劳动厌恶参数κ设定区间范围为1-2之间，本文取1.2。参考于经典的金融加速器模型，家庭部门风险规避系数σ设为0.9。两种商品偏好j都取为0.2。根据2020年4月到2021年2月的住房市场数据 ，首付比例设为0.2833，由首套房贷款利率在稳态下通过模拟矩估计计算金融中介对家庭购房贷款利率溢价水平为〖E(ξ〗_h)≈0.0357。使用1991~2015年中央银行对金融机构再贷款一年期利率和GDP数据估计得 ρ_R≈0.9929,ρ_Y≈0.0071。
\indent The parameters $\beta$ and $\beta_G$ denote the subjective discount rates of residents and local governments respectively, and are set to 0.98 according to the classical DSGE model. The depreciation rate $\delta_k$ for capital goods is taken to be about 0.01. The labour supply elasticity $\kappa$ is taken to be 1.2 in this paper. Referring to the classical financial accelerator model, the household sector risk aversion coefficient $\sigma$ is set to 0.9. $j_h$  and $j_f$ are taken as 0.2 both and labour preference $\kappa=0.1$. Personal income tax rate $T_g$ and corporate income tax rate $T_q$ are both taken as 0.05. Based on housing market data from April 2020 to February 2021, the down payment ratio is set to 0.2833 and calculated from the first home loan interest rate in steady state by simulating moment estimation of financial. The level of the interest rate premium of intermediaries on home purchase loans is $E(\xi_{h,t})\approx 0.0357$. The preference of government investment $\gamma_d$ is 0.5. Using the one-year interest rate on refinancing loans from the central bank to financial institutions and GDP data from 1991 to 2015, we estimate $\rho_R \approx 0.9929$,$\rho_Y \approx 0.0071$.\newline
%非贷款品部门生产函数中资本品份额ϕ_f为0.3，土地份额ψ_f为0.5，贷款品部门土地份额ψ_h为0.2，资本品份额ϕ_h为0.4。住房抵押贷款比例以及利率溢价系数在给定一阶最优条件下构造矩条件并使用Bayes公式和基准贷款利率、首付比例以及溢价水平的先验分布估计目标参数的后验参数，具体结果列示与表3中。
\indent The non-real estate sector production function has a capital goods share $\phi_{f}$ of 0.5 and a land share $\psi_{f}$ of 0.2, a land share $\psi_{h}$ of 0.3 and a capital goods share $\phi_{h}$ of 0.4 for the real estate sector. The home mortgage ratio and the mortgage rate premium coefficient are constructed given first-order optimal conditions and the moment conditions are estimated using the Bayes formula and the prior distributions of the benchmark loan rate.The posterior parameters of the target parameters are estimated using the Bayes formula and the prior distributions of the benchmark loan rate, the down payment ratio and the premium level, and the results are presented in Table 2.
\begin{table}[]
    \centering
\resizebox{\textwidth}{!}{
    \begin{tabular}{cccc}
    \hline
        Params & Prior distribution&Prior Mean& Prior Std. 
        \\
        \hline 
        $R$ & Beta&3.2$\%$& 1.092$\%$ \\
        $\theta$& Beta&0.2833&0.0471  \\
        $f_h(\omega_{h,t})$ & Gamma & 1.1275 & 0.0694  \\
        \hline
        Params & Posterior distribution & Prior Mean& Prior Std.  
        \\
        \hline
        $\mu$ &  Beta&3.36$\%$& 1.19$\%$ \\
        $\xi_h$&Beta& 3.22$\%$& 1.53$\%$  \\
        \hline
    \end{tabular}}
    \caption{Bayes Estimation Result of Important Parameters}
    \label{tab:my_label}
\end{table}
%由于上述方程组是非线性方程，一般使用投影法或者扰动法求解。Mikkel & Christian(2021)证明投影法和使用VAR预测的脉冲响应函数是渐进一致的。因此本文生成变量R_t、θ_t、μ_t和ξ_t的随机冲击，然后根据上述一般均衡条件求解出各个内生模型变量，通过随机模拟生成N期数据，然后对生成数据进行VAR回归，根据BIC准则选择最优滞后阶数。最后使用VAR模型推导脉冲响应函数和进行预测误差分解。

\indent As the above set of equations are non-linear, they are typically solved using either the projection method or the perturbation method. Mikkel \& Christian (2021)\citep{plagborg2021local}demonstrate that the impulse response functions projected by the projection method and using the VAR are asymptotically consistent. This paper therefore generates stochastic shocks for the variables $R_t, \theta_t, \mu_t$ and $\xi_t$, then solves for each endogenous model variable according to the general equilibrium conditions described above, generates N periods of data through stochastic simulation, and then performs a VAR regression on the generated data to select the optimal lag order according to the BIC criterion. Finally, the VAR model is used to derive the impulse response function and to perform the forecast error decomposition.
%section:货币蓄水池的溢价来源
\section{Premium source of monetary reservoir}
%使用%(x)表示变量x的增长率，比较房地产部门(贷款品)和非房地产部门的市值，房地产部门市值相对于非房地产部门的增幅溢价主要由以下方程决定：
\indent Using $\%(x)$ to denote the growth rate of variable $x$ and comparing the market value of the real estate(Leveraged commodities) sector with that of the non-real estate sector, the increase premium of the market value of the real estate sector relative to the non-real estate sector is mainly determined by the following equation.
\begin{equation}
    \%(\frac{P_{h,t}Y_{h,t}}{P_{f,t}Y_{f,t}})=\%(C^\sigma_t\omega_{h,t}\frac{(1+R_t)*j_h}{R_t})
\end{equation}
%简言之，造成房地产部门相对于非房地产部门的市值溢价主要是由家庭部门购买房地产部门产品所使用的杠杆导致的，杠杆增幅越大，房地产部门的市值相对增幅越大。由ω_(h,t)=(1-θ_t )^(ξ_t/(1-ξ_t )-1) e^((-1)/(1-ξ_t )) R_t^(-1)，房贷杠杆主要由首付比例θ_(h,t)、利率R_t以及房贷利率系数ξ_(h,t)决定，首付比例可以认为在短期是相对稳定的，但是利率和房贷利率溢价系数却可以由中央银行和金融中介按不同情况决定，因此ω_(h,t)在短期的增长率可以分解为房贷利率溢价系数和基准利率两个来源：
\indent In short, the market value premium of the real estate sector relative to the non-real estate sector is caused mainly by the leverage used by the household sector to purchase products in the real estate sector, and the larger the increase in leverage, the larger the relative increase in the market value of the real estate sector. By $\omega_(h,t)\approx [R_t*{(1-\theta_{h,t})}]^{\frac{-1}{1+\xi_{t}}}$, mortgage leverage is mainly determined by the down payment ratio $\theta_{h,t}$, the interest rate $R_t$, and the mortgage interest rate coefficient $\xi_{h,t}$. The down payment ratio can be considered to be relatively stable in the short run, but the interest rate and the mortgage interest rate premium coefficient however, can be determined by the central bank and financial intermediaries on a case-by-case basis, so that the growth rate of $\omega_{h,t}$ in the short run can be decomposed into two sources: the mortgage rate premium coefficient and the prime rate.
\begin{equation}
    \%(\omega_{h,t})=\%[(1-\theta_t)^{\frac{\xi_t}{1+\xi_t}-1}e^{-1-\frac{\xi_t}{1-\xi_t}R^{-\frac{1}{1+\xi_t}}_t}]
\end{equation}
%上述公式的经济意义是，房贷杠杆将上升并推动房地产部门市值上涨，这解释了货币蓄水池资产价格的市值膨胀现象。简而言之，一种资产具有货币蓄水池功能的必要条件是交易制度存在杠杆，即贷款品。住房和股票都具有此种特征，住房资产由于其高价值必然要求首付加定期还款的交易方式，天然具有杠杆，股票市场的交易活动虽然在极大多数国家均为全额结算，但是机构和专业投资者能够凭借其资金优势充分运用多种金融工具达到杠杆交易的目的，例如美国股市以机构和专业投资者为主，因此其股票市场成为继房地产市场之后的又一个货币蓄水池，而中国股市仍以散户投资者为主，此外，房地产市场仍有更低成本的杠杆交易制度(房贷合同)，所以中国房市而非中国股市成为货币蓄水池。
\indent The economic implication of the above equation is that  mortgage leverage will rise and drive up the market value of the real estate sector, which explains the phenomenon of market value premium in the price of a monetary reservoir asset. In short, a necessary condition for an asset to function as a monetary reservoir is the presence of leverage in the trading system. Both housing and equities have this characteristic, with housing assets being naturally leveraged due to their high value, which inevitably requires a down payment plus regular repayments, and equity markets, where trading activity is fully settled in most countries, but institutional and professional personal investors are able to use their capital advantage to fully leverage a wide range of financial instruments, such as the US stock market, which is dominated by institutional and professional investors. For example, the US stock market is dominated by institutional and professional personal investors, thus making its stock market another money reservoir after the real estate market, whereas the Chinese stock market is still dominated by retail investors and therefore there is merely the minority of investors can be accessible to leverage their equities. In addition, the real estate market still has a lower cost leveraged trading system (mortgage contracts) compared to stock market, thus making the Chinese housing market, rather than the Chinese stock market, a monetary reservoir.
\begin{equation}
\frac{P_{h,t}}{P_{f,t}}\propto R^{\frac{\phi_h-\phi_f}{1-\xi_t}}_t\times L^{\psi_f-\psi_h}_t\times (C^\sigma_t\omega_{h,t}\frac{(1+R_t)}{R_t})^{1-\phi_h-\psi_h+\psi_f-\psi_h}
\end{equation}
\begin{table}
    \centering
    \resizebox{\textwidth}{!}{
    \begin{tabular}{ccc}
    \hline
        Signals &Factors&Coefficients  \\
        \hline
         $\%(\omega_{h,t})$&Percentage change in leverage of leveraged commodities&$1-\phi_h-\psi_h+\psi_f-\psi_h$\\
         $\%R_t$&Percentage change in baseline interest rate &$\frac{\phi_h-\phi_f}{1-\xi_t}$\\
         $\%L_t$&Percentage change in land supply&$\psi_f-\psi_h$\\
         $\%E(\sum^{+\infty}_{k=t}{\frac{\beta^k\ C^{-\sigma}_k}{(1+R_k)^k}}$&%预期消费边际效用折现值
Percentage change in land price&$\phi_h+\psi_h-\psi_f+\psi_h-1$\\
         \hline
    \end{tabular}}
    \caption{Decomposition of the weighting of the currency reservoir sector premium factor}
    \label{tab:my_label}
\end{table}
%根据公式对两部门价格之比的动态变化进行分解，货币蓄水池资产相对于非房地产部门的溢价涨幅由货币蓄水池资产杠杆、基准利率、劳动供给以及土地供给四个因素决定。根据幂指数的性质，本文将各因素变动率对溢价涨幅百分比的影响整理在表3，
\indent Decomposing the dynamics of the ratio of prices between the two sectors according to the formula, the premium increase of money pool assets relative to the non-real estate sector is determined by four factors: money pool asset leverage, benchmark interest rate, labor supply, and land supply. Based on the nature of the power index, the effect of the rate of change of each factor on the percentage of premium increase is organized in Table 3, in this paper.
%μ_t主要由金融中介可贷资金决定， ξ_(t)可以作为数量型货币政策的代理变量，这为中国房市货币蓄水池的形成提供了一种解释，当中央银行超发货币(如输入型通胀或财政赤字货币化)，金融中介将适度调高住房抵押贷款比例以获取更高利润，当期房贷杠杆将上升并推动房地产部门市值膨胀，这为房地产部门持续繁荣的现象提供一种解释。
\indent $\mu_t$ is mainly determined by the loanable funds of financial intermediaries. $\xi_{t}$ can be used as a proxy variable for quantitative monetary policy, which provides an explanation for the formation of a monetary reservoir in the Chinese housing market. When the central bank over-issues money (e.g. imported inflation or monetisation of the fiscal deficit), financial intermediaries will moderately increase the proportion of housing mortgages to obtain higher profits, and the current mortgage leverage will rise and stimulate more premium of monetary reservoir assets, which provides an explanation for the continued boom in the real estate sector.\newline
\begin{table}[]
    \centering
    \begin{threeparttable}
    \resizebox{\textwidth}{!}{
    \begin{tabular}{cccc}
    \hline
    Year&$I_h/GDP$&$A_h/A$&$\omega$\\
\hline
       2010 & 11.71\% & 55.56\% & 27.3 \\
2011 & 12.65\% & 53.82\% & 27.9 \\
2012 & 13.33\% & 55.94\% & 30   \\
2013 & 14.51\% & 53.99\% & 33.5 \\
2014 & 14.77\% & 53.60\% & 36   \\
2015 & 13.93\% & 54.70\% & 39.2 \\
2016 & 13.74\% & 52.78\% & 44.7\\
\hline
Annual Growth Rate& 2.83\% & -0.81\% & 8.63\%\\
\hline
   Year&$LL_h/LL$&$NFL/LL$&$IH/LL_h$\\
   \hline
    2014 & 11.38\% & 23.27\% & 19.08\% \\
2015 & 11.76\% & 22.33\% & 17.54\% \\
2016 & 13.53\% & 21.91\% & 14.12\% \\
2017 & 15.07\% & 23.29\% & 12.50\% \\
2018 & 16.16\% & 23.97\% & 11.52\% \\
2019 & 17.21\% & 24.40\% & 10.62\% \\
2020 & 18.52\% & 25.38\% & 9.19\% \\
\hline
Annual Growth Rate& 8.52\% & -1.52\% & -11.35\%\\
\hline
    \end{tabular}}
    \centering
    \caption{Contribution of the real estate sector to economic growth and the level of credit to the household sector}
    \label{tab:my_label}
    \begin{tablenotes}
    \normalsize
    \item[1]{Data source: National Bureau of Statistics, National Balance Sheet Research Centre, Guotaian database, China Financial Yearbook (2020)}
    \item[2]{$I_h/GDP$ denotes Investment in property development
/GDP.$A_h/A$ is the residential sector housing assets
/Total Assets.$LL_h/LL$ represents the Medium and long-term loans to the residential sector/total loans to financial institutions,while $NFL/LL$ is the Medium and long-term loans to non-financial institutions/total loans to financial institutions.$IH/LL_h$ denotes Disposable income per urban household/medium and long-term loans to the residential sector.$\omega$ denotes residential sector leverage.}
    \end{tablenotes}
    \end{threeparttable}
\end{table}
%表4中数据所呈现的趋势特征与本文分析的结论基本一致。2010年以来城镇居民家庭人均收入可支配收入与中长期贷款的比例逐年下降，金融机构对居民部门中长期贷款的比例增长率也远高于对非金融部门，家庭部门正更多地透支其未来收入并提高杠杆率，其中住房资产占居民部门总资产的比例也稳定在50%左右，考虑到快速增长的居民部门杠杆率以及日益增加的房地产开发投资额占GDP比重，可以得出结论：家庭部门正扩大对住房资产的融资，而融资来源正是金融中介。住房资产购买的杠杆交易制度为房地产市场成为货币蓄水池提供基础条件，地方政府通过财政融资为家庭部门和金融中介提供信贷资金来源，金融中介的贷款溢价对可贷资金的依赖程度下降，为货币蓄水池资产价格和市值的逆周期繁荣提供动力，例如2015年12月至2019年6月美国联邦储备局将基准利率从0.25%提高到2.5%,但是同期的美国股市以及间接受美元流动性影响的中国房市价格水平都保持较高的增幅。根据上述模型大部分的繁荣周期动力来源于金融中介的贷款偏好和地方政府的信用扩张，金融中介产生这种信贷偏好的原因有多种，但最主要的来源于三方面：一是历史数据的确表明住房资产价格涨幅具有“一枝独秀”的特征 ，持有更多该种资产能保证贷款资产的增值性(Dong,2019)；二是住房贷款本质是抵押贷款，金融中介必定偏好抵押贷款而非信用贷款；三是住房资产作为固定资产的折旧和贬值风险在所有类型的固定资产中占有明显优势，比如住房资产主要对区位和与其绑定的公共福利等因素敏感，这也是一二线城市房地产市场比经济欠发达地区更繁荣的重要原因。金融资产也具有不折旧、杠杆性等，但是中国股票市场仍以散户投资者为主，衍生工具普及程度低，股票市场的杠杆性受到极大约束，并且股票市场并不具有第三方效用即不计入当地经济增长，地方政府并无意愿为其融资，因此吸纳的货币全部来源于中央银行的宽松货币政策，这解释了为何房地产部门是主要的货币蓄水池资产而股票市场呈现周期性繁荣。
\indent The data in Table 4 show similar trends that are generally consistent with the option in this paper: the ratio of disposable income per urban household to medium- and long-term loans has been declining year on year since 2010, and the proportion of medium- and long-term loans by financial institutions to the residential sector has been growing at a much higher rate than to the non-financial sector. Considering the rapidly growing leverage of the residential sector and the increasing amount of investment in property development as a share of GDP, it can be derived that the household sector is expanding its financing of housing assets, and that the source of financing is none other than financial intermediation. The leveraged trading system for housing asset purchases provides the underlying conditions for the real estate market to become a monetary reservoir. Local governments provide a source of credit funding for the household sector and financial intermediaries through fiscal financing. Financial intermediaries' loan premiums become less dependent on loanable funds, providing the impetus for a countercyclical boom in the prices and market value of money reservoir assets, such as the December 2015 to June 2019 US Federal Reserve raised its benchmark interest rate from 0.25$\%$ to 2.5$\%$, but both the US stock market and the Chinese housing market, maintained higher price levels over the same period. \newline
%许多学者研究土地财政对地方经济发展的融资作用(付敏杰等,2017; 黄少安,2012;刘勇政,2021)，因此土地财政本质上是一种地方政府信用扩张的政策工具，多地高价则债务多发，少地低价则债务少发。土地融资的直接抵押资产是土地但核心资产仍是税收收入。根据地方政府的预算约束公式(16)和(18)可知，通过地方政府的信用扩张ω_(h,t)，GDP总值将远高于一般水平，这为中国的高速发展提供了一种解释，即地方政府通过土地和未来税收担保为房地产这一高价值高杠杆的生产部门提供供需两方的融资，房贷合同等同于家庭部门对地方政府的支付承诺，因此房地产部门为地方政府提供预先 “兑现”经济增长的功能，由此也解释土地价格推动房价上升现象(Wu,2016)，地方政府使用财政融资推动高速增长，为了偿还债务和继续融资，必然推动土地价格上涨，土地成本的上升最终进一步推高了房价(梅冬州和温兴春,2021),这种发展方式的结果是居民部门的杠杆率和房地产开发投资额占GDP比重同时快速增长(表4)。
\indent Many Chinese scholars have studied the role of land finance in financing local economic development [Fu et al.(2017)\citep{Fu2017Evo};Huang(2012)\citep{Huang2012Renttax}], so land finance is essentially a policy tool for local government credit expansion, where more land at higher prices results in more fiscal income and fewer debt. The direct collateral asset for land finance is land but the essential asset is still tax revenue. According to the local government budget constraint equations (16) and (18), the total value of GDP will be much higher than average through local government credit expansion$\omega_{h,t}$, which provides an explanation for China's high growth rate, with mortgage contracts amounting to household sector commitments to the real estate sector through land and future tax guarantees. Therefore it provides the local government with the function of 'cashing in' on economic growth in advance, thus also explaining the phenomenon of land prices driving up house prices [Wu(2016)\citep{wu2016evaluating}], where the local government uses fiscal financing to drive high growth, which inevitably drives up land prices in order to service debt and continue financing. The rise in land costs eventually pushed up house prices further Mei(2021)\citep{Mei2021Housing}, and the result of this development approach was a rapid increase in the leverage of the residential sector and the share of real estate development investment in GDP at the same time (see Table 4).
%section:货币蓄水池对货币政策有效性的影响分析
\section{Impact by monetary reservoir on the efficiency of monetary policy}
%本文使用脉冲响应和误差方差分解两种形式分析货币蓄水池-财政融资投资的货币传导机制下价格型货币政策代理变量R_t、数量型货币政策代理变量μ_t以及住房市场主要变量如何对宏观经济产生影响。其中脉冲响应分析均采取一个标准差的上行冲击，图4描述了利率、住房再融资比例、首付比例以及房贷利率溢价系数四种外生冲击的脉冲响应分析结果。消费受到利率、住房再融资比例、首付比例、房贷利率溢价四种冲击的影响都与理论预期有所出入，其中利率和房贷利率溢价的冲击效应最显著，而再融资比例和首付比例的冲击效应则比较细微，主要原因是支付房贷的资金成本挤压了家庭部门的预算约束，而另外两种冲击则分别通过房贷杠杆和中央银行货币政策调整规则间接地影响消费。作为传统货币政策中介变量的利率，其冲击效应效应则分为家庭部门的消费和房贷杠杆两条传导机制。消费传导机制遵循利率-消费-工资-劳动力供应-生产部门的传导链条，而房贷杠杆渠道则是通过房贷杠杆-财政融资-财政转移渠道对经济系统产生冲击效应。与传统货币政策传导机制的不同之处在于，首付比例、房贷利率溢价和再融资比例等房地产市场的重要政策变量通过货币蓄水池资产的杠杆ω_(h,t)对房地产部门外的宏观经济系统产生冲击效应，这与国内外通过信贷渠道影响宏观经济的实证经验一致。
\indent This paper uses both impulse response and variance decomposition to analyse how the price-based monetary policy, the quantity-based monetary policy proxy and the main housing market variables affect the macroeconomy under a monetary transmission mechanism of monetary reservoir-fiscal financing-investment mechanism. The impulse response analyses all take one standard deviation of the upward shock. Figure 4-7 depict the results of the impulse response analysis for four exogenous shocks: interest rate, housing refinancing ratio, down payment ratio and mortgage interest rate premium coefficient. Consumption is affected by all four shocks - interest rate, housing refinancing ratio, down payment ratio and mortgage interest rate premium - in a way that deviates from theoretical expectations, with the shock effects of interest rate and mortgage interest rate premium being the most significant. Those of refinancing ratio and down payment ratio are more subtle mainly because the cost of funding mortgage payments squeezes the budget constraint of the household sector, while the other two shocks are indirectly affected through mortgage leverage, and monetary policy adjustment by central bank indirectly affect consumption. The effect of shocks on interest rates, which are the traditional monetary policy mediating variable, is divided into two transmission mechanisms: consumption in the household sector and mortgage leverage. The consumption transmission mechanism follows the interest rate-consumption-wage-labour supply-production sector transmission chain, while the mortgage leverage channel generates a shock effect on the economic system through the mortgage leverage - fiscal financing - fiscal transfer channel. The difference from the traditional monetary policy transmission mechanism is that important policy variables in the real estate market such as the down payment ratio, mortgage interest rate premium and refinancing ratio have effect on the macroeconomic system outside the real estate sector through the leverage of the monetary reservoir assets $\omega_{h,t}$, which is consistent with the empirical experience.
%R
\begin{figure}
    \centering
    \subfigure{
    \begin{minipage}[b]{0.25\linewidth}
    \includegraphics[width=3cm,height=2cm]{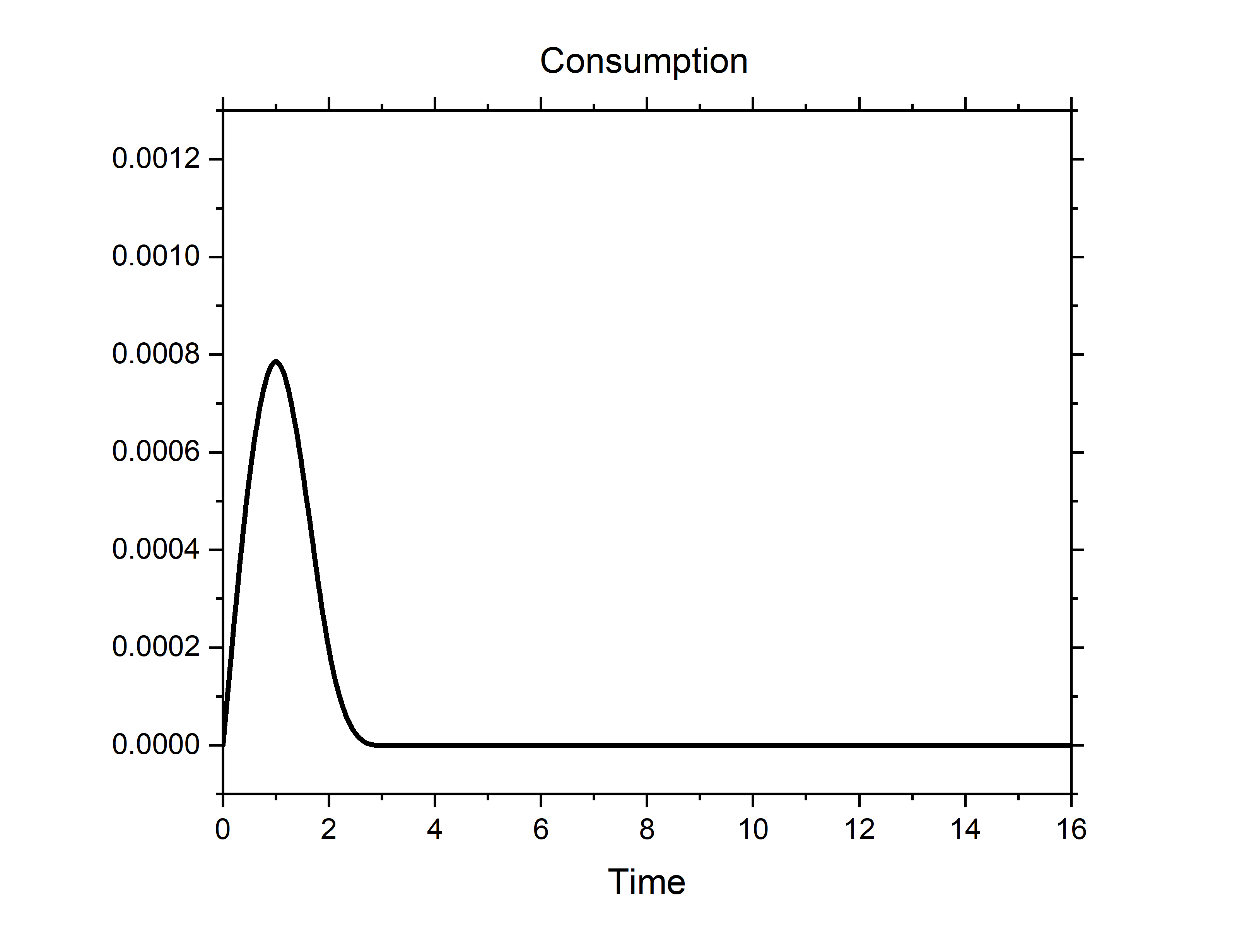}\hspace{-2cm}
    \caption*{Consumption}
    \end{minipage}
    }
\quad
 \subfigure{
    \begin{minipage}[b]{0.25\linewidth}
    \includegraphics[width=3cm,height=2cm]{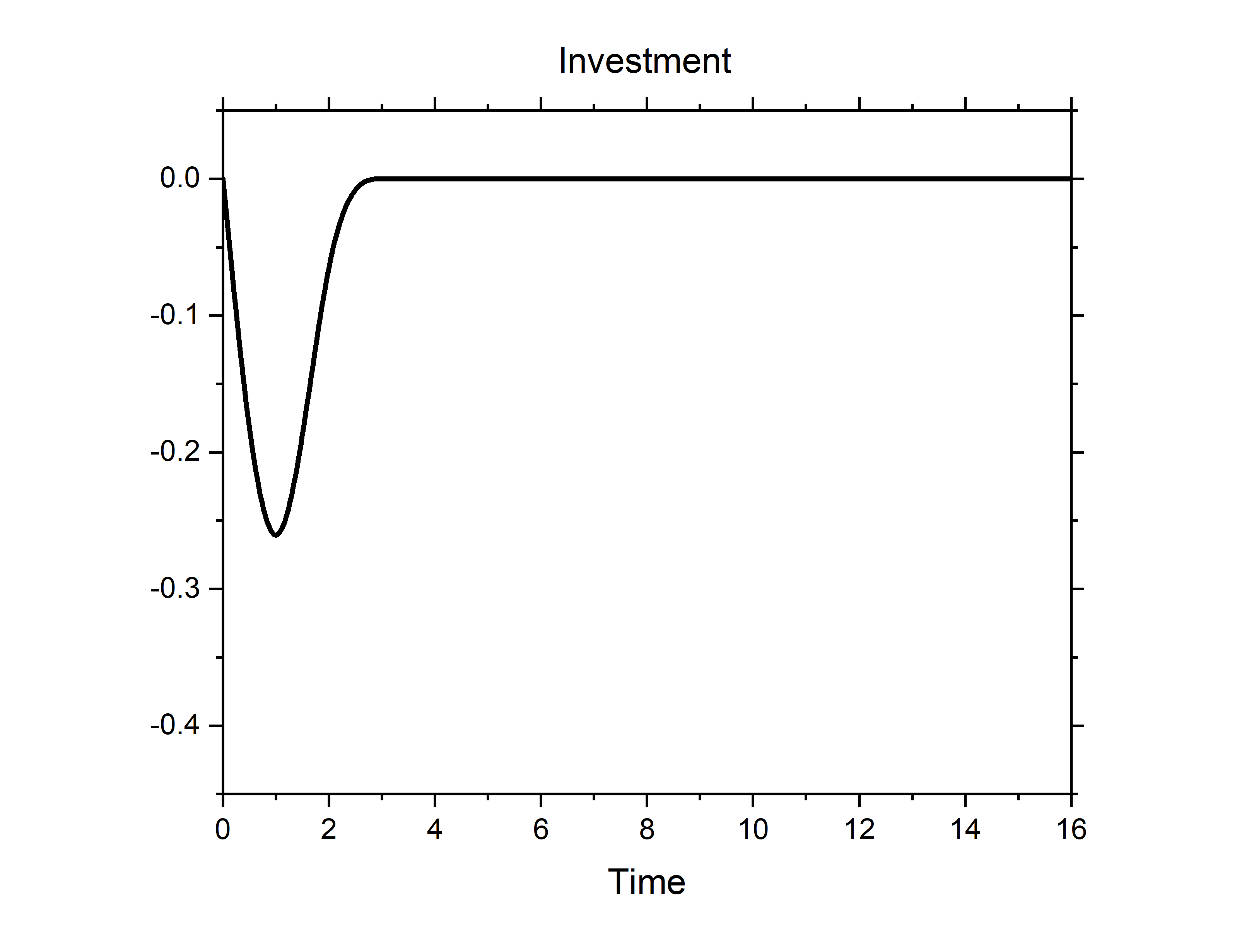}\hspace{-2cm}
    \caption*{Investment}
    \end{minipage}}
\quad
   \subfigure{
    \begin{minipage}[b]{0.25\linewidth}
    \includegraphics[width=3cm,height=2cm]{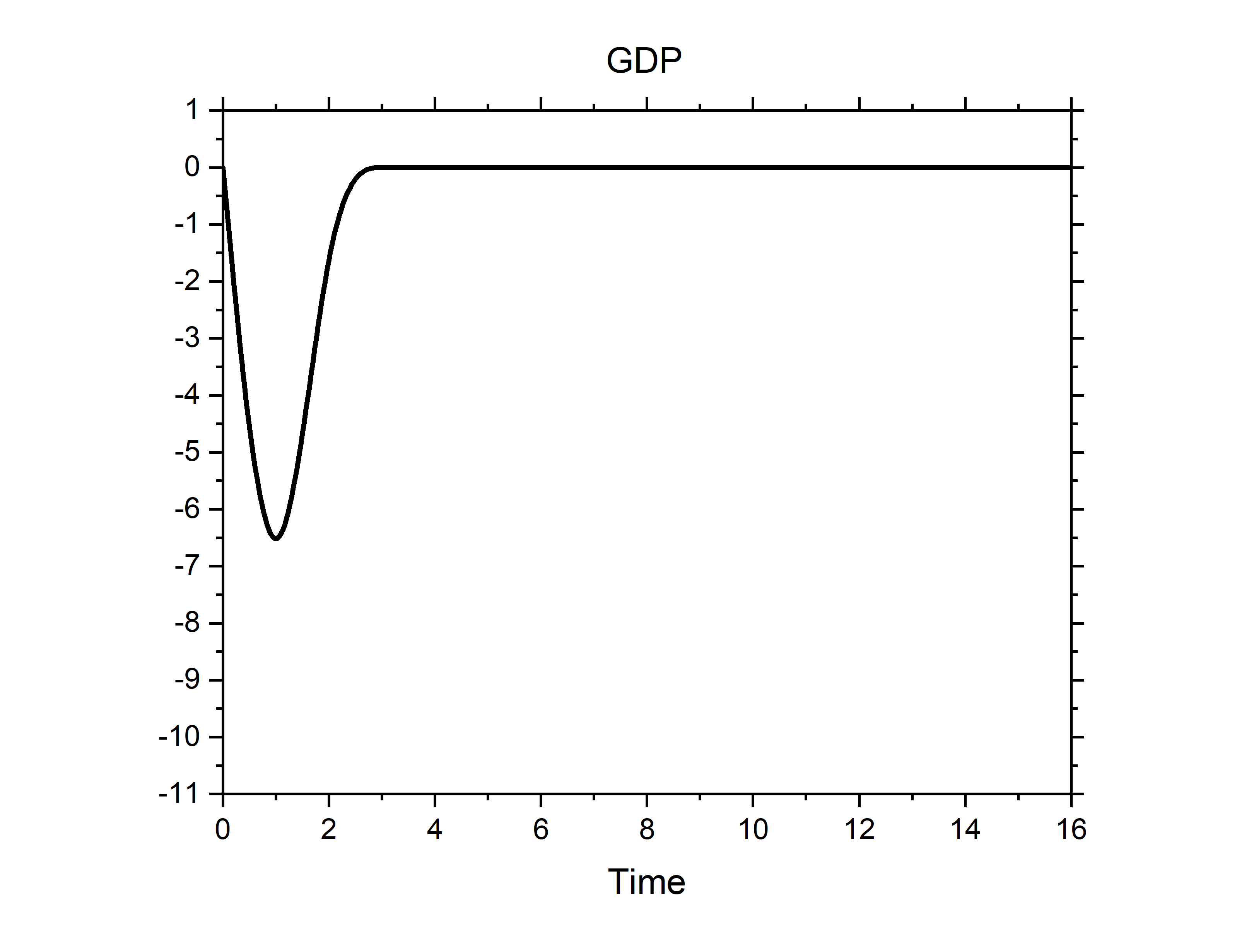}\hspace{-2cm}
    \caption*{GDP}
    \end{minipage}
    }
\quad
    \subfigure{
    \begin{minipage}[b]{0.25\linewidth}
    \includegraphics[width=3cm,height=2cm]{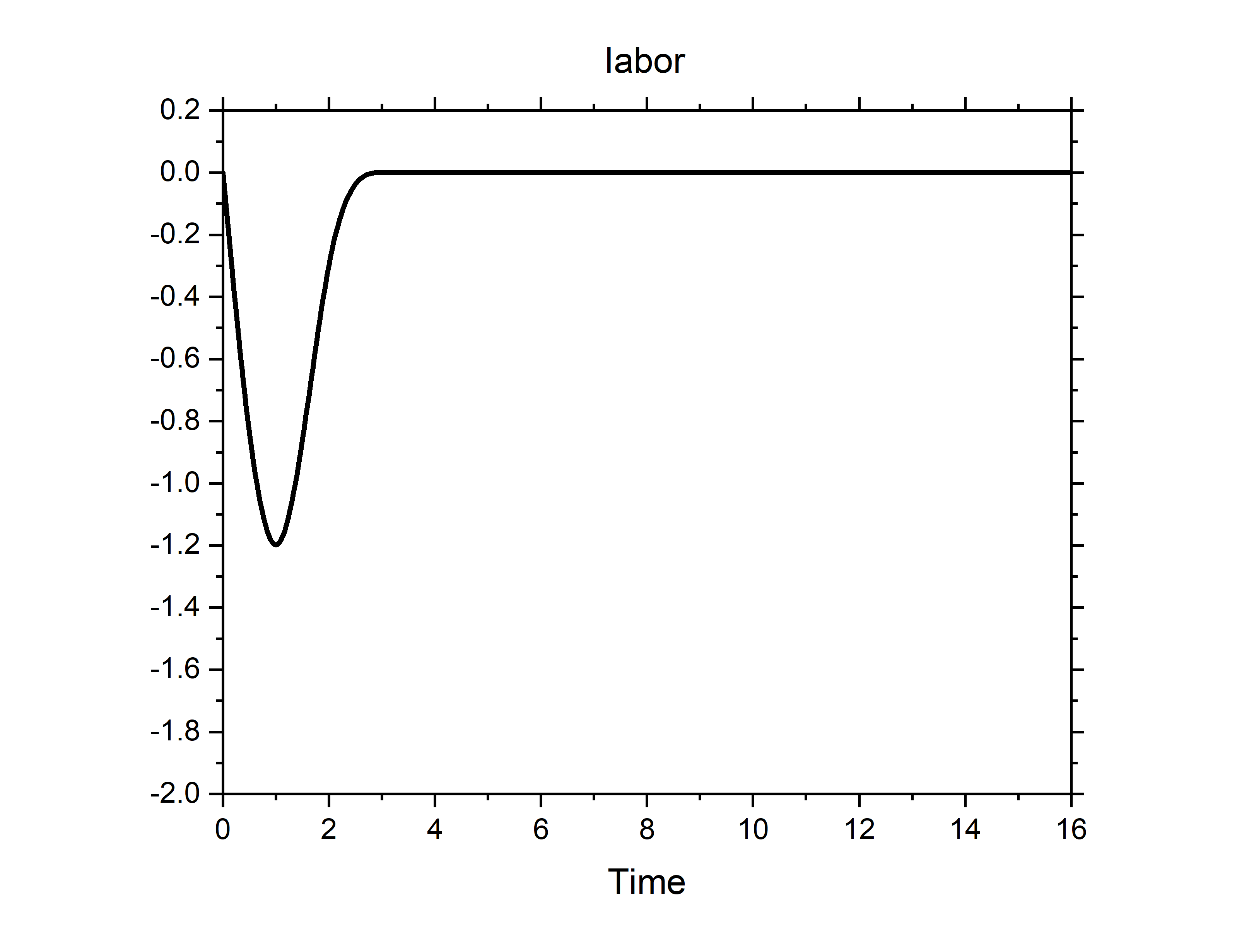}\hspace{-2cm}
    \caption*{Labor}
    \end{minipage}
    }
\quad
\subfigure{
    \begin{minipage}[b]{0.25\linewidth}
    \includegraphics[width=3cm,height=2cm]{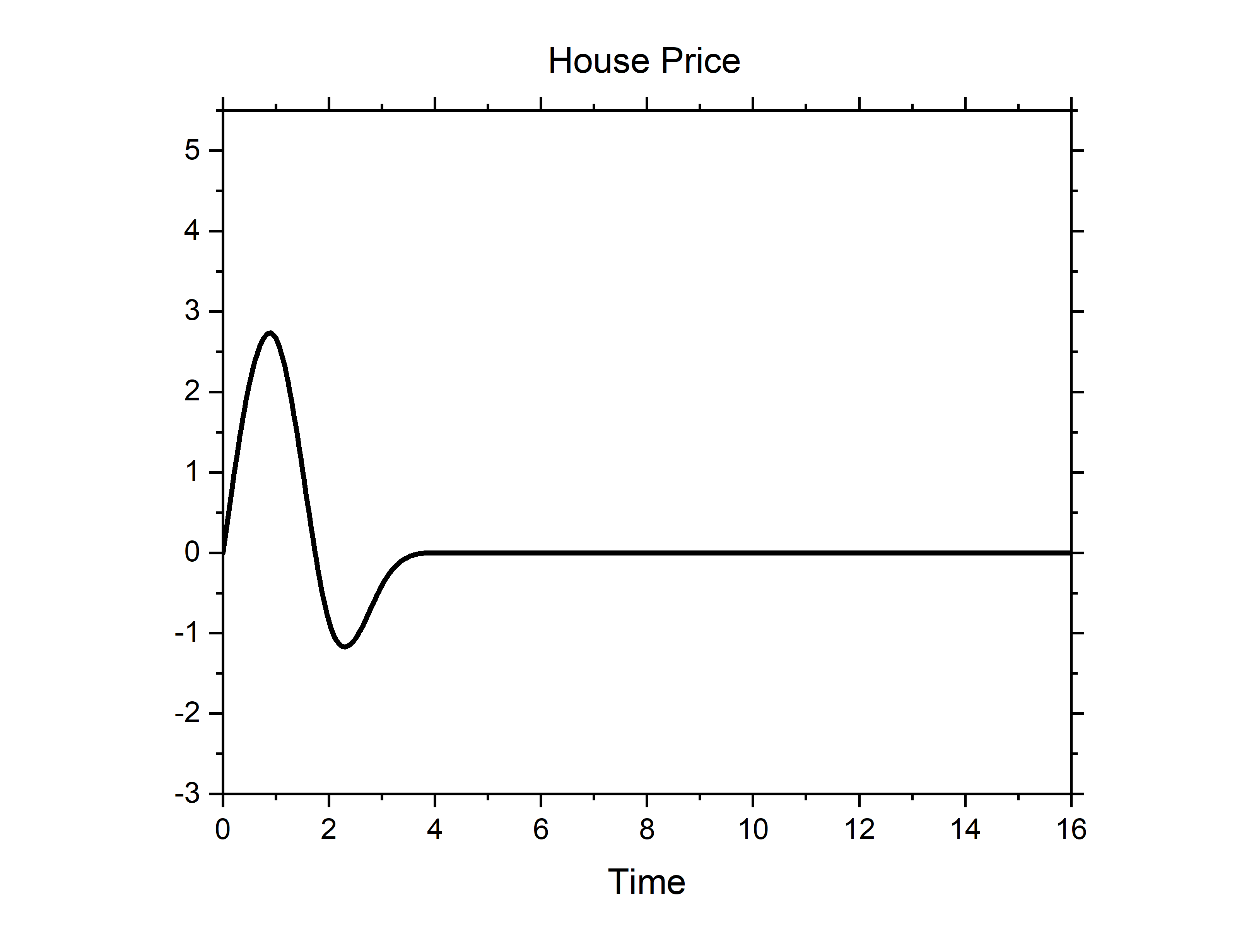}\hspace{-2cm}
    \caption*{Price of real estate}
\end{minipage}
}
\quad
\subfigure{
\begin{minipage}[b]{0.25\linewidth}
     \includegraphics[width=3cm,height=2cm]{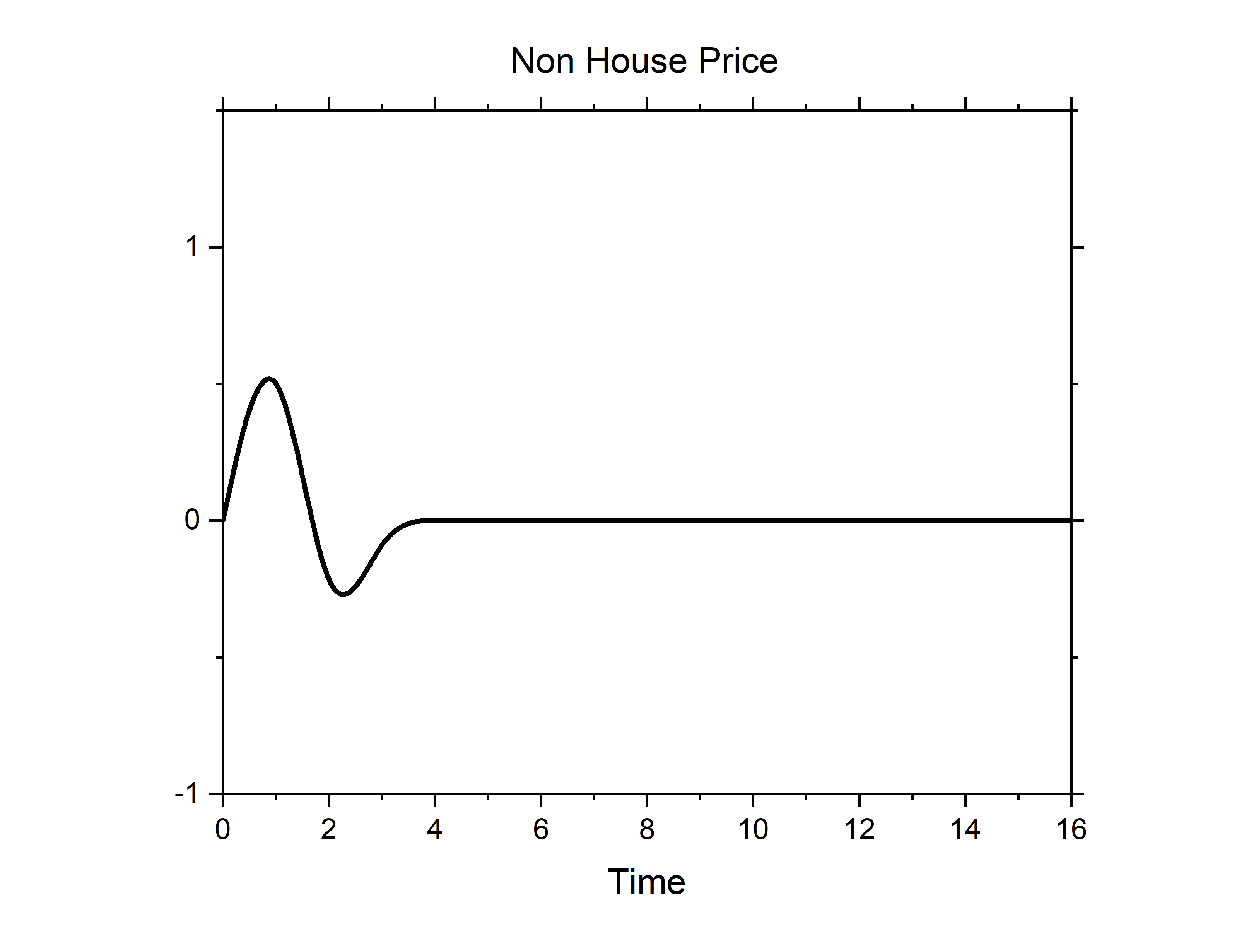}\hspace{-2cm}
    \caption*{Price of non real estate}
    \end{minipage}
    }
\quad
    \subfigure{
    \begin{minipage}[b]{0.25\linewidth}
    \includegraphics[width=3cm,height=2cm]{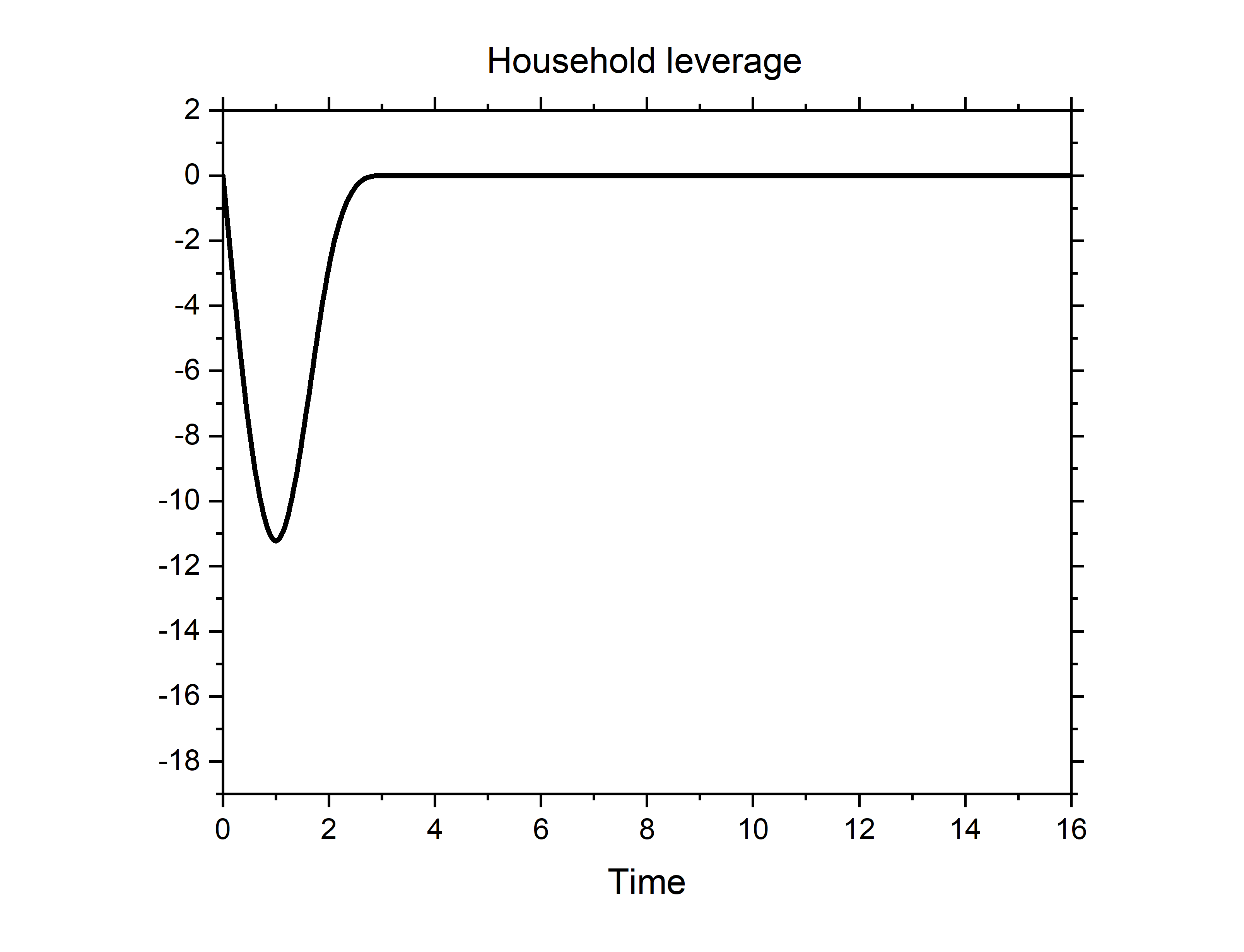}\hspace{-2cm}
    \caption*{Household leverage}
\end{minipage}}
\quad
\centering
    \subfigure{
    \begin{minipage}[b]{0.25\linewidth}
    \includegraphics[width=3cm,height=2cm]{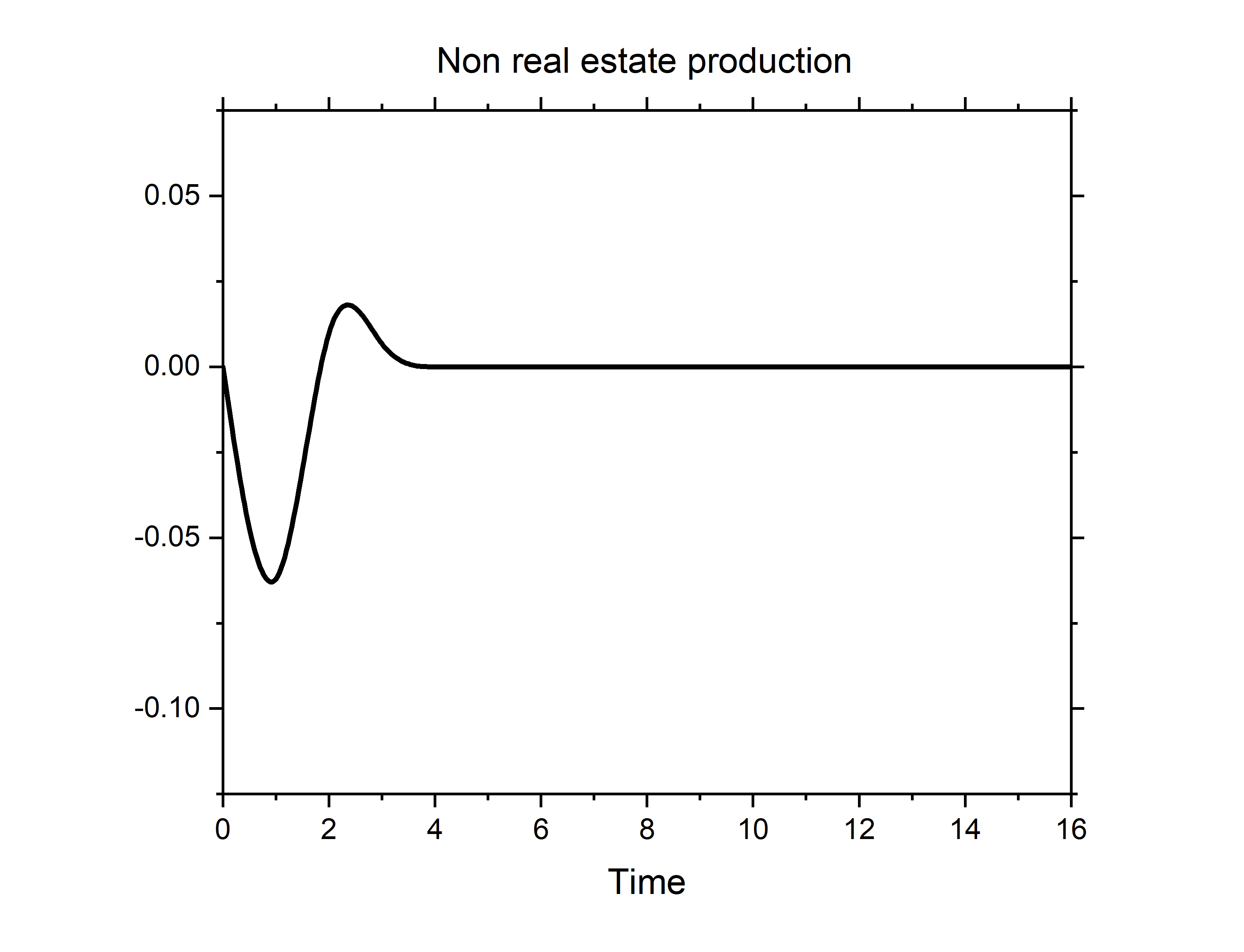}\hspace{-2cm}
    \caption*{Production of non real estate}
\end{minipage}}
\quad
 \subfigure{
    \begin{minipage}[b]{0.25\linewidth}
   \includegraphics[width=3cm,height=2cm]{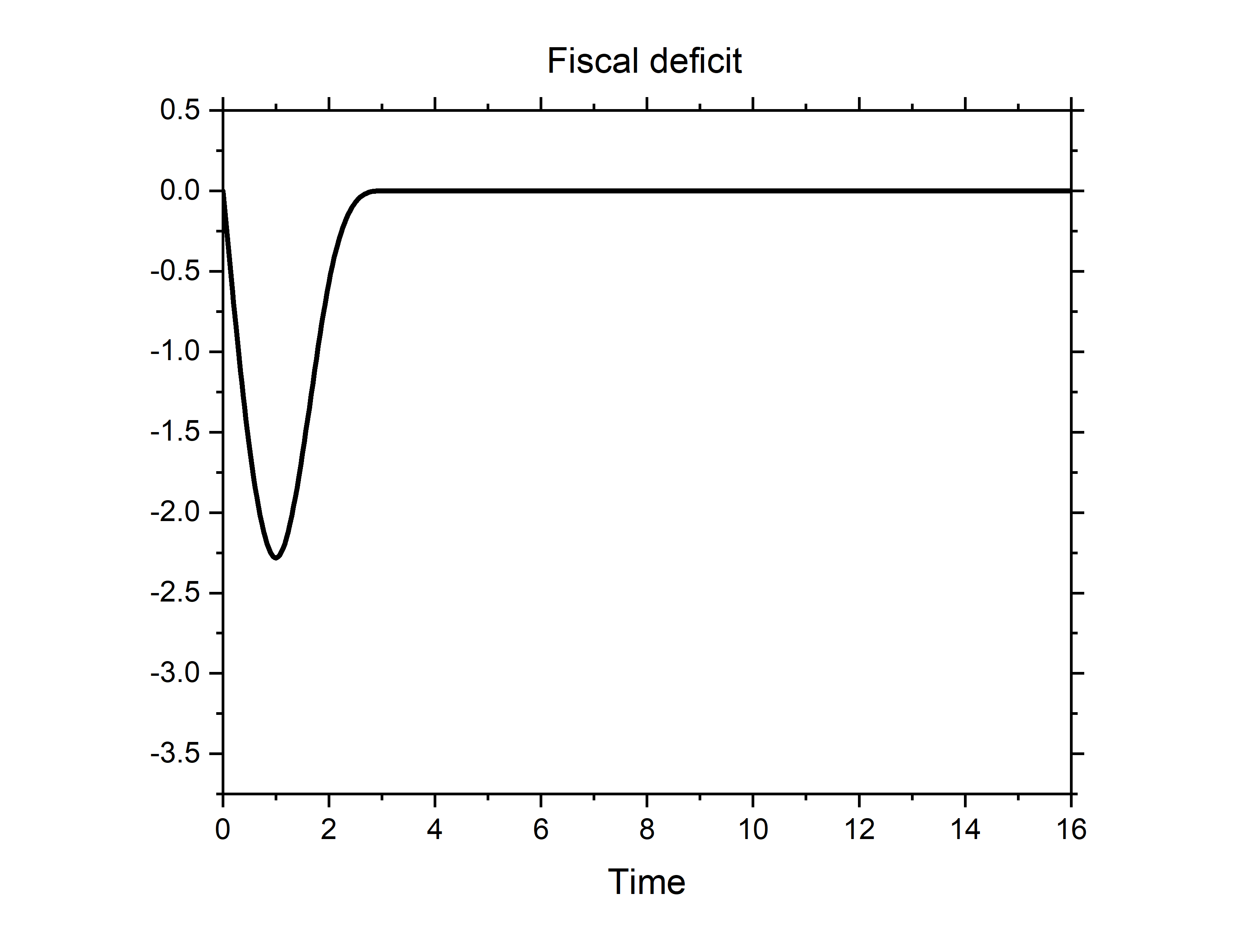}\hspace{-2cm}
    \caption*{Fiscal deficit}
    \end{minipage}
    }
    \caption{Impact Response Analysis by $R_t$}
\end{figure}
%mu
\begin{figure}
    \centering

    \subfigure{
    \begin{minipage}[b]{0.25\linewidth}
    \includegraphics[width=3cm,height=2cm]{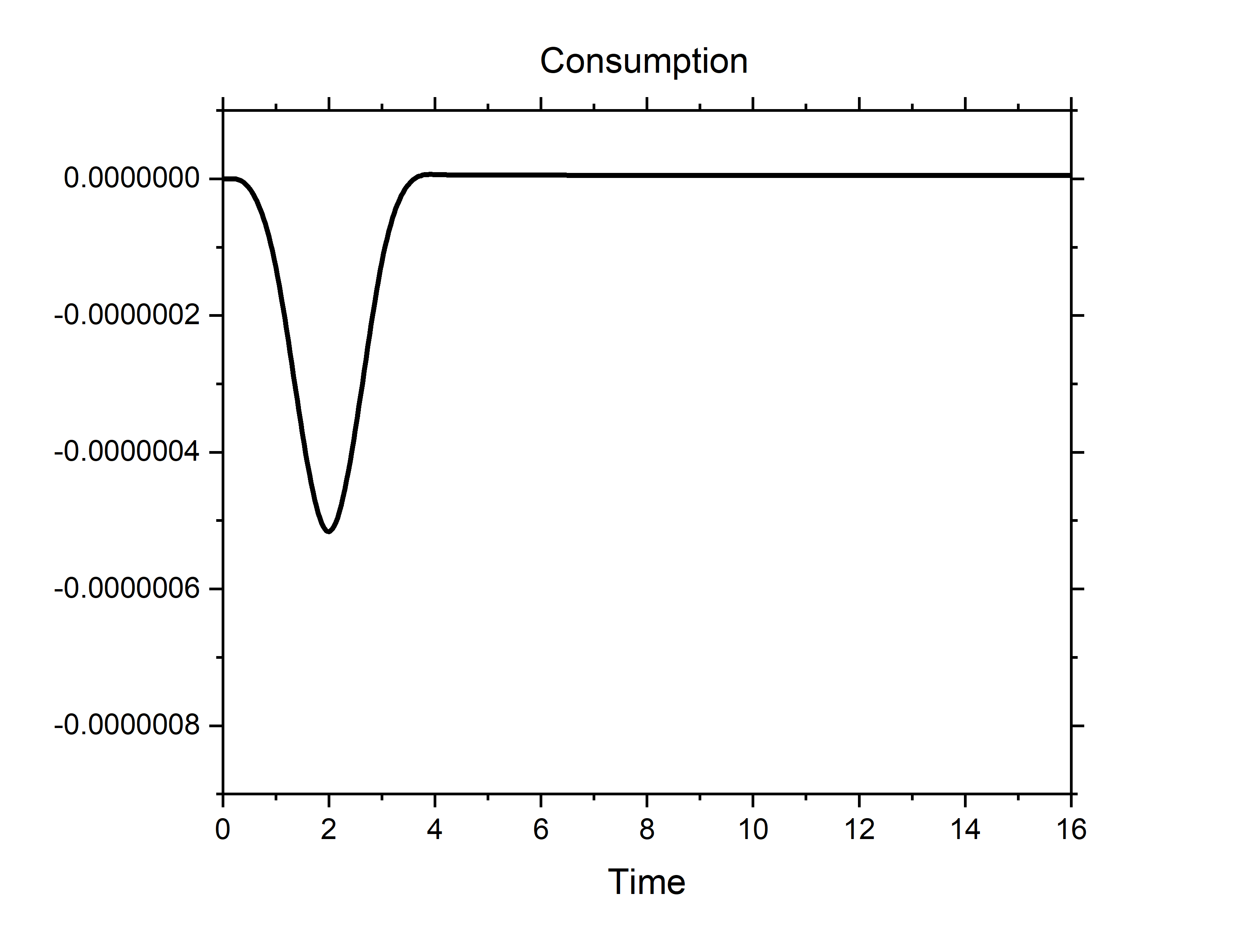}\hspace{-2cm}
    \caption*{Consumption}
    \end{minipage}
    }
\quad
 \subfigure{
    \begin{minipage}[b]{0.25\linewidth}
    \includegraphics[width=3cm,height=2cm]{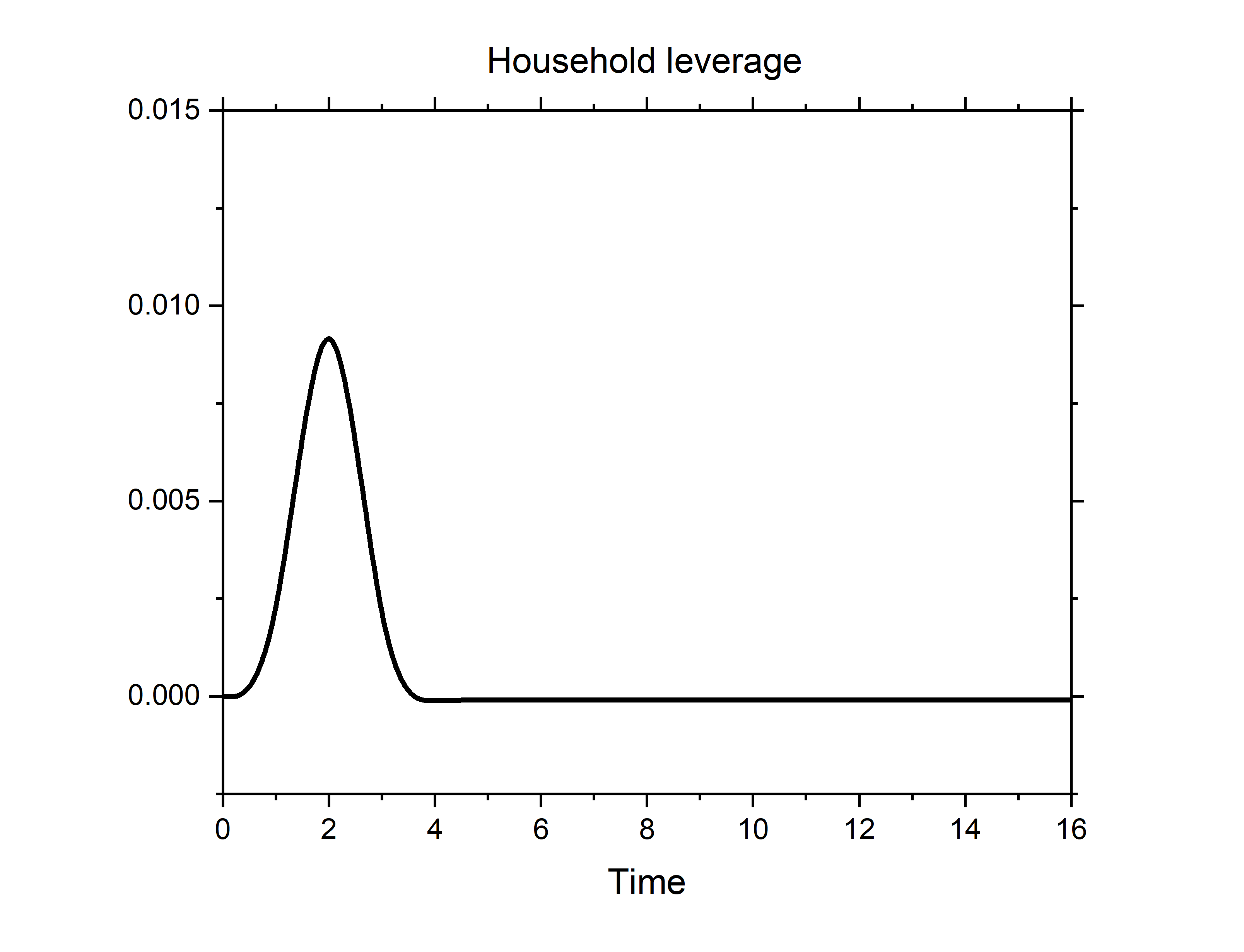}\hspace{-2cm}
    \caption*{Investment}
    \end{minipage}}
\quad
   \subfigure{
    \begin{minipage}[b]{0.25\linewidth}
    \includegraphics[width=3cm,height=2cm]{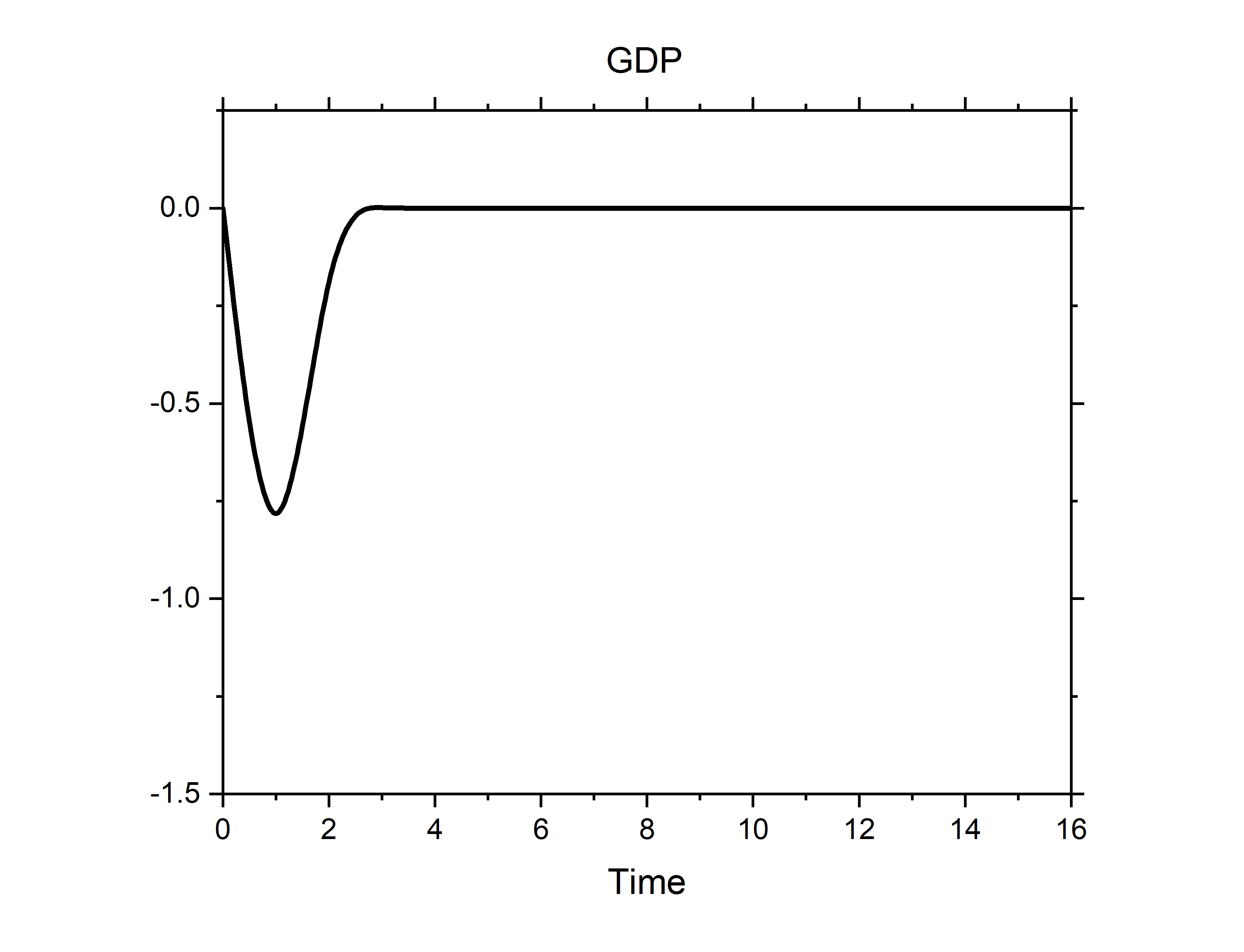}\hspace{-2cm}
    \caption*{GDP}
    \end{minipage}
    }
\quad
    \subfigure{
    \begin{minipage}[b]{0.25\linewidth}
    \includegraphics[width=3cm,height=2cm]{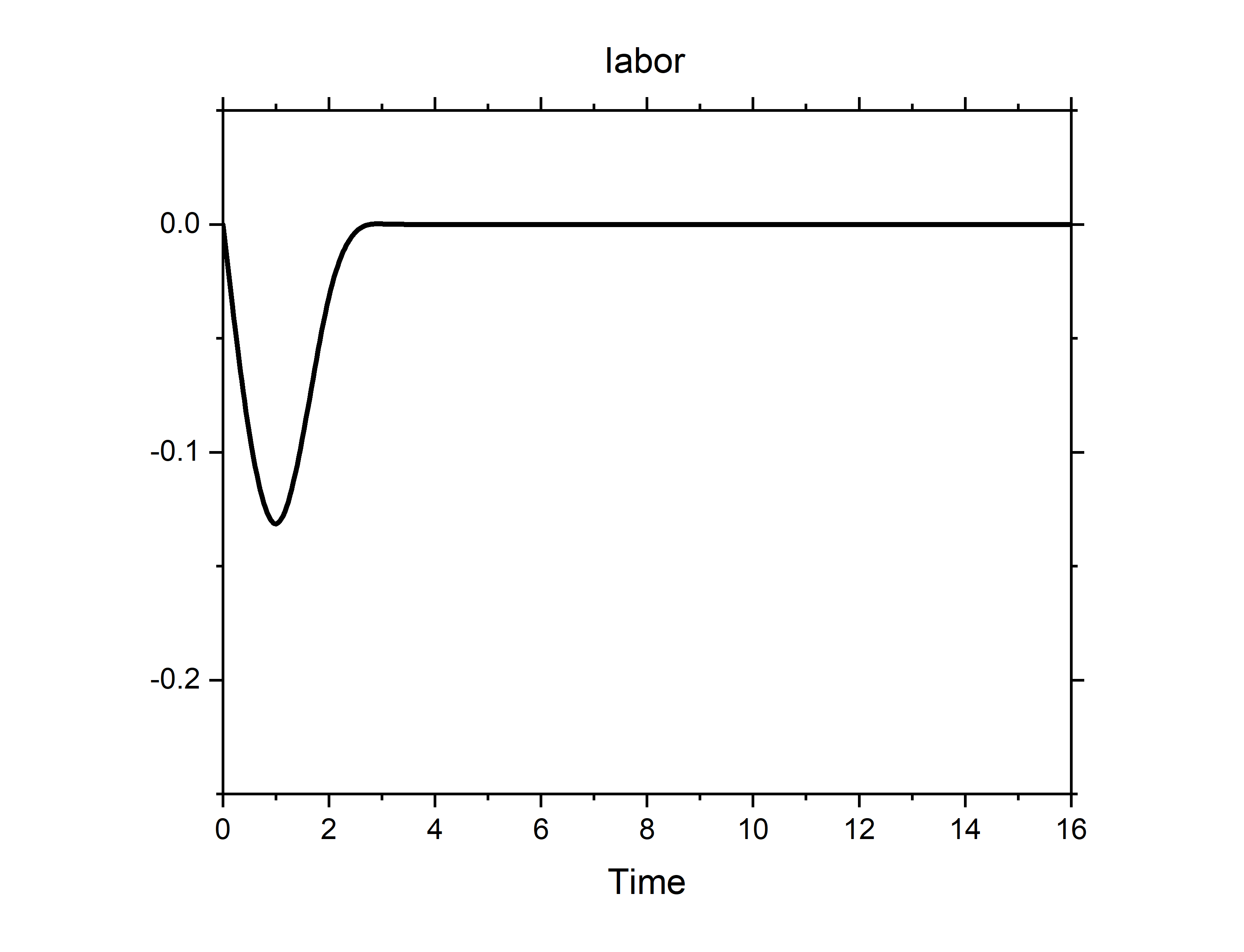}\hspace{-2cm}
    \caption*{Labor}
    \end{minipage}
    }
\quad
\subfigure{
    \begin{minipage}[b]{0.25\linewidth}
    \includegraphics[width=3cm,height=2cm]{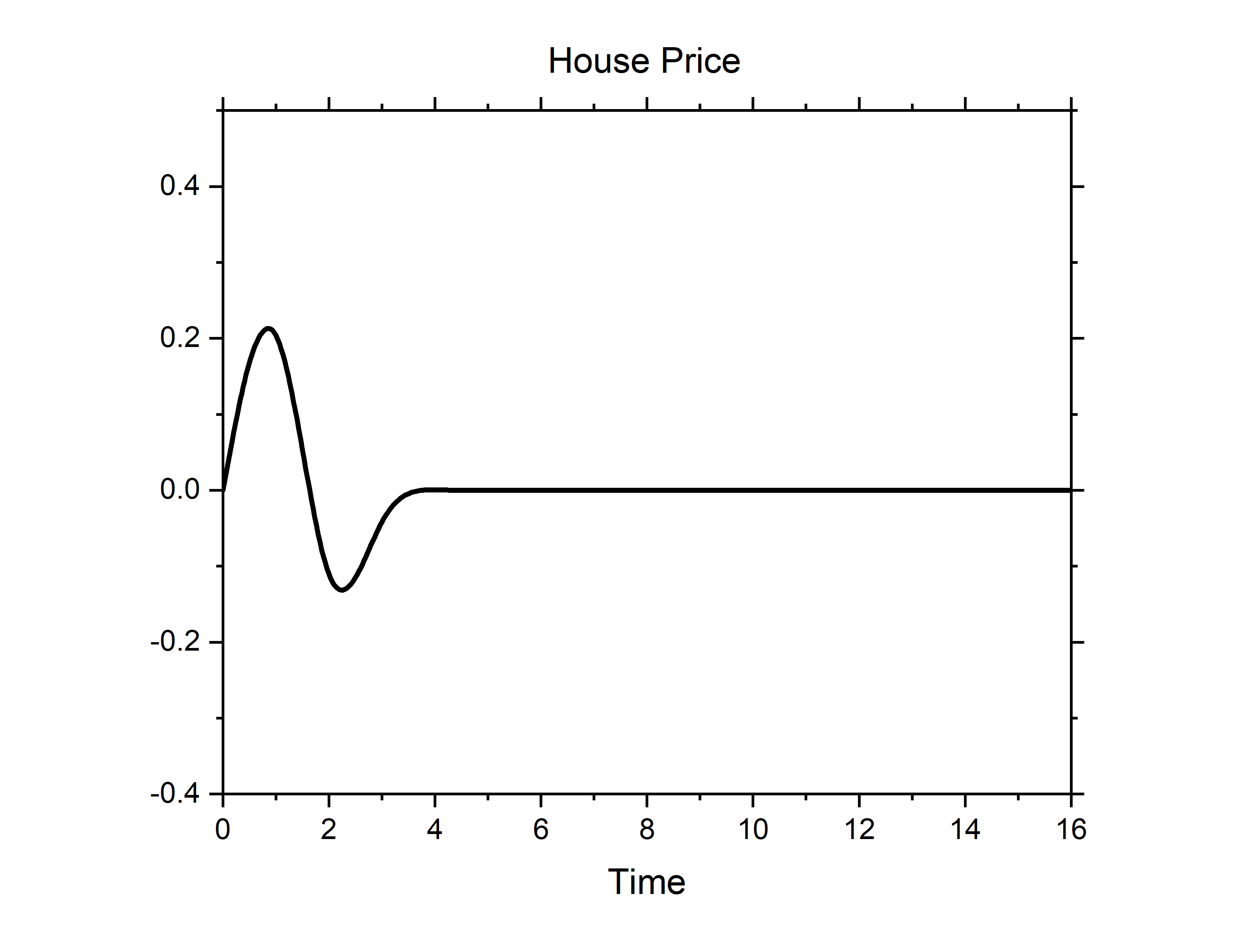}\hspace{-2cm}
    \caption*{Price of real estate}
\end{minipage}
}
\quad
\subfigure{
\begin{minipage}[b]{0.25\linewidth}
     \includegraphics[width=3cm,height=2cm]{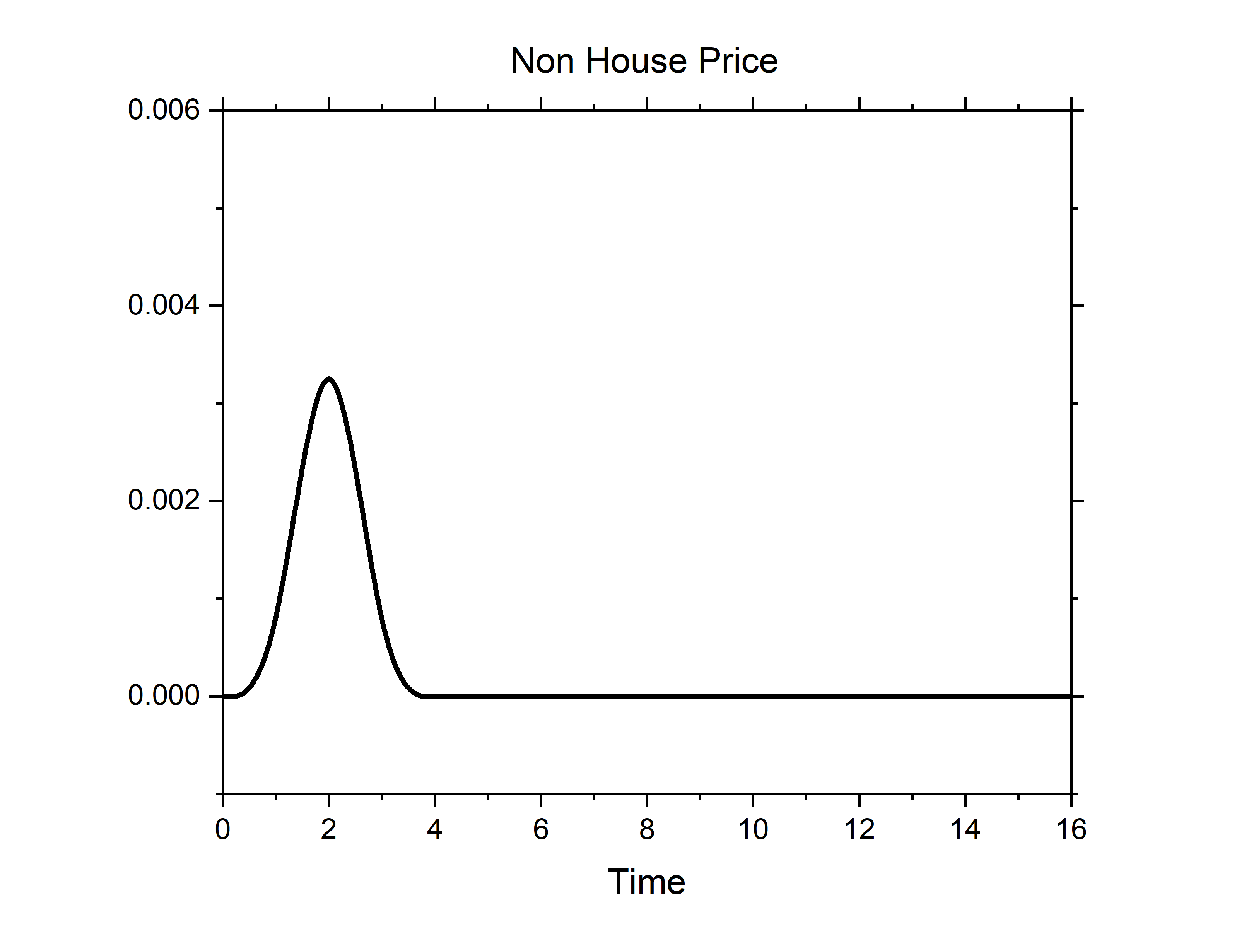}\hspace{-2cm}
    \caption*{Price of non real estate}
    \end{minipage}
    }
\quad
    \subfigure{
    \begin{minipage}[b]{0.25\linewidth}
    \includegraphics[width=3cm,height=2cm]{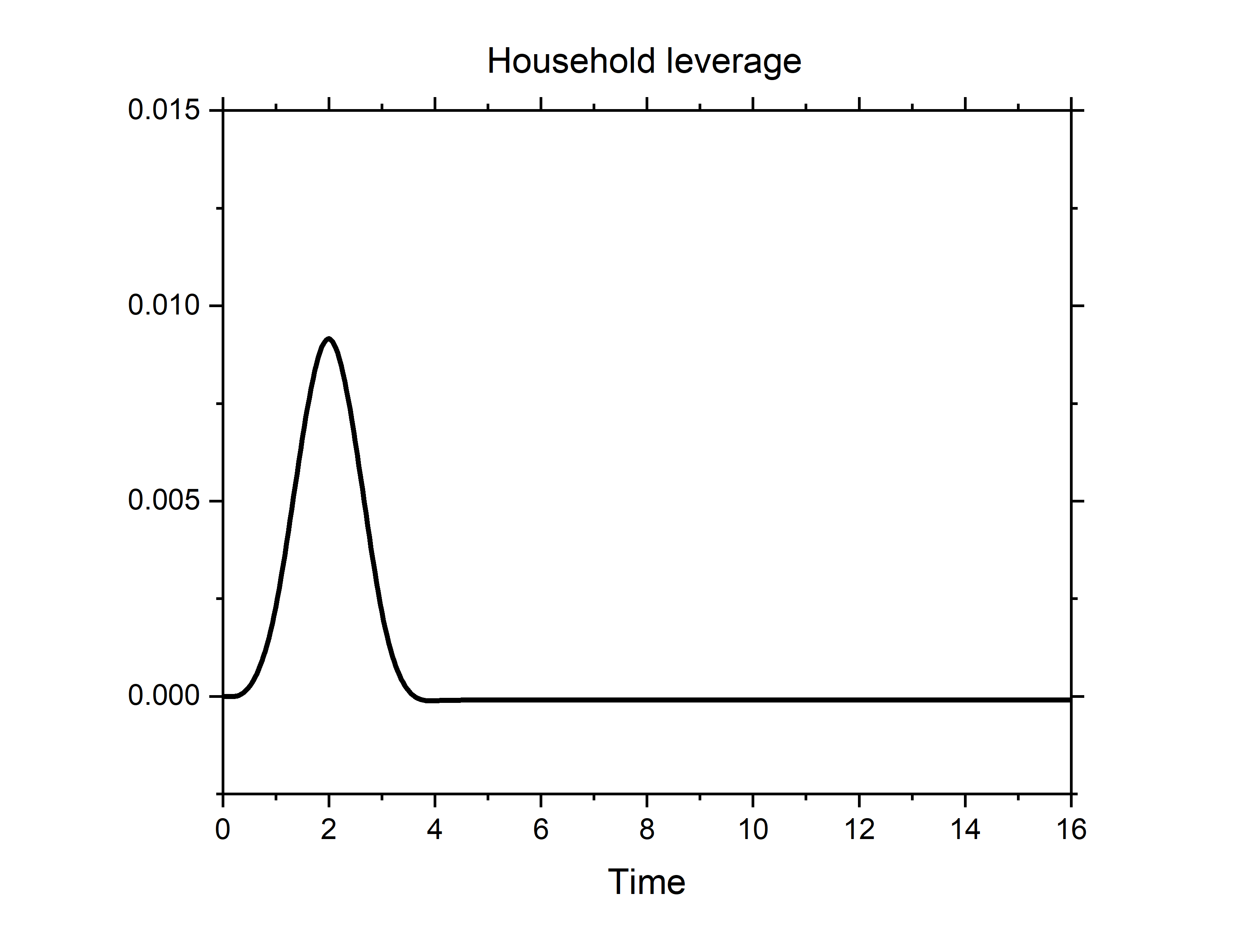}\hspace{-2cm}
    \caption*{Household leverage}
\end{minipage}}
\quad
    \subfigure{
    \begin{minipage}[b]{0.25\linewidth}
    \includegraphics[width=3cm,height=2cm]{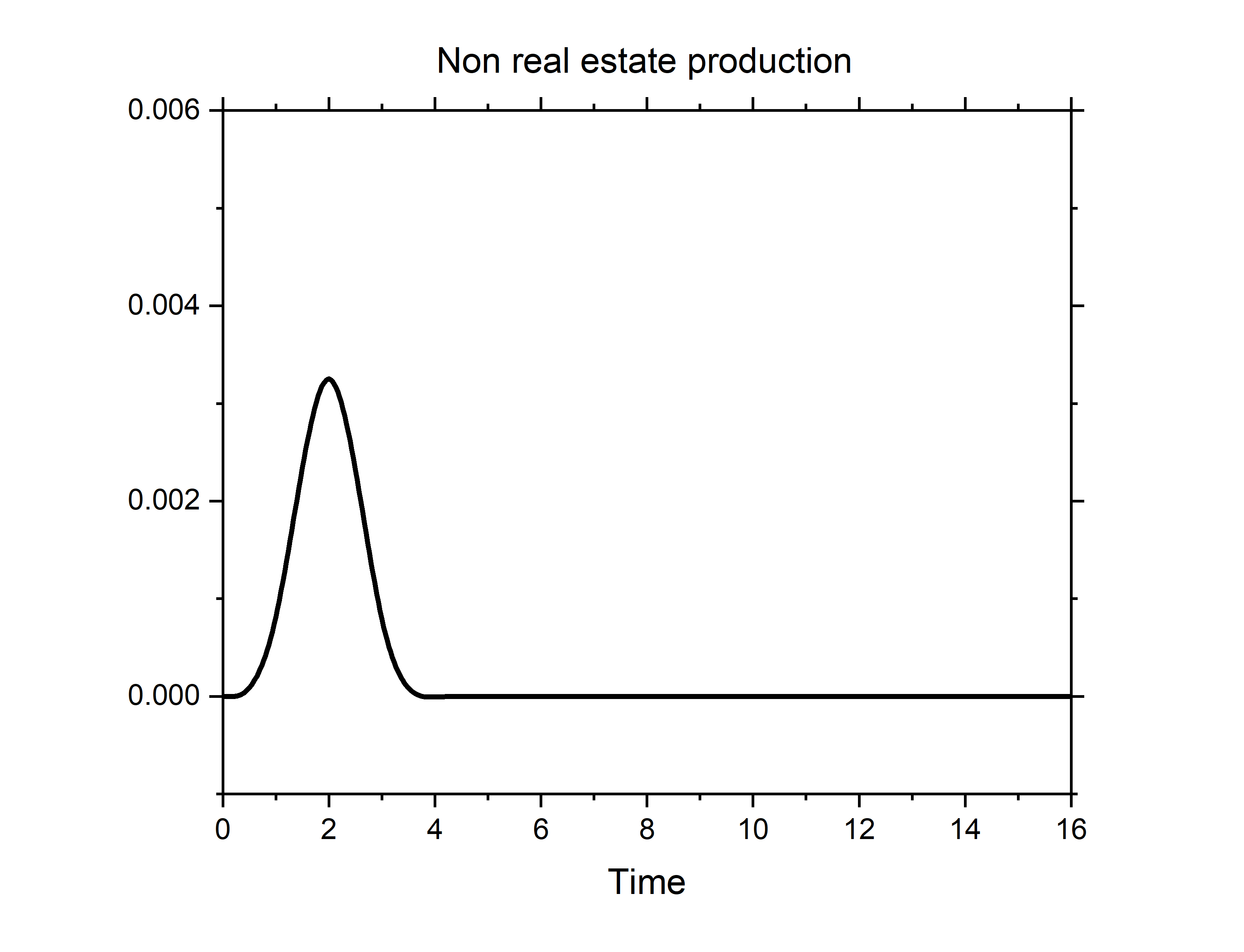}\hspace{-2cm}
    \caption*{Production of non real estate}
\end{minipage}}
\quad
 \subfigure{
    \begin{minipage}[b]{0.25\linewidth}
   \includegraphics[width=3cm,height=2cm]{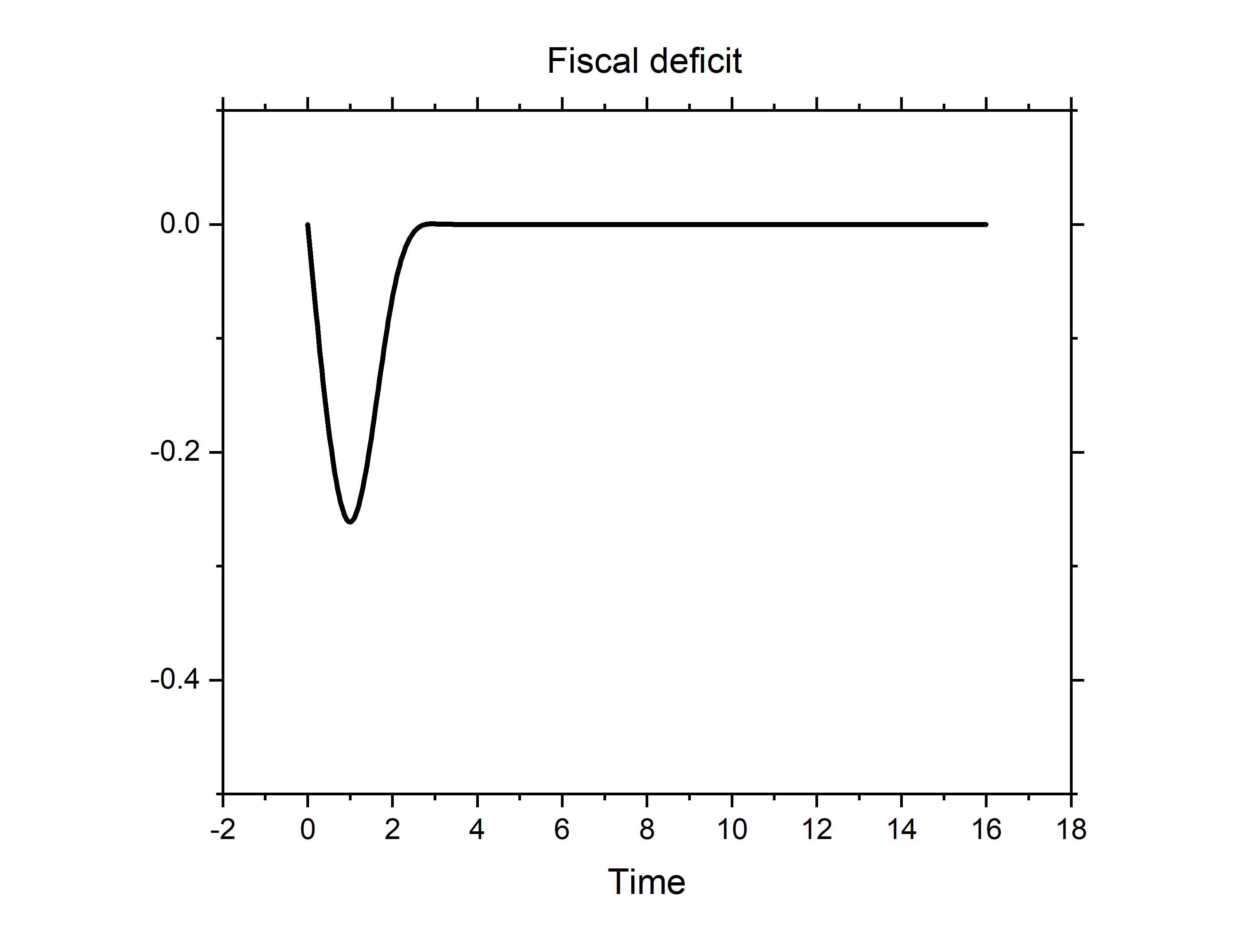}\hspace{-2cm}
    \caption*{Fiscal deficit}
    \end{minipage}
    }
    \caption{Impact Response Analysis by $\mu_t$}
\end{figure}
%S
\begin{figure}
    \centering

    \subfigure{
    \begin{minipage}[b]{0.25\linewidth}
    \includegraphics[width=3cm,height=2cm]{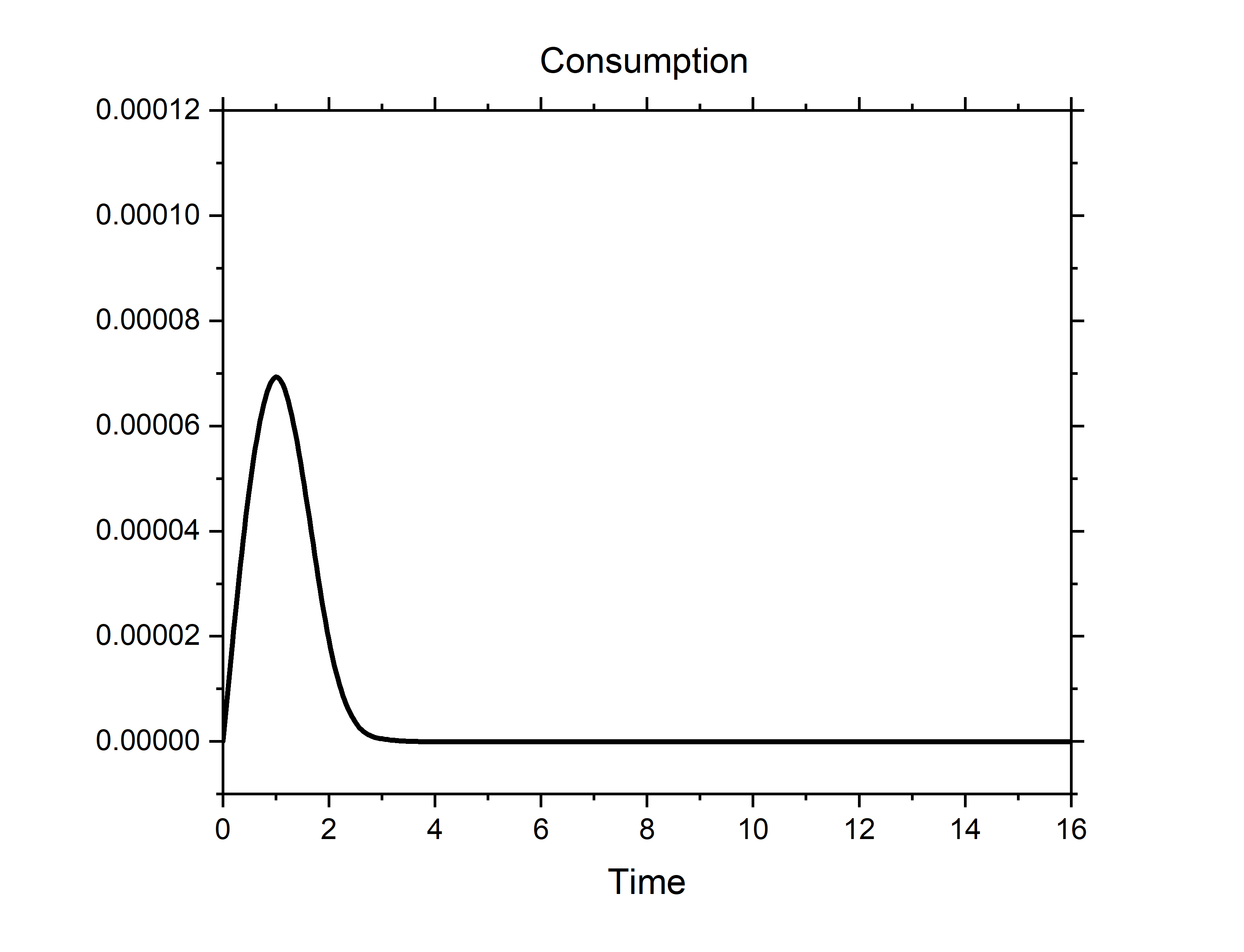}\hspace{-2cm}
    \caption*{Consumption}
    \end{minipage}
    }
\quad
 \subfigure{
    \begin{minipage}[b]{0.25\linewidth}
    \includegraphics[width=3cm,height=2cm]{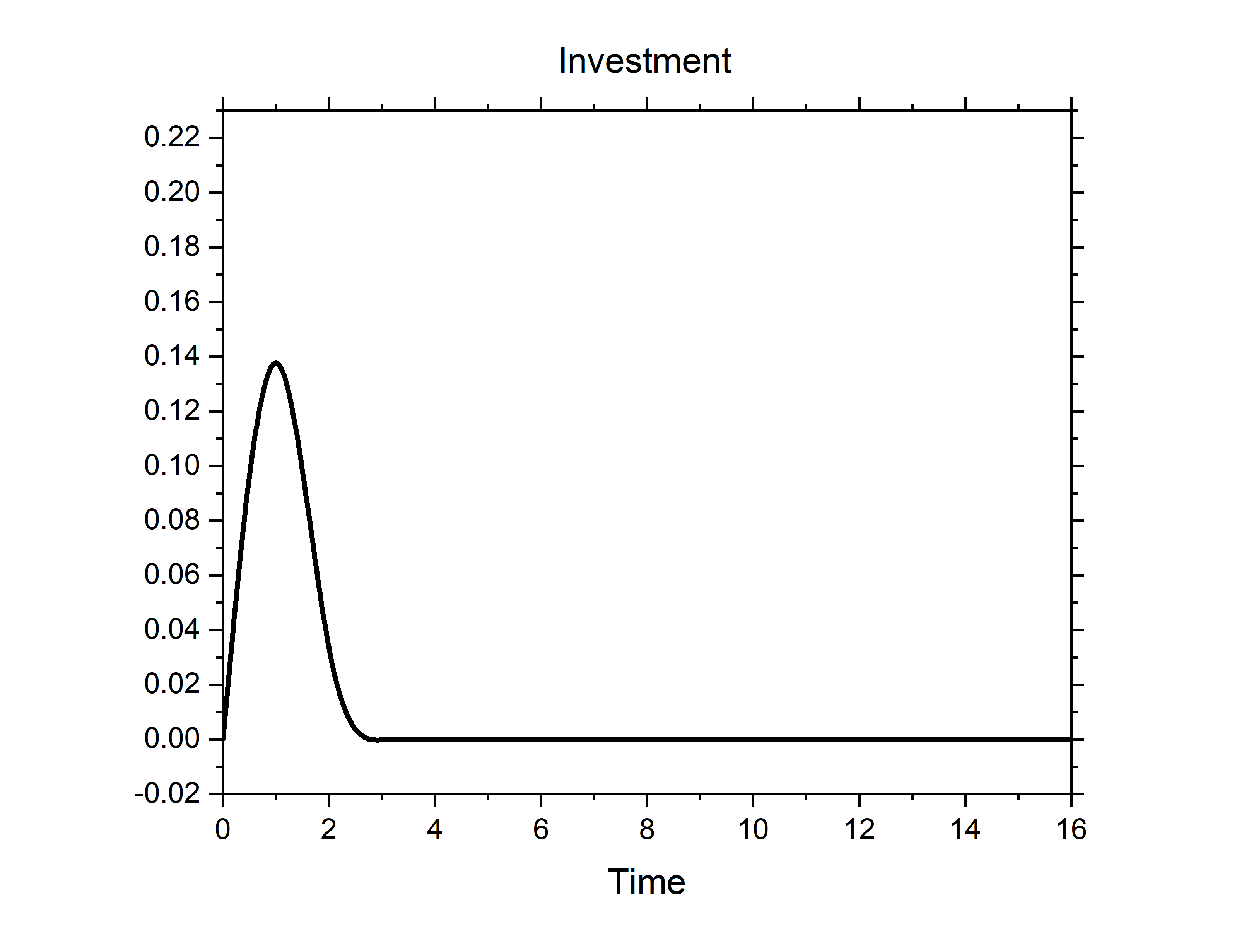}\hspace{-2cm}
    \caption*{Investment}
    \end{minipage}}
\quad
   \subfigure{
    \begin{minipage}[b]{0.25\linewidth}
    \includegraphics[width=3cm,height=2cm]{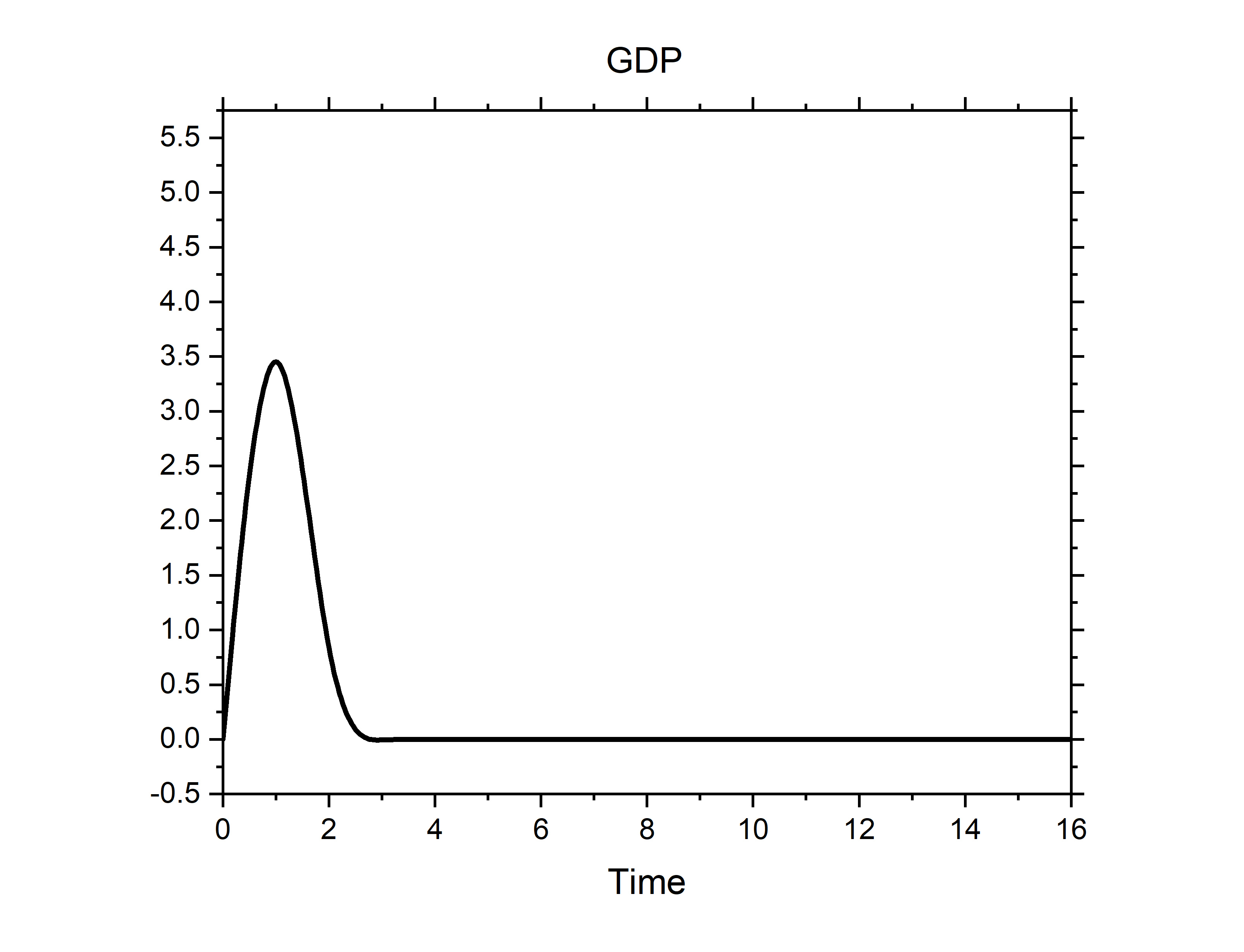}\hspace{-2cm}
    \caption*{GDP}
    \end{minipage}
    }
\quad
    \subfigure{
    \begin{minipage}[b]{0.25\linewidth}
    \includegraphics[width=3cm,height=2cm]{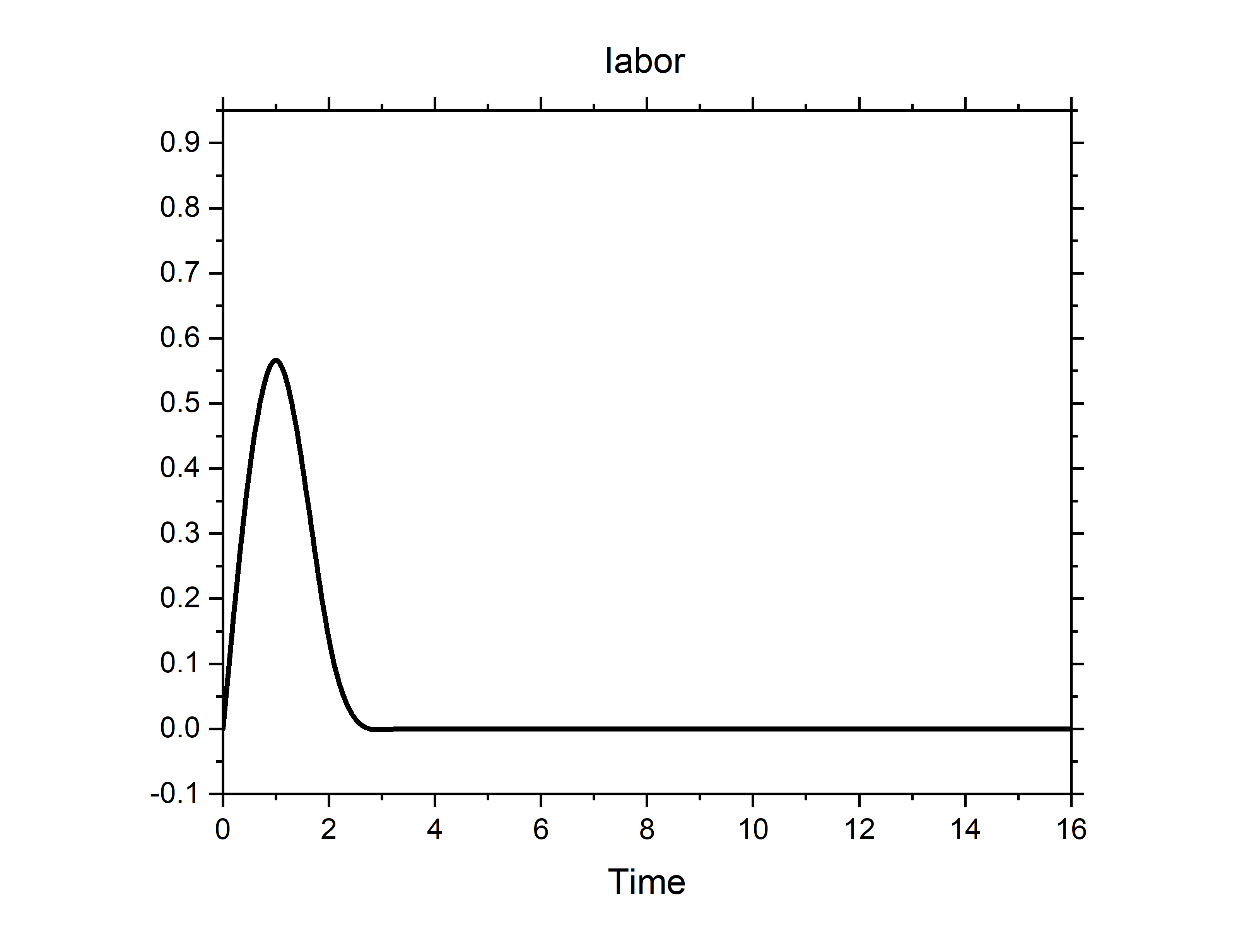}\hspace{-2cm}
    \caption*{Labor}
    \end{minipage}
    }
\quad
\subfigure{
    \begin{minipage}[b]{0.25\linewidth}
    \includegraphics[width=3cm,height=2cm]{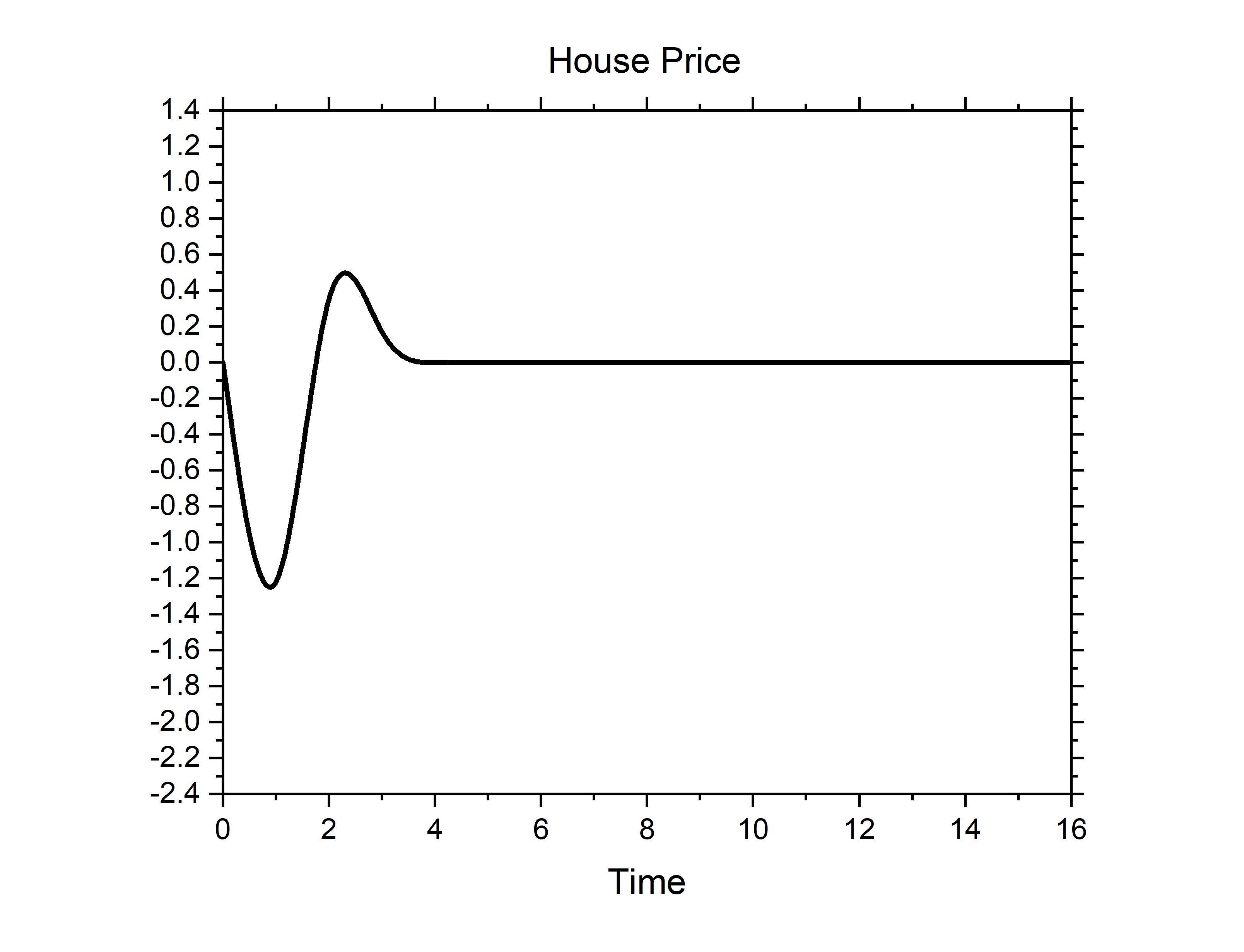}\hspace{-2cm}
    \caption*{Price of real estate}
\end{minipage}
}
\quad
\subfigure{
\begin{minipage}[b]{0.25\linewidth}
     \includegraphics[width=3cm,height=2cm]{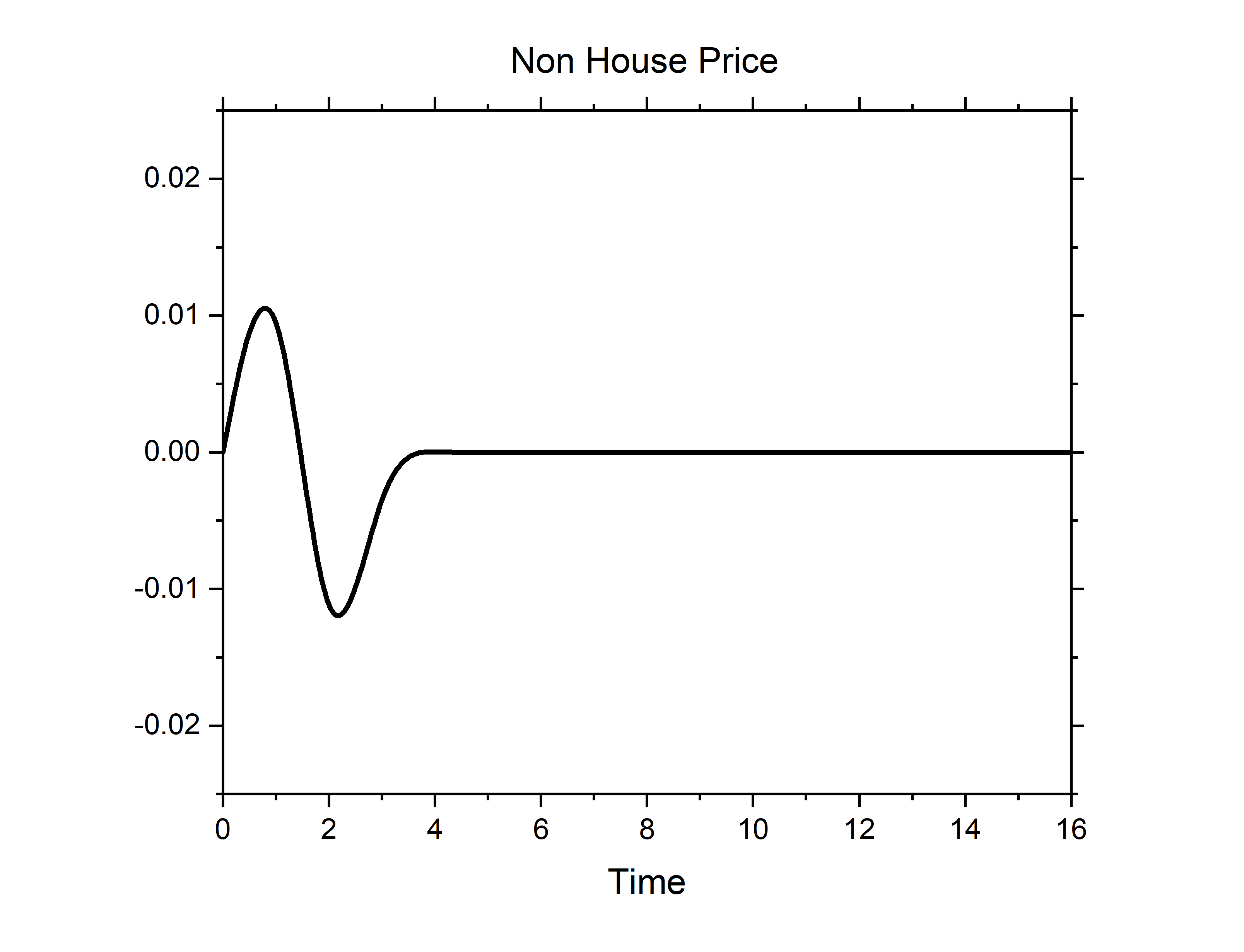}\hspace{-2cm}
    \caption*{Price of non real estate}
    \end{minipage}
    }
\quad
    \subfigure{
    \begin{minipage}[b]{0.25\linewidth}
    \includegraphics[width=3cm,height=2cm]{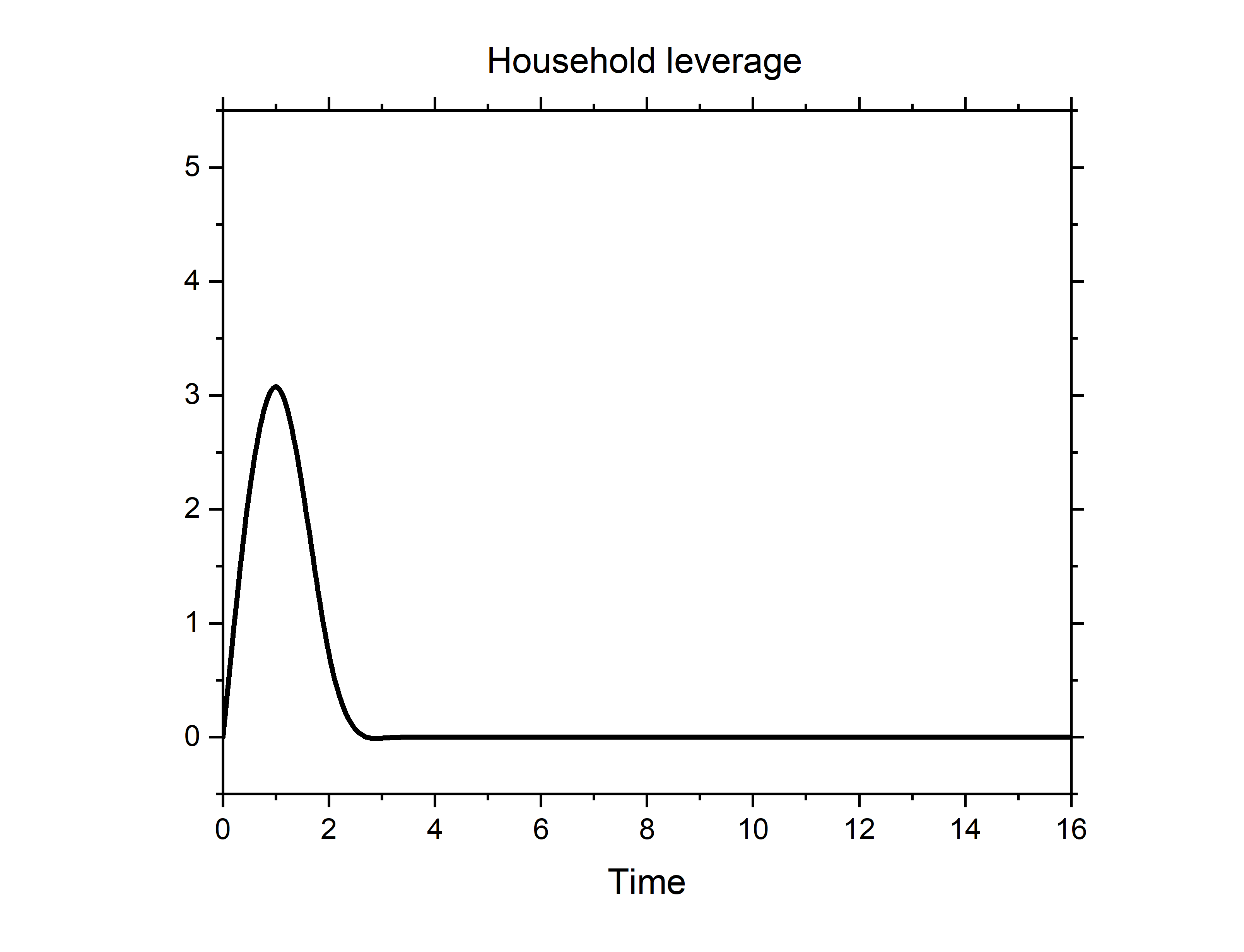}\hspace{-2cm}
    \caption*{Household leverage}
\end{minipage}}
\quad
    \subfigure{
    \begin{minipage}[b]{0.25\linewidth}
    \includegraphics[width=3cm,height=2cm]{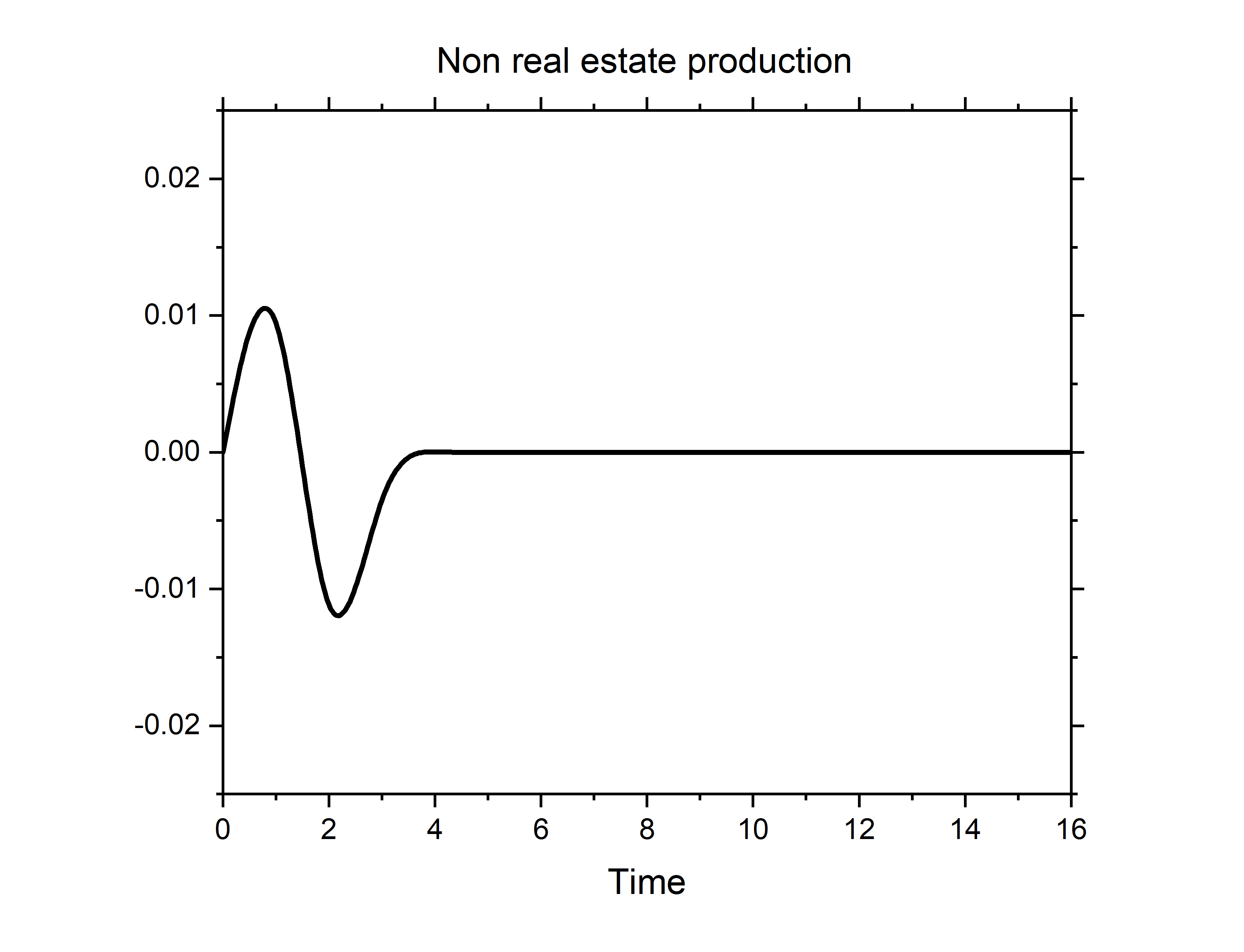}\hspace{-2cm}
    \caption*{Production of non real estate}
\end{minipage}}
\quad
 \subfigure{
    \begin{minipage}[b]{0.25\linewidth}
   \includegraphics[width=3cm,height=2cm]{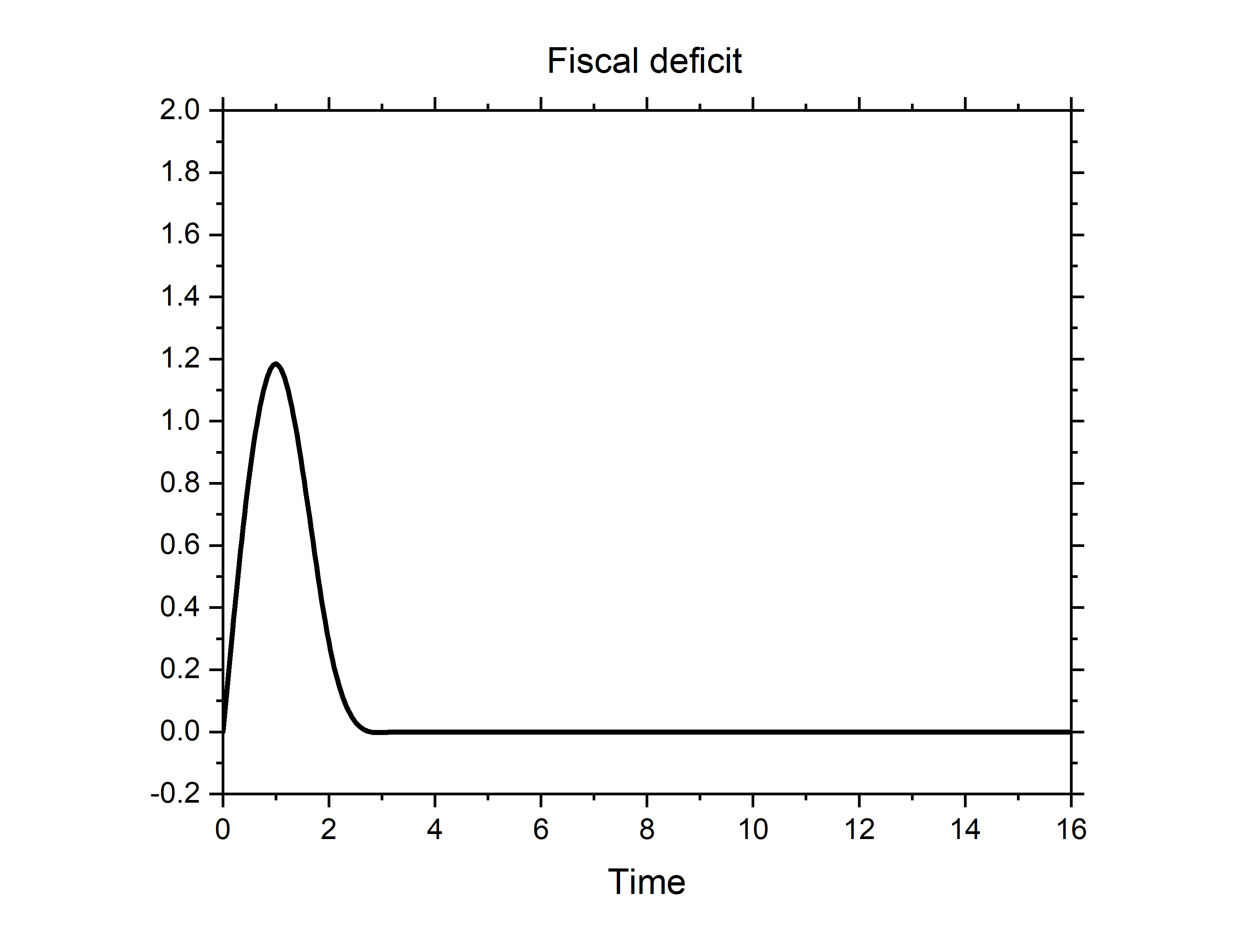}\hspace{-2cm}
    \caption*{Fiscal deficit}
    \end{minipage}
    }
    \caption{Impact Response Analysis by $\theta_t$}
\end{figure}
%xi
\begin{figure}
    \centering

    \subfigure{
    \begin{minipage}[b]{0.25\linewidth}
    \includegraphics[width=3cm,height=2cm]{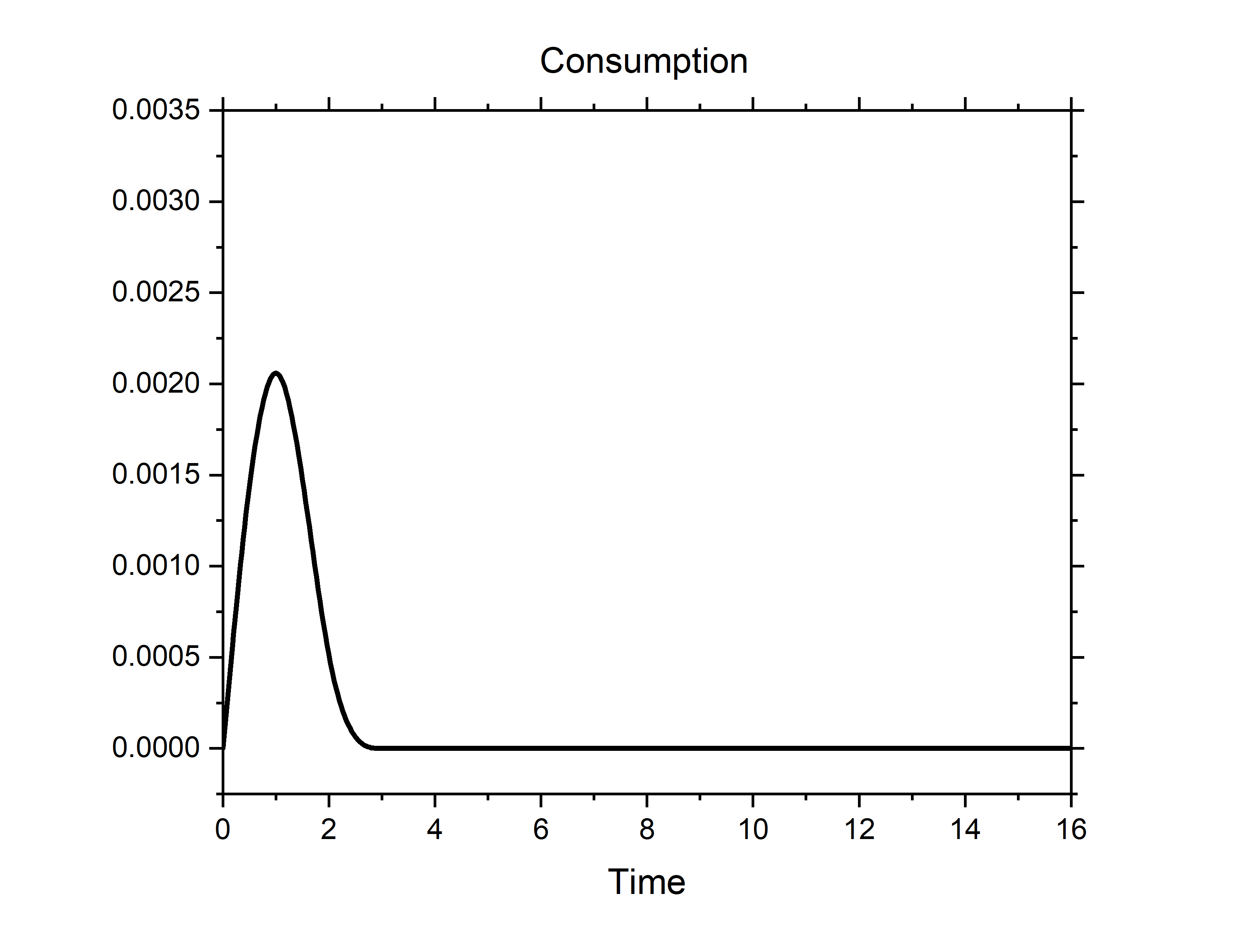}\hspace{-2cm}
    \caption*{Consumption}
    \end{minipage}
    }
\quad
 \subfigure{
    \begin{minipage}[b]{0.25\linewidth}
    \includegraphics[width=3cm,height=2cm]{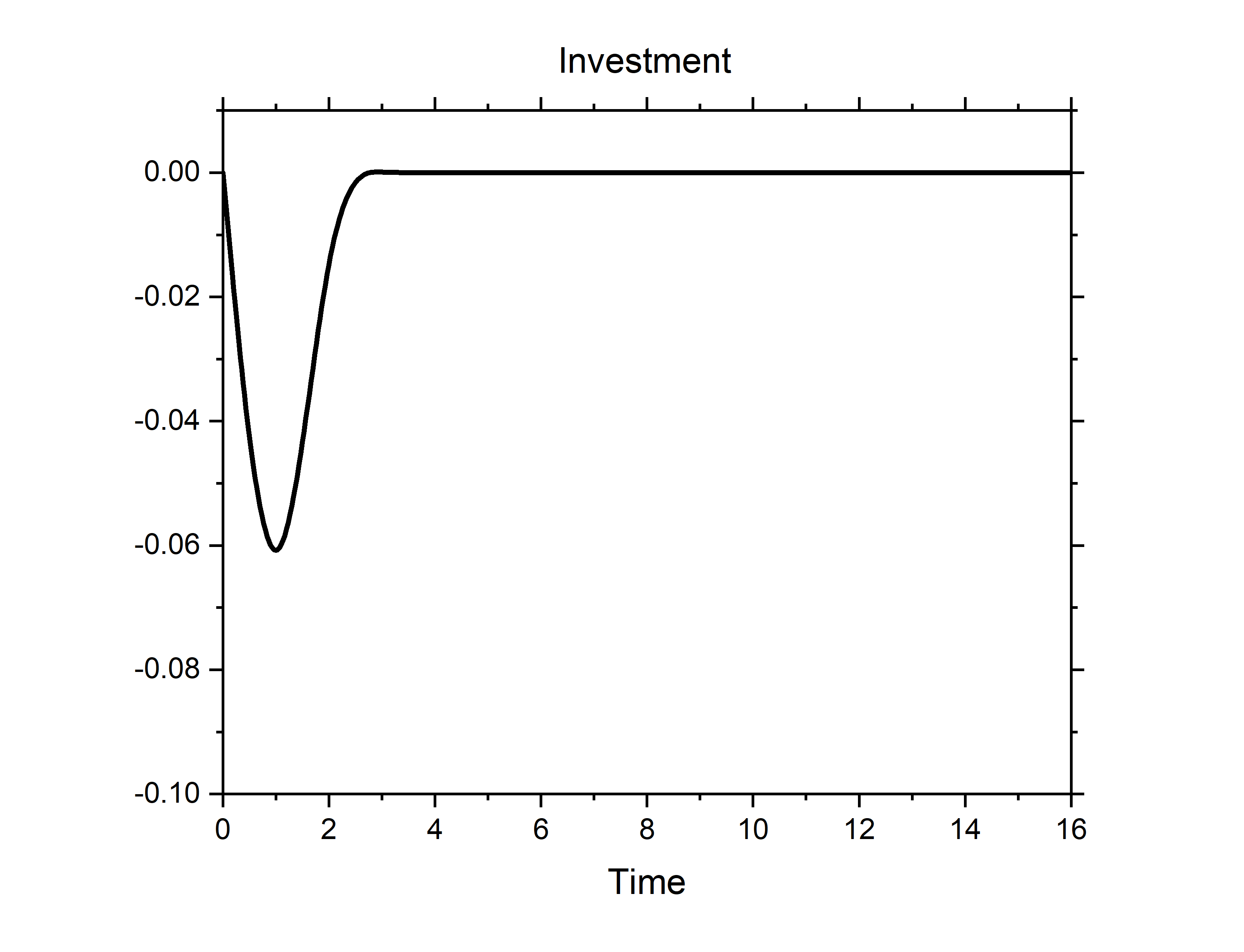}\hspace{-2cm}
    \caption*{Investment}
    \end{minipage}}
\quad
   \subfigure{
    \begin{minipage}[b]{0.25\linewidth}
    \includegraphics[width=3cm,height=2cm]{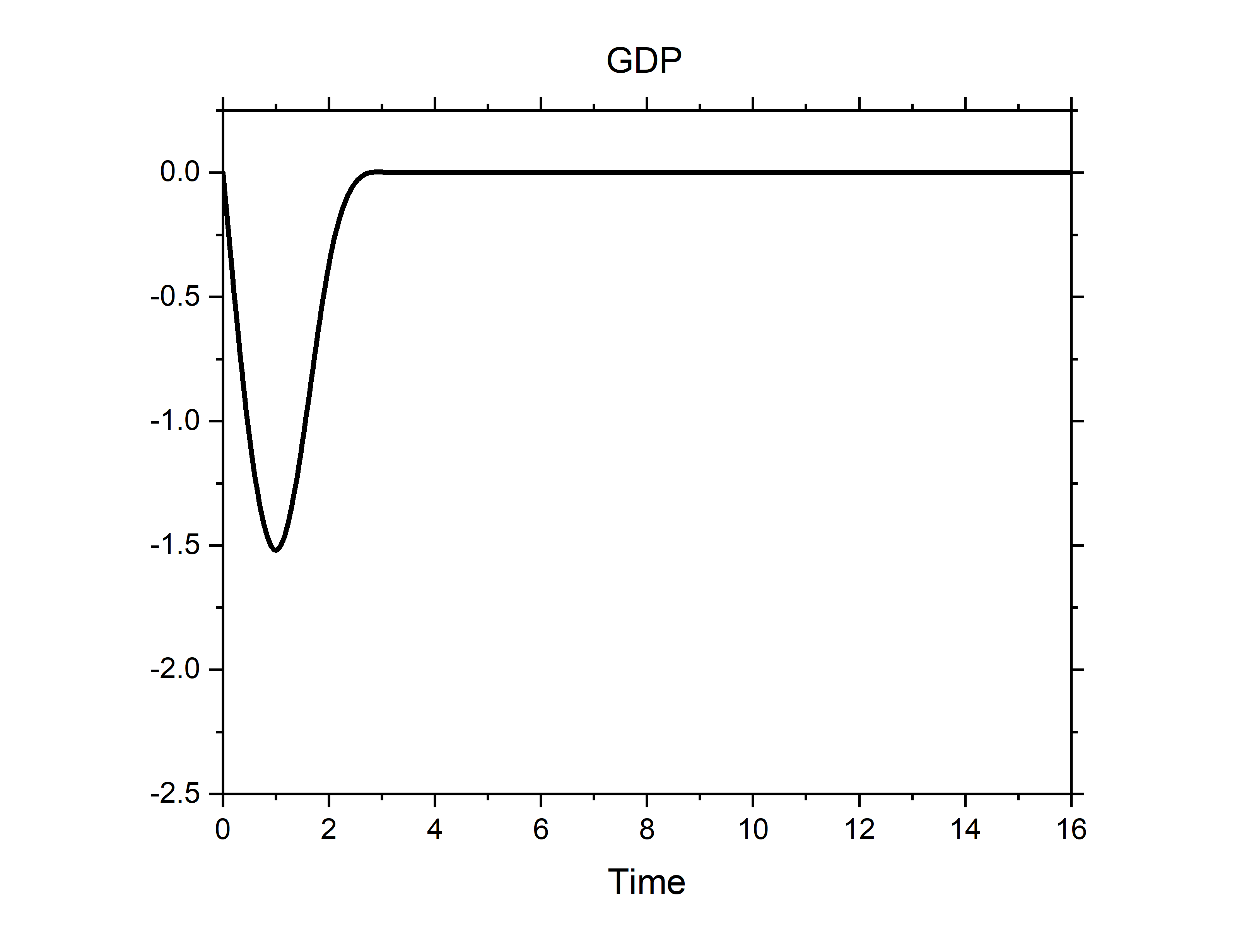}\hspace{-2cm}
    \caption*{GDP}
    \end{minipage}
    }
\quad
    \subfigure{
    \begin{minipage}[b]{0.25\linewidth}
    \includegraphics[width=3cm,height=2cm]{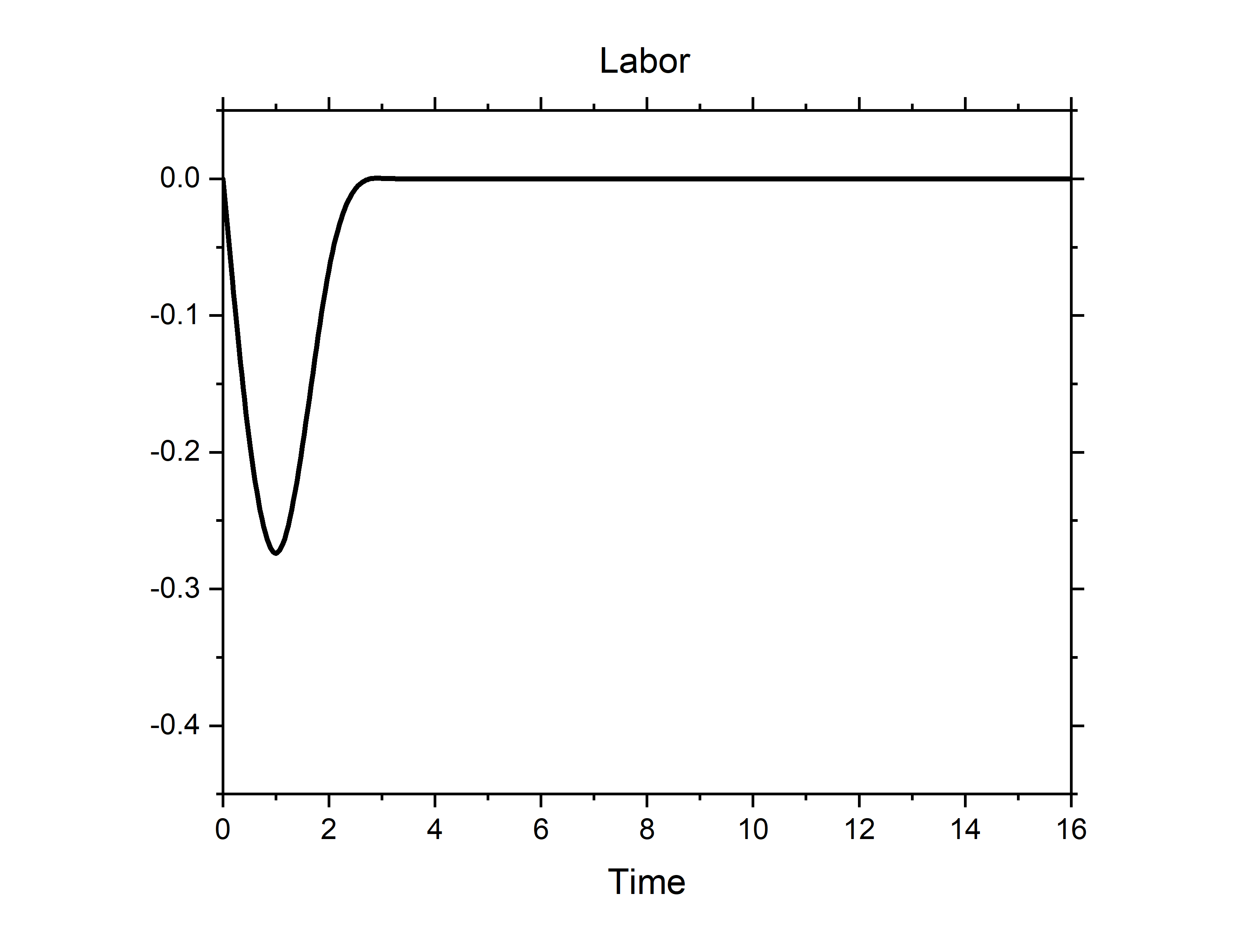}\hspace{-2cm}
    \caption*{Labor}
    \end{minipage}
    }
\quad
\subfigure{
    \begin{minipage}[b]{0.25\linewidth}
    \includegraphics[width=3cm,height=2cm]{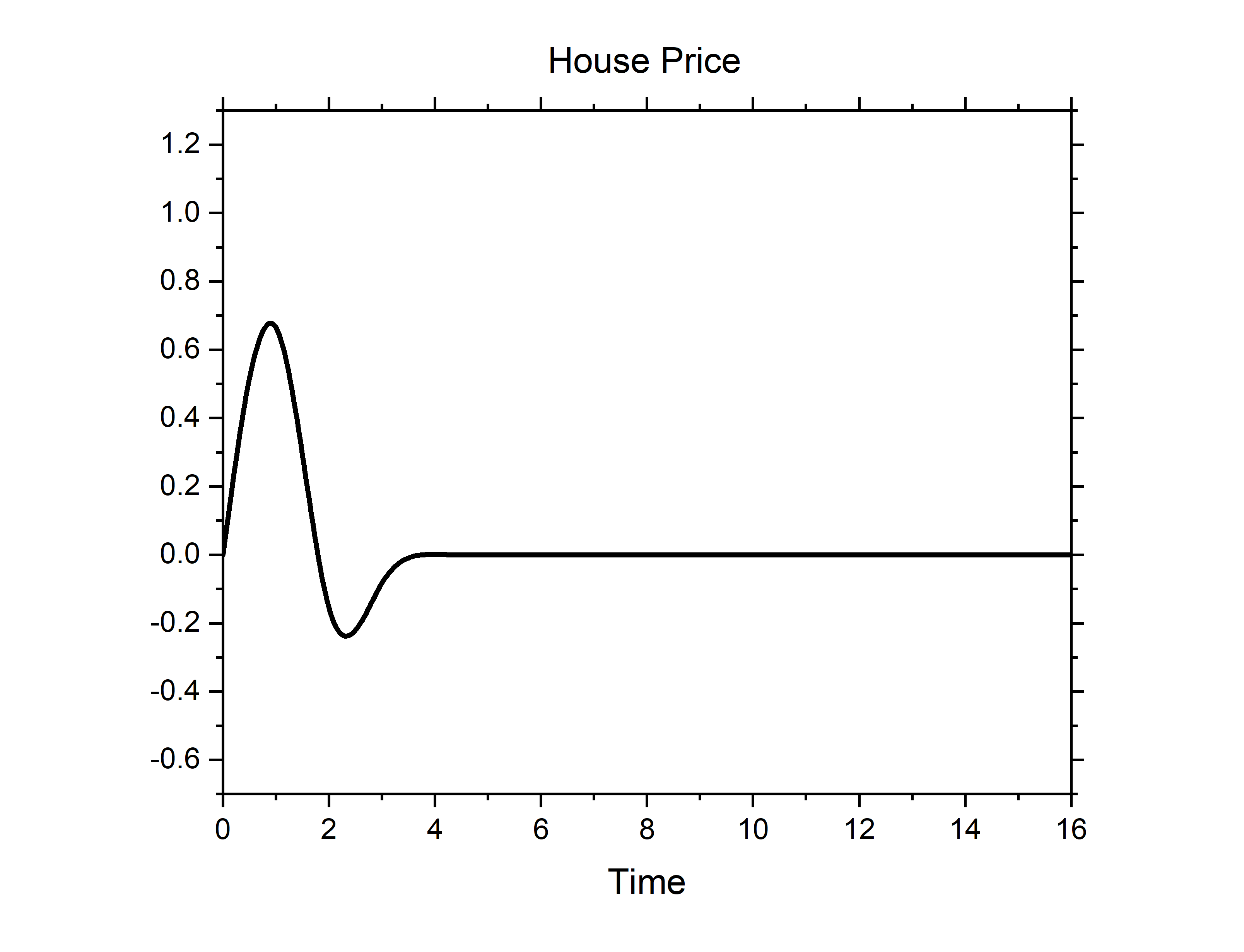}\hspace{-2cm}
    \caption*{Price of real estate}
\end{minipage}
}
\quad
\subfigure{
\begin{minipage}[b]{0.25\linewidth}
     \includegraphics[width=3cm,height=2cm]{R-pnH.png}\hspace{-2cm}
    \caption*{Price of non real estate}
    \end{minipage}
    }
\quad
    \subfigure{
    \begin{minipage}[b]{0.25\linewidth}
    \includegraphics[width=3cm,height=2cm]{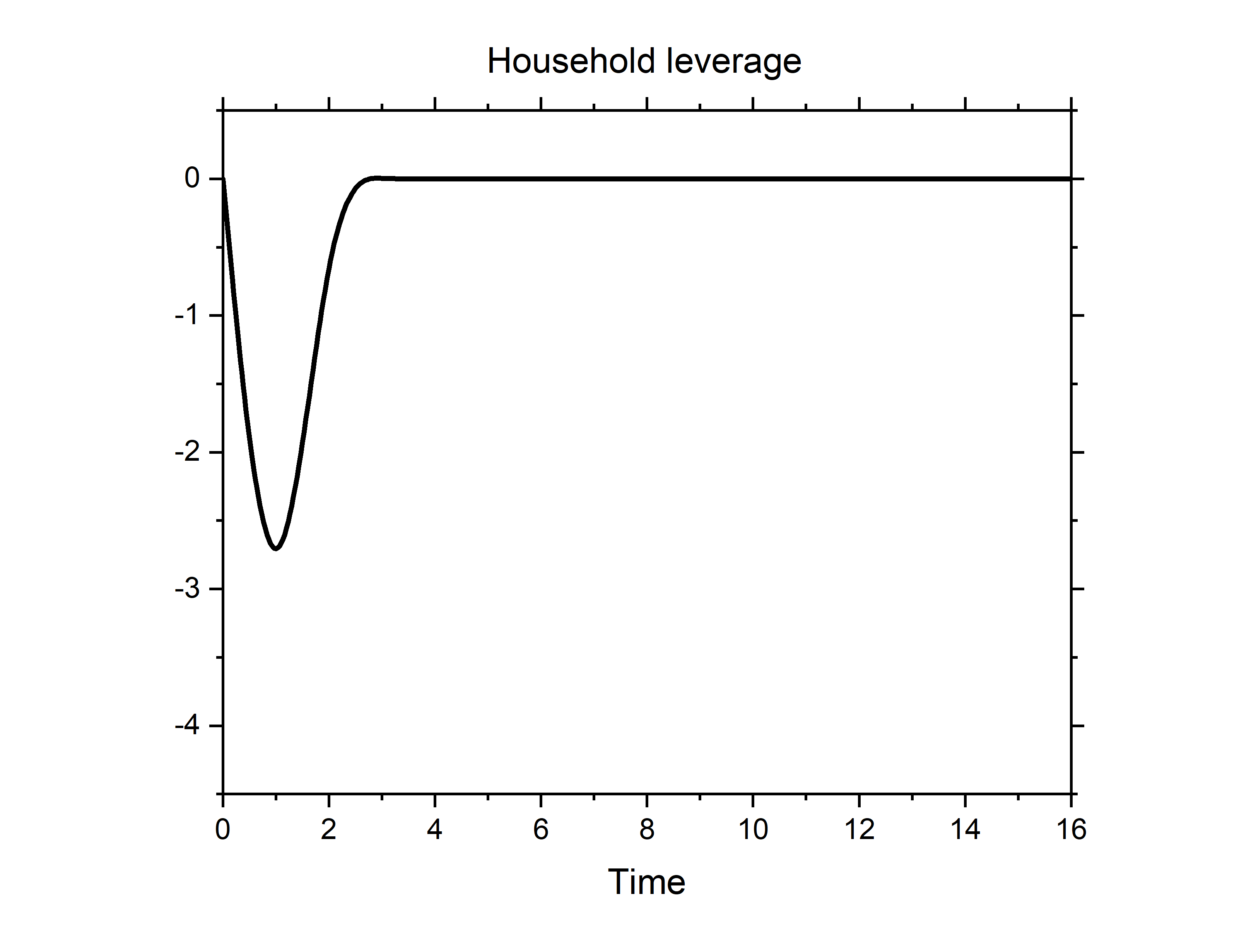}\hspace{-2cm}
    \caption*{Household leverage}
\end{minipage}}
\quad
    \subfigure{
    \begin{minipage}[b]{0.25\linewidth}
    \includegraphics[width=3cm,height=2cm]{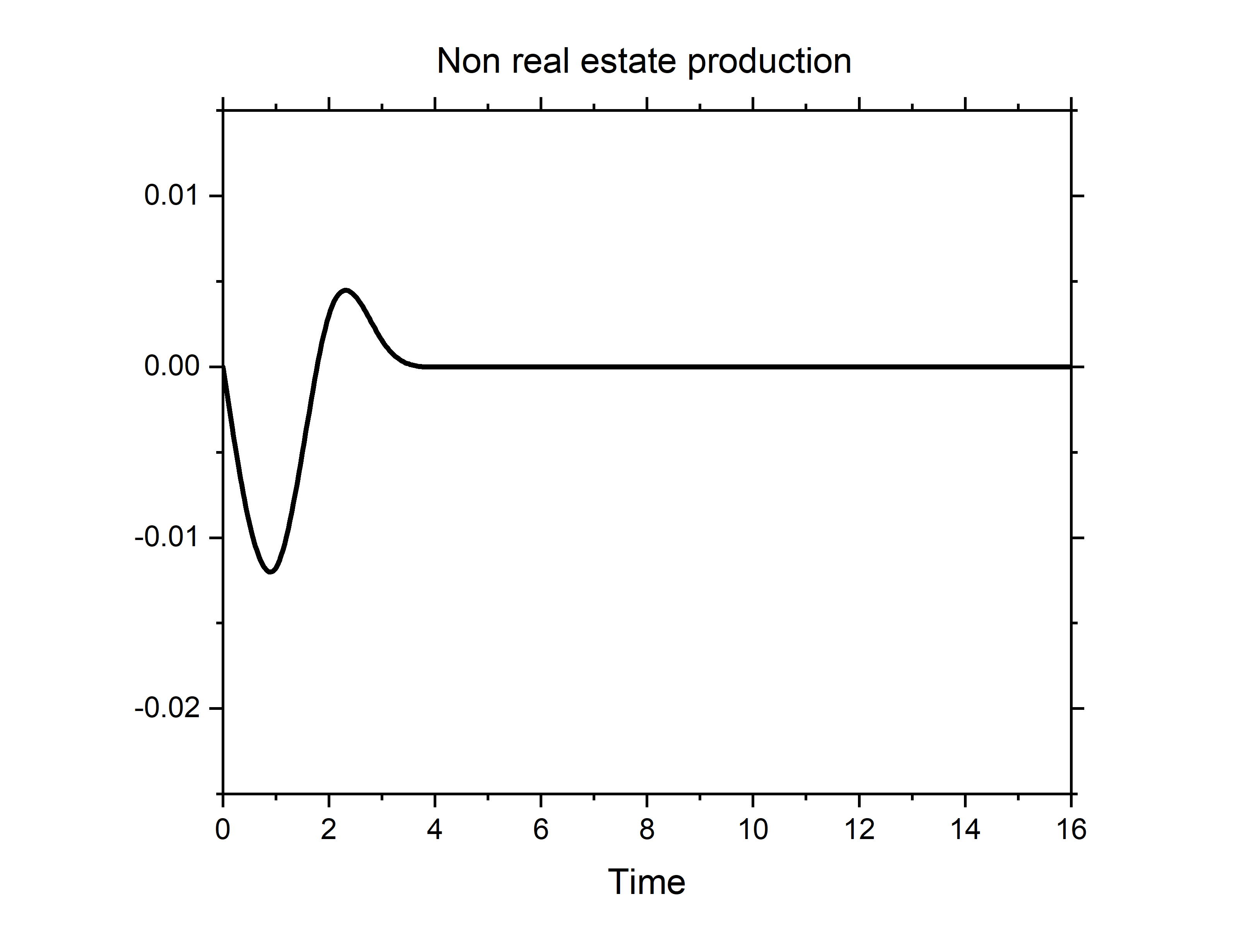}\hspace{-2cm}
    \caption*{Production of non real estate}
\end{minipage}}
\quad
 \subfigure{
    \begin{minipage}[b]{0.25\linewidth}
   \includegraphics[width=3cm,height=2cm]{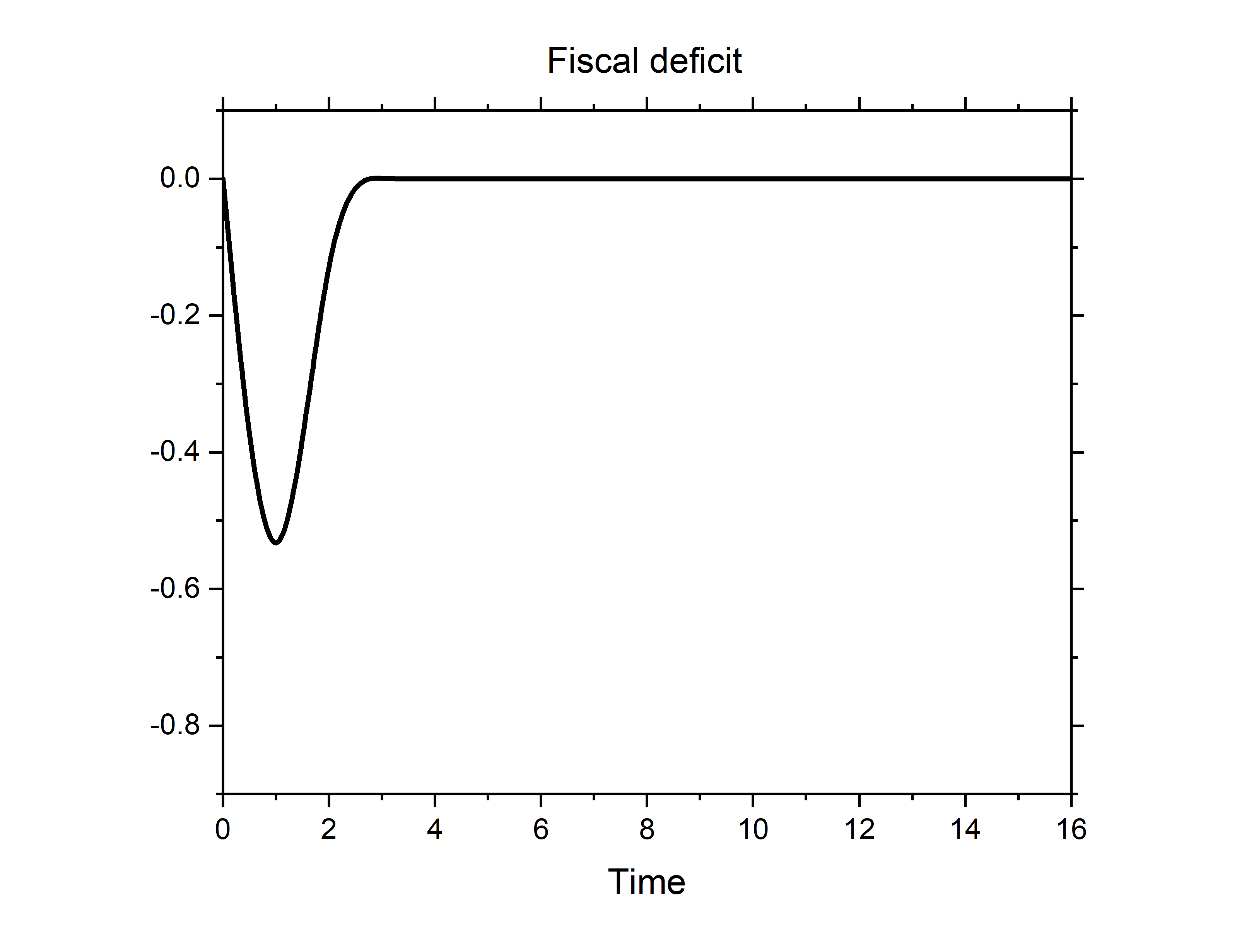}\hspace{-2cm}
    \caption*{Fiscal deficit}
    \end{minipage}
    }
    \caption{Impact Response Analysis by $\xi_{h,t}$}
\end{figure}
%表5描述方差分解的结果。方差分解的结果表明条件分解和无条件分解两种情形下非传统货币政策变量如住房再融资比例、首付比例以及房贷利率溢价都具有不弱于基准利率的调控效果，其中家庭部门的购房行为、房贷杠杆和土地财政起到传导桥梁的作用。主要经济变量如消费、投资、各部门价格、生产产出受四种因素影响的强弱排序在短期和长期都没有明显的区别，财政融资和家庭杠杆率在短期主要受利率的影响，但是在长期，首付比例的影响程度最高，均占50%以上，贷款利率溢价的长期影响仅次于首付比例，这说明货币蓄水池资产杠杆(信贷)渠道在长期的重要作用。
\indent Table 5 depicts the result of the variance decomposition. The result of the variance decomposition shows that non-traditional monetary policy variables such as housing refinancing ratio, down payment ratio and mortgage interest rate premium have a moderating effect that is no weaker than the base interest rate in both the conditional and unconditional decomposition scenarios, with household sector home purchase behaviour, mortgage leverage and land finance acting as a transmission bridge. There is no significant difference in the ranking of the strength of economic variables such as consumption, investment, prices across sectors and production output affected by the four factors in both the short and long run. Financial financing and household leverage are mainly affected by interest rates in the short run, but in the long run, the highest degree of influence is seen in the down payment ratio, which accounts for over 50$\%$ of the total, and the long run impact of the loan interest rate premium is second only to the down payment ratio, suggesting that the asset leverage of the money storage pool channel plays an important role in the long run.\newline
\begin{table}
    \centering
    \resizebox{\textwidth}{!}{
    \begin{tabular}{ccccc}
    \hline
        &$R$&$\mu$&$\theta$&$\xi_h$\\
        &Con./Uncon. &Con./Uncon. &Con./Uncon. &Con./Uncon.\\
        \hline
        Consumption&7.29\% /14.31\% &3.18\% /22.94\%&80.90\%/51.37\% &8.62\%/11.37\% \\
        Investment&71.38\% /14.13\% &11.52\% /3.99\%&0.00\%/51.69\% &17.10\%/30.19\% \\
        Price of real estate sector&60.54\% /57.84\% &6.09\% /8.94\%&29.84\%/29.56\% &3.53\%/3.66\% \\
        Price of non real estate sector&60.77\% /58.29\% &2.09\% /2.27\%&6.59\%/9.14\% &30.54\%/30.29\% \\
        GDP&71.37\% /71.37\% &11.07\% /11.07\%&0.00\%/0.00\% &17.59\%/17.56\% \\
        Production of non real estate sector&33.90\% /33.58\% &4.73\% /4.87\%&8.60\%/9.70\% &52.77\%/51.85\% \\
        Fiscal financing&72.12\% /14.12\% &11.15\% /3.98\%&0.00\%/51.84\% &16.73\%/30.05\% \\
        Home leverage&85.78\% /16.26\% &4.05\% /2.59\%&2.06\%/54.24\% &8.11\%/26.91\% \\
         \hline
    \end{tabular}}
    \caption{Decomposition result by four different shocks}
    \label{tab:my_label}
\end{table}
%总而言之，以往认为是房地产吸纳中央银行超发货币并阻挡了超发货币向一般消费品和生产品部门传递的观点并不成立，原因有二：一是货币蓄水池资产的市值膨胀和溢价涨幅现象的核心原因是资产杠杆，信用扩张导致的通胀现象通过杠杆交易制度放大了部分资产的市值和价格波动，表现出结构性分化“通胀”的局面；二是货币蓄水池资产的杠杆交易制度提供家庭部门的支付承诺，也即劳动收入质押融资，本质上就是一种非典型证券，如果该种资产具有第三方效用(例如地方政府通过推动房地产发展可以获得政绩和增长效应)，那么该种资产有充足的动力自我信用扩张而不受金融中介可贷资金的限制，而风险成本由第三方或者第三方的下一对手方承担。以土地财政为例，地方政府官员受GDP锦标赛的激励，透支未来税收为公共品投资融资产生正外部性推动经济增长都可以理解为一种第三方效用，土地财政及其引发的地方债务问题本质上是地方政府信用扩张的后果。同理2008年以前的美国房地产市场由于第三方金融机构通过转售房地产抵押支持证券能获得丰厚利润，房地产部门开始迅速的自我信用扩张直至资产泡沫过大而造成金融危机。
\indent In sum, the previous view that real estate absorbed the central bank's excess money and blocked its transmission to the general consumer goods and production sectors is not valid for two reasons: first, the core cause of the phenomenon of market value inflation and premium increases in money pool assets is asset leverage, and the inflationary phenomenon caused by credit expansion amplifies the market value and price volatility of some assets through a leveraged trading system, showing Second, the leveraged trading system of monetary pool assets provides payment commitments to the household sector, i.e., pledged financing of labor income, which is essentially an atypical security if it has third-party utility (e.g., local governments can obtain performance and growth effects by promoting real estate development) If the asset has third-party utility (e.g., local governments can obtain performance and growth effects by promoting real estate development), then the asset has ample incentive to expand self-credit without the constraints of lendable funds from financial intermediaries, while the cost of risk is borne by the third party or the next counterparty to the third party. In the case of land finance, for example, local government officials are incentivized by GDP tournaments to overdraw future taxes to finance investment in public goods to generate positive externalities to drive economic growth, which can be interpreted as a third-party utility, and land finance and the resulting local debt problems are essentially the consequences of local government credit expansion. Similarly, in the U.S. real estate market before 2008, as third-party financial institutions were able to make huge profits by reselling real estate mortgage-backed securities, the real estate sector began to rapidly expand its own credit until the asset bubble became too large and caused a financial crisis.
%section:货币蓄水池的长期效应：增长效应或抑制效应？
\section{The short term effect of monetary reservoir}
%贷款品杠杆率：短期高速增长的动力
\subsection{Leverage of loan good:power of economic growth in short term}
%将两个部门的产值做增长率分解
\indent Decompose the output of the two sectors into growth rates
\begin{equation}
    \%(P_{h,t}Y_{h,t}+P_{f,t}Y_{f,t})=\%(C^\sigma_t\omega_{h,t}\frac{(1+R_t)*j_h}{R_t}+j_f)
\end{equation}
%通过解关于利率R_t的差分方程可以解出C_t。因此当基准利率不变时，C_t^σ基本保持不变。P_(h,t) Y_(h,t)+P_(f,t) Y_(f,t)的增长率主要由ω_(h,t)的增长率决定。因此，短期经济增长的主要动力来自于贷款品杠杆。
\indent $C_t$ can be solved by solving the difference equation on the interest rate $R_t$. Thus when the base rate is constant, $C_t^\sigma$ remains essentially constant. the growth rate of $P_{h,t} Y_{h,t} + P_{f,t} Y_{f,t}$ is largely determined by the growth rate of $\omega_{h,t}$. Thus, the main driver of short-term economic growth comes from the leverage of loan goods.
%由于可贷资金的限制，金融中介不可能为贷款品部门提供持续的融资，因此出于刺激短期增长的目的，地方政府为贷款品部门提供新增融资。如图3所示，地方政府债务的增加，等同于银行债权资产的扩大，这增加了金融中介的风险偏好，降低金融中介的利率溢价系数ξ_t，其他的中介变量包括贷款品资产再融资比例μ_t和贷款品资产首付比例θ_t。由于贷款品资产再融资比例μ_t和贷款品资产首付比例θ_t具有非常明显的取值限制，地方政府主要通过财政融资扩大银行资产负债表的方式为银行提供更充足的贷款资金。货币蓄水池资产作为贷款品具有更高的溢价水平，因此金融中介对贷款品部门具有更高的借贷偏好，因此更多的贷款流向贷款品部门，创造更多的经济增长。
\indent Financial intermediaries are unlikely to provide continuous financing for the loan goods sector due to the constraints on loanable funds, so for the purpose of stimulating short-term growth, local governments provide additional financing for the loan goods sector. As shown in Figure 3, an increase in local government debt equates to an expansion in bank debt assets, which increases the risk appetite of financial intermediaries and reduces their interest rate premium coefficient $\epsilon_t$. Other intermediation variables include the refinancing ratio of loan goods assets $\mu_t$ and the down payment ratio of loan goods assets $\theta_t$. Since the refinancing ratio of loan goods assets $\mu_t$ and the down payment ratio of loan goods assets $\theta_t$ have very significant local governments provide banks with more adequate loan funds mainly through fiscal financing to expand banks' balance sheets. Money pool assets have a higher level of premium as loan goods, and therefore financial intermediaries have a higher preference for lending to the loan goods sector, and therefore more loans flow to the loan goods sector, creating more economic growth.
%然而，由于金融中介对借贷资金的比例调整，地方政府的财政杠杆也必然随着贷款品部门的杠杆增长而增长。由金融中介部分的结论可知，地方政府和贷款品部门的杠杆具有联动性。两者的主要区别在于，贷款品的交易机制存在首付比例和再融资机制，因此两者之比只与θ_t和μ_t相关。即便金融中介对地方政府和贷款品部门的贷款风险偏好不完全一致，即对贷款品部门的溢价系数ξ_(h,t)高于对地方政府的利率溢价系数ξ_(G,t)，也并不影响两者的联动性。
\indent However, the fiscal leverage of local governments also necessarily grows in line with the leverage of the loan product sector due to the adjustment of the proportion of borrowed funds by financial intermediaries. The findings from the financial intermediation section show that the leverage of local governments and the loan product sector is linked. The main difference between the two is that there is a down-payment ratio and a refinancing mechanism for the loan product transaction mechanism, so that the ratio between the two is only related to $\theta_t$ and $\mu_t$. Even if financial intermediaries' risk preferences for lending to local governments and the loan product sector are not perfectly aligned, i.e. the premium coefficient $\epsilon_{h,t}$ for the loan product sector is higher than the interest rate premium coefficient $\epsilon_{G,t}$ for local governments, this does not affect the linkage between the two.
\begin{equation}
    \omega_{h,t}=(1-\theta_t)^{-\frac{1}{1+\xi_t}}e^{-\frac{1}{1-\xi_t}}R^{-\frac{1}{1+\xi_t}}_t(1-\mu_t)^{-\frac{\xi_t}{1+\xi_t}}
\end{equation}
\begin{equation}
    \omega_{G,t}=e^{-\frac{1}{1-\xi_t}}R^{-\frac{1}{1+\xi_t}}_t
\end{equation}
%因此地方政府实际是贷款部门的杠杆率增加的重要推手，虽然在资产负债表上，地方政府并非贷款品部门的实际债权人，这种隐含的债权债务关系被金融中介的第三方借贷融资关系所掩盖。地方政府成立融资平台吸收社会资金，形成政府债务，通过扩大财政支出和财政资金注入的方式为贷款品部门提供融资便利，而地方政府的行为动机是刺激经济增长以获取政绩或者维持就业。由于委托代理问题的存在，每任地方官员并不对下一任官员和未来地方财政状况负责，因此几乎每任地方官员都会选择短平快的增加财政杠杆推动经济增长的方式，而不会选择见效慢的有利于长期增长的方式。
\indent Local governments are therefore in fact an important driver of increased leverage in the lending sector, although on the balance sheet they are not actual creditors of the lending sector, and this implied debt relationship is masked by third-party lending and financing relationships with financial intermediaries. Local governments set up financing platforms to absorb social capital, form government debt and facilitate financing for the lending sector by expanding fiscal expenditure and injecting fiscal funds, while local governments act with the motive of stimulating economic growth to obtain political performance or maintain employment. Because of the principal-agent problem, each local official is not responsible for the next official and the future local financial situation, so almost every local official will choose the short and quick way to increase fiscal leverage to promote economic growth, rather than the slow way to facilitate long-term growth.
%资源歧视性配置
\subsection{Discriminatory allocation of resources}
%货币蓄水池机制将通过扭曲资源配置极大地削弱宏观政策的有效性。将K_(h,t)/K_(f,t) 、L_(h,t)/L_(f,t) 、n_(h,t)/n_(f,t) 和G_t的增长要素进行分解：
%经济增长结构失衡:
\indent The monetary reservoir mechanism will greatly reduce the effectiveness of macro policy by distorting resource allocation. Decomposing the growth factors of $\frac{K_{h,t}}{K_{f,t}}$, $\frac{L_{h,t}}{L_{f,t}}$, $\frac{n_{h,t}}{n_{f,t}}$ and $G_t$.
\begin{equation}
\begin{aligned}
    \%\frac{K_{h,t}}{K_{}f,t}&=\%[C^\sigma_t\omega_{h,t}(1+R_t)\frac{j_h\phi_h}{R^{1+\frac{1}{1+\xi_t}}_t}[(1-\theta_t)(1-\mu_t)*e]^{\frac{-\xi_t}{1+\xi_t}}\div \frac{j_f\phi_f}{R^{\frac{1}{1+\xi_t}}_te^{\frac{-\xi_t}{1+\xi_t}}}]\\
    &\propto\%[\frac{\omega_{h,t}(1+R_t)}{[(1-\theta_t)(1-\mu_t)]^{\frac{\xi_t}{1+\xi_t}}R_t}]
\end{aligned}
\end{equation}
%假设基准利率不变，因此(-1)/(1+ξ_t )%R_t=0，
\indent Assuming that the base rate is unchanged and therefore $\frac{-1}{1+\xi_t}\%R_t=0$.
\begin{equation}
    \%\frac{K_{h,t}}{K_{f,t}}\propto\%(\omega_{h,t}-\frac{\xi_t}{1+\xi_t}\%(1-\theta_t))-\frac{\xi_t}{1+\xi_t}\%(1-\mu_t)
\end{equation}
%贷款品和非贷款品两部门的借贷资本分配比例之比变化由μ_t、θ_t和ξ_t三个因素决定(ω_(h,t)也是这三个因素的函数)。金融中介放松借贷约束，增加两部门的借贷资本，但是也扩大了两部门的信贷资源配置分化程度。同理对两部门的土地和劳动力配置比例之比作增长率分解可得
\indent The change in the ratio of the allocation of borrowed capital to the two sectors, lending and non-lending, is determined by three factors, $\mu_t$, $\theta_t$ and $\epsilon_t$ ($\omega_{h,t}$ is also a function of these three factors). The easing of the borrowing constraint by financial intermediaries increases borrowing capital in both sectors, but also widens the degree of divergence in the allocation of credit resources between the two sectors. Similarly a growth rate decomposition of the ratio of land and labour allocation in the two sectors yields the following equation.
\begin{equation}
    \%(\frac{L_{h,t}}{L_{f,t}})\propto\%(\omega_{h,t}=\%[(1-\theta_t)^{\frac{-1}{1+\xi_t}}e^{\frac{-1}{1-\xi_t}}R^{-\frac{1}{1+\xi_t}}_t(1-\mu_t)^{\frac{-\xi_t}{1+\xi_t}}])
\end{equation}
\begin{equation}
    \%(\frac{n_{h,t}}{n_{f,t}})\propto\%(\omega_{h,t}=\%[(1-\theta_t)^{\frac{-1}{1+\xi_t}}e^{\frac{-1}{1-\xi_t}}R^{-\frac{1}{1+\xi_t}}_t(1-\mu_t)^{\frac{-\xi_t}{1+\xi_t}}])
\end{equation}
%金融中介放松对贷款品部门的信贷偏好也加剧其他生产要素在两部门之间配置的分化程度。由于土地和劳动并没有财政融资供给新增可贷资金，因此在给定供给数量下，金融中介对贷款品部门的信贷偏好反而挤出了其他部门的生产要素，地方政府通过财政融资投资和金融中介向贷款品部门间接贷款的过程也是贷款品部门在要素市场挤出非贷款品部门的过程。由于有地方政府的隐形财政支持和金融中介的信贷偏好，贷款品部门能够付出更高的土地出让金和工资福利，自然能购买更多土地使用权和劳动力。
\indent The easing of credit preferences by financial intermediaries for the loan goods sector also increases the degree of divergence in the allocation of other factors of production between the two sectors. Since land and labour are not financially available to supply new loanable capital, the credit preference of financial intermediaries for the lending sector crowds out other factors of production in other sectors for a given quantity of supply, and the process of indirect lending by local governments to the lending sector through financially-financed investment and financial intermediaries is also a process by which the lending sector crowds out the non-lending sector in the factor market. With the invisible financial support of local governments and the credit preference of financial intermediaries, the lending goods sector is able to pay higher land premiums and wages and benefits, and is naturally able to purchase more land use rights and labour.
%因此，地方政府和贷款品部门的杠杆率上升的同时，越来越多的生产要素被歧视性配置地过多分配到推动短期经济总量增长而无益于长期内生增长的贷款品部门。
\indent As a result, leverage in the local government and loan goods sectors has risen while an increasing number of factors of production have been discriminatorily over-allocated to the loan goods sector, which drives short-term aggregate economic growth without contributing to long-term endogenous growth.
\subsection{Structural imbalances in economic growth}
%在不改变总土地供给、劳动供给和货币政策的前提下，政府对经济的调控主要通过影响资本K_(h,t)、K_(f,t)和G_t三种中介变量达到政策目的。但是非贷款品部门的借贷资本只与金融中介的利率溢价系数的变化有关，如果金融中介和地方政府不改变贷款品的首付比例和再融资比例，新增借贷的增长率仅由金融中介的利率溢价系数决定，结合货币蓄水池溢价来源的因素分解可以得到方程。
\indent Without changing aggregate land supply, labour supply and monetary policy, government regulation of the economy achieves its policy objectives mainly by influencing the three intermediary variables of capital $K_{h,t}$, $K_{f,t}$ and $G_t$. However, the borrowed capital in the non-loan goods sector is only related to the change in the interest rate premium coefficient of the financial intermediary. If the financial intermediary and the local government do not change the down payment ratio and refinancing ratio of loan goods, the growth rate of new borrowing is only determined by the interest rate premium coefficient of the financial intermediary, and the factor decomposition of the source of the monetary reservoir premium can be combined to obtain the following equation.
\begin{equation}
\begin{aligned}
    \%K_{h,t}&=\%\omega_{h,t}-\frac{\xi_t}{1+\xi_t}[\%(1-\theta_t)+\%(1-\mu_t)-\Delta\frac{\xi_t}{1+\xi_t}]\\
    &\propto\%\omega_{h,t}-\Delta\frac{\xi_t}{1+\Delta_t}=-\Delta(\frac{1}{1-\xi_t})-\Delta\frac{\xi_t}{1+\xi_t}\\
    &=\Delta[\frac{1}{1-\xi_t}-\frac{\xi_t}{1+\xi_t}]=\Delta[\frac{1+\xi^2_t}{1-\xi^2_t}]
\end{aligned}
\end{equation}
\begin{equation}
    \%I_t=\%K_t-\%K_{t-1}=\Delta[\frac{1+\xi^2_t}{1-\xi^2_{t}}]-\Delta[\frac{1+\xi^2_{t-1}}{1-\xi^2_{t-1}}]
\end{equation}
%上述方程是三阶差分方程，通过模拟分析可知，随着模拟期数的增加，为了保持投资增长率固定，，ξ_t的曲线逐渐趋向于0，这预示更高的地方政府财政杠杆率和贷款品部门杠杆率,金融危机发生的概率逐渐上升。因此地方政府和金融中介都认为维持适当的利率溢价有利于经济稳定，从而主要通过扩大财政开支和增加地方政府杠杆率维持增长。这种增长结构依旧造成地方政府的债务危机。对财政投资的增长率再做分解可得以下方程。
\indent The above equation is a third-order difference equation and the simulation analysis shows that as the number of simulation periods increases, the curve of $\epsilon_t$ gradually tends to zero in order to maintain a certain rate of investment growth $\%I_t=\bar{d}$, which predicts higher local government fiscal leverage and lending goods sector leverage, and therefore a gradually increasing probability of financial crisis. Both local governments and financial intermediaries therefore perceive that maintaining an appropriate interest rate premium is conducive to economic stability and thus maintain growth mainly by expanding fiscal spending and increasing local government leverage. This growth structure continues to contribute to the debt crisis of local governments.A further decomposition of the growth rate of financial investment yields the following equation.
\begin{equation}
\begin{aligned}
    \%G_t&=\%[\{C^\sigma_t\omega_{h,t}(1+R_t)\frac{j_h}{R_t}\phi_h+j_f\phi_f\}(\omega_{G,t}-\frac{1+i_t}{i_t}(1-(1+i_t)^{-\omega_{G,t}}))\\
    &\propto\{\%\omega_{h,t}\}+\%[\omega_{h,t}-\frac{1-(1+i_t)^{-\omega_{G,t}}}{Log[1+i_t]}]\\
    &\approx \%[\omega_{h,t}]+\%[\omega_{G,t}]
\end{aligned}
\end{equation}
%在货币蓄水池机制中，当基准利率不变时，金融中介的利率溢价系数ξ_t将发挥主要的调控作用。第二重要的调控路径是首付比例θ_t和贷款品再融资比例μ_t。这三个非传统的政策变量构成了货币蓄水池机制的核心架构，地方政府维持较高的财政杠杆，通过财政投资为金融中介提供更多可贷资金，降低利率溢价系数ξ_t吸引投资或者直接扩大财政支出。首付比例θ_t和贷款品再融资比例μ_t两者也存在替代关系，当首付比例较高时，金融中介通过提高再融资比例也能达到事实上的降首付比例和提高贷款品杠杆的效果，具体操作如下：贷款品购买者寻求过桥资金按首付比例θ_t买入资产，获得资产所有权后再向金融中介抵押贷款品获得μ_t比例的资金偿还过桥资金。这正是中国房地产市场中常见的“赎楼贷”金融产品的主要运作方式。这种产品为购买者提供了更便利的加杠杆服务，也增强了货币蓄水池机制对宏观经济的影响。
\indent In the monetary reservoir mechanism, the interest rate premium coefficient $\epsilon_t$ of financial intermediation will play a major regulatory role when the benchmark interest rate is unchanged. The second important regulatory path is the downpayment ratio $\theta_t$ and the loan refinancing ratio $\mu_t$. These three non-traditional policy variables form the core structure of the monetary pool mechanism, with local governments maintaining high fiscal leverage, providing more loanable funds to financial intermediaries through fiscal investment, lowering the interest rate premium coefficient $\epsilon_t$ to attract investment or directly expanding fiscal spending. There is also a substitution relationship between the down payment ratio $\theta_t$ and the refinancing ratio $\mu_t$ of loan products. When the down payment ratio is high, the financial intermediary can also achieve the effect of de facto lowering the down payment ratio and increasing the leverage of loan products by increasing the refinancing ratio, as follows: the loan product purchaser seeks bridging funds to buy the asset at the down payment ratio $\theta_t$, acquires ownership of the asset and then pledges the loan product to the financial intermediary to obtain $mu_t$ ratio to repay the bridging funds. This is the main modus operandi of the "foreclosure" financial product commonly found in the Chinese real estate market. Such products provide purchasers with a more convenient way to leverage and enhance the macroeconomic impact of the monetary reservoir mechanism.
%财政投资依赖性强
\subsection{Strong reliance on financial investment}
%扩张信用形成财政盈余的重要途径是政府对公共品部门的大额投资。财政投资产生正外部性推动社会总体生产率的提升，为地方政府带来更多的所得税额以及潜在福利，例如人口增长和科技创新。生产部门无需支付公共品投资也能获得更高的收入当期公共品的正外部性大幅提高经济产值形成税收返还，财政赤字得到缓解甚至达到盈余。这也解释了21世纪初中国各地高速发展的另一重要原因，主要经验是给予地方政府一定融资自主权，基于货币蓄水池资产自行信用扩张为公共品部门融资，并在公共品正外部性的支持下反哺非货币蓄水池资产的部门达到共同高速增长。
\indent In this paper, the process by which fiscal investment leads to a reduction in total factor productivity is referred to as the growth effect and vice versa as the dampening effect. An important way to generate fiscal surpluses from credit expansion is through large government investments in the public goods sector. Fiscal investment generates positive externalities that boost overall social productivity, generating more income tax revenue for local governments and potential benefits such as population growth and technological innovation. The productive sector earns higher revenues without having to pay for public goods investments. The positive externalities of public goods in the current period significantly increase economic output to form tax rebates, and fiscal deficits are mitigated or even reach surpluses. This also explains another important reason for the rapid growth across China in the early 21st century. The main lesson is that local governments were given some financing autonomy to finance the public goods sector based on their own credit expansion of monetary reservoir assets, and to feed the non-monetary reservoir assets sector with the support of positive externalities of public goods to achieve joint high growth.\newline
%根据公式23，λ_t^P大于0时表示政府投资产生净收益(增长效应)，否则产生净亏损(抑制效应)。增长效应高于抑制效应的结果是由公共品提高的生产效率增长与信用扩张并存，当公共品部门的投资效率下降，过度信用扩张造成货币增发，新增政府投资并没有提高产出并且也没有增加财政收入，在长期必然造成财政赤字，财政赤字也通过中央银行购买地方债务或者中央财政向地方政府转移支付的方式解决，这两种方式都必然导致愈加宽松的货币政策。
\indent According to Equation 23, $\lambda_t^P$ is greater than zero when government investment generates a net gain (growth effect) and a net loss (disincentive effect) otherwise. When the efficiency of investment in the public goods sector decreases, excessive credit expansion causes an increase in money issuance, new government investment does not increase output and does not increase fiscal revenue, which inevitably leads to a fiscal deficit in the long run, which is also solved by the central bank purchasing local debt or the central government transferring money to local governments. The fiscal deficit is also solved by the central bank's purchase of local debt or the central government's transfer payments to local governments, both of which inevitably lead to an increasingly accommodative monetary policy.\newline
%根据宏观均衡的公式29，为了维持产品市场出清，公共品部门的外部性参数λ_t^P需要满足公式30(见附录A)。公式(30)清楚地体现公共品效率的主要决定因素和长期变动趋势。首先要维持公共品部门的正外部性，即财政投资的增长效应而非抑制效应，政府投资需要确保总是大于折历史已投资公共品总额的折旧额，即财政投资应该是净投资，G_t>δ_k ∑_(i=0)^(t-1) G_i 。
\indent According to equation 29 of the macro equilibrium, in order to maintain product market clearing, the externality parameter $\lambda_t^P$ in the public goods sector needs to satisfy equation 30 (see Appendix A). Equation (30) clearly captures the main determinants of public goods efficiency and the trend of long-term changes. First, to maintain positive externalities in the public goods sector, i.e., the growth effect rather than the dampening effect of fiscal investment, government investment needs to ensure that it is always greater than the depreciation of the total amount of historical invested public goods, i.e., fiscal investment should be a net investment
\begin{equation}
    \lambda^P_t=\frac{Log[C_t+I_t+G_t]-Log[K^{\rho_h}_{h,t}L^{\psi_h}_{h,t}n^{1-\rho_h-\psi_h}_{h,t}+K^{\rho_f}_{f,t}L^{\psi_f}_{f,t}n^{1-\rho_f-\psi_f}_{f,t}]}{G_t-\delta_k\sum^{t-1}_i{G_i}}
\end{equation}
%然而当δ_k>0时，长期∑_(i=0)^(t-1)▒G_i 是随时间快速增长的函数，因此G_t-δ_k ∑_(i=0)^(t-1) G_i必然存在明显的顶峰，越过峰值点后，λ_t^P开始下降并且必将逐渐小于0。λ_t^P<0意味着财政投资对产出的弹性为负：Φ_t λ_t^P<0，同等要素投入下产出降低，这推动长期实际物价水平的提高。
%另一方面，降低贷款品部门的杠杆，刺激消费和投资将有利于使得财政投资的边际效益转为正。对此结论的经济学现实解释是，当以往财政投资(主要是基础设施)已经过度投资时，转变以(财政)投资和增加微观部门杠杆率ω_(h,t)和G_t=f(ω_(G,t))为主的经济增长方式，转为依赖消费和投资的经济增长方式，这有利于降低产出减少带来的经济负效应。该结论符合当前中国地方政府高杠杆和政府投资拉动的经济现象以及地方政府、房地产部门和家庭部门杠杆率节节攀升的困境。
\indent However, when $\delta_k>0$, long-run $\sum^{t-1}_i{G_i}$ is a fast-growing function with time, so $G_t-\delta_k\sum^{t-1}_i{G_i}$ must have a clear peak, after which $\lambda_t^P$  starts to fall and must gradually become less than 0. $\lambda_t^P<0$ means that the elasticity of fiscal investment with respect to output is negative: $\Phi_t \times\lambda_t^P<0$ and output decreases for the same factor inputs, which drives up the long-run real price level.
On the other hand, reducing the leverage of the loan sector and stimulating consumption and investment will help to make the marginal benefit of fiscal investment turn positive. The economically realistic interpretation of this conclusion is that when previous fiscal investment (mainly in infrastructure) has been overinvested, shifting from an economic growth approach dominated by (fiscal) investment and increased micro-sector leverage $\omega_(h,t)$ and $G_t=f(\omega_(G,t))$ to one that relies on consumption and investment is beneficial in reducing the negative economic effects of reduced output. This finding is consistent with the current economic phenomenon of high local government leverage and government investment pull in China and the dilemma of rising leverage in the local government, real estate sector and household sector.\newline
%许多中国学者对中国高投资和政府驱动下的公共品投资有不少研究(刘勇政等,2021;陈斌开等,2016;范子英等,2018)，中国政府投资的正外部性主要来源于提供财政补贴所吸收的技术、管理以及人力资本积累，也即传统生产要素之外的增长，然而技术等非传统生产要素在不发生革命性创新条件下是边际递减的，人力资本积累在长期也依赖于人口增长，另外对某个部门的信贷偏好也将损害其他部门的生产效率(罗知,2015;张杰,2016;彭俞超,2018)，因此λ_t^P在长期将逐渐减少，上述发展模式在长期必然导致货币增发，这一点在技术、人力资本等生产要素吸收不足的地区更为明显。因此，当政府乐于投资教育和科技等几乎无折旧和损耗的公共品时，上述推论并不成立。因为当δ_k趋近于零时，∑_(i=0)^(t-1)▒G_i 的长期增长并不要求新增财政投资覆盖并不存在的财政“折旧”。但是现实中，大多数中国的地方政府投资公共基础设施的比例远大于教育和科技，对长期增长造成显著的负作用，虽然在短期它推动经济增长和为官员带来经济绩效，这是一种特殊的效用。这也是地方政府更多投资于基础设施的原因，因此δ_k>0。
\indent Many Chinese scholars have conducted a number of studies on high investment and government-driven investment in public goods in China (Liu Yongzheng et al,2021; Chen Binkai et al,2016; Fan Ziying et al,2018), the positive externalities of Chinese government investment are mainly derived from the accumulation of technology, management, and human capital absorbed by the provision of financial subsidies, i.e., growth beyond traditional factors of production, however, non-traditional factors of production such as technology are marginal decreasing in the absence of occur under the condition of revolutionary innovation is marginal decreasing, human capital accumulation in the long run also depends on population growth, in addition credit preference for one sector will also harm the productivity of other sectors (Luo Zhi,2015; Zhang Jie,2016; Peng Yu Chao,2018), therefore $\lambda_t^P$ will gradually decrease in the long run, the above development model in the long run will inevitably lead to increased monetary issuance, which is in the technology, human capital and other factors of production are not absorbed enough in the region is more obvious. Therefore, the above corollary does not hold when the government is happy to invest in public goods with almost no depreciation and depletion, such as education and technology. Because when $\delta_k$ tends to zero, the long-run growth of $\sum^{t-1}_i{G_i}$ does not require additional fiscal investment to cover the non-existent fiscal "depreciation". In reality, however, most Chinese local governments invest much more in public infrastructure than in education and science and technology, which has a significant negative effect on long-run growth, although in the short run it has a particular utility in driving economic growth and economic performance for officials. This is the reason why local governments invest more in infrastructure, so $\delta_k>0$.\newline
%其次，新增投资、当期消费以及政府投资应该大于两大部门的总产出。在不改变总土地供给和劳动供给的前提下，政府对经济的调控主要通过影响资本K_(h,t)、K_(f,t)和G_t三种中介变量达到政策目的。其中I_t也是K_(h,t)和K_(f,t)的函数。将K_(h,t)、K_(f,t)和G_t的增长要素进行分解：
\indent Second, new investment, current consumption, and government investment should be greater than the total output of the two major sectors. Without changing the aggregate land supply and labor supply, government regulation of the economy mainly achieves policy objectives by influencing three mediating variables, capital $K_{h,t}, K_{f,t}, and G_t$. Where $I_t$ is also a function of $K_{h,t}$ and $K_{f,t}$. Decomposing the growth elements of $K_{h,t}, K_{f,t}, and G_t$.
\begin{equation}
    \%K_{h,t}\propto \%\omega_{h,t}+\frac{1}{1-\xi_t}\%R_t-\%(\sum^{+\infty}_{k=t}{\frac{\beta^k\ C^{-\sigma}_k}{(1+R_k)^k}}
\end{equation}
\begin{equation}
    \%K_{f,t}\propto \frac{1}{1-\xi_t}\%R_t
\end{equation}
\begin{equation}
\begin{aligned}
    \%G_{t}&\propto \frac{1}{1-\xi_t}\%\omega_{h,t}-\%E(\sum^{+\infty}_{k=t}{\frac{\beta^k\ C^{-\sigma}_k}{(1+R_k)^k}}+\%[\omega_{h,t}-\frac{1-(1+i_t)^{-\omega_{G,t}}}{Log(1+i_t)}]\\
    &\approx \%\omega_{h,t}-\%E(\sum^{+\infty}_{k=t}{\frac{\beta^k\ C^{-\sigma}_k}{(1+R_k)^k}}+\%[\omega_{G,t}]
\end{aligned}
\end{equation}
%通过对三个变量的分解可知，它们共同的增长要素是R_t、ω_(h,t)、ω_(G,t)以及 E(∑_(k=t)^(+∞)▒〖(〖β/ρ)〗^k C_k^(-σ) 〗)。在货币蓄水池机制中，当基准利率不变时，金融中介的利率溢价系数ξ_t和首付比例θ_t将发挥主要的调控作用。第二重要的调控路径时E(∑_(k=t)^(+∞)▒〖(〖β/ρ)〗^k λ_k 〗)中包含的贷款品再融资比例μ_t。因此，这三个非传统的政策变量构成了货币蓄水池的核心架构，地方政府维持较高的财政杠杆，挤出了其他部门的信贷资金，推高了利率溢价系数ξ_t。首付比例和再融资比例两者存在替代关系，当首付比例较高时，金融中介通过提高再融资比例也能达到事实上的降首付比例和提高贷款品杠杆的效果，具体操作如下：贷款品购买者寻求过桥资金按首付比例θ_t买入资产，获得资产所有权后再向金融中介抵押贷款品获得μ_t比例的资金偿还过桥资金。这正是中国房地产市场中常见的“赎楼贷”金融产品的主要运作方式。这种产品为购买者提供了更便利的加杠杆服务，也增强了货币蓄水池机制对宏观经济的影响。
\indent The decomposition of the three variables shows that their common growth factors are $R_t, \omega_(h,t), \omega_(G,t)$, and $E(\sum^{+\infty}_{k=t}{\frac{\beta^k\ C^{-\sigma}_k}{(1+R_k)^k}}$. In the monetary reservoir mechanism, the interest rate premium coefficient $\xi_t$ and the down payment ratio $\theta_t$ of the financial intermediary will play the main regulatory role when the benchmark interest rate is unchanged. The second important regulatory path is the refinancing ratio $\mu_t$ of loans included in $E(\sum^{+\infty}_{k=t}{\frac{\beta^k\ C^{-\sigma}_k}{(1+R_k)^k}}$. Thus, these three non-traditional policy variables form the core structure of the monetary reservoir, with local governments maintaining high fiscal leverage, squeezing out credit funds from other sectors, and pushing up the interest rate premium coefficient $\xi_t$. The down payment ratio and the refinancing ratio are substitutes. There is a substitution relationship between the two. When the down payment ratio is high, the financial intermediary can also achieve the de facto effect of lowering the down payment ratio and increasing the leverage of the loan product by increasing the refinancing ratio as follows: the loan product purchaser seeks bridging funds to buy the asset at the down payment ratio $\theta_t$, obtains ownership of the asset and then pledges the loan product to the financial intermediary to obtain $mu_t$ proportion of funds to repay the bridging funds. This is the primary mode of operation of the "foreclosure" financial product common in the Chinese real estate market. This product provides a more convenient leveraging service for buyers and enhances the macroeconomic impact of the monetary reservoir mechanism.\newline
%总结以上过程，货币蓄水池-财政融资投资的货币传导机制是一种自发性的信用扩张，属于非传统的货币传导机制的一部分。这种过程在长期必然转化为中央银行的货币超发，其主要原因在于两点：一、该机制下的经济增长模式依赖于地方政府、贷款品部门和家庭部门的高杠杆，在长期高杠杆必然意味着财政赤字或者普遍的债务危机，这将导致货币政策不得不增发货币通过通货膨胀稀释债务，因此该机制的运行就是储蓄“未来超发货币”的过程，因此名为“货币蓄水池”；二、货币蓄水池机制与金融中介之间的相辅相成的关系导致贷款品部门和非贷款品部门之间的信贷不平衡。即便货币政策不变，金融中介也通过各种金融工具为贷款品部门提供新增融资，而非贷款品部门则无缘于此。
\indent To summarize the above process, the monetary transmission mechanism of monetary reservoir-fiscal financing investment is a spontaneous credit expansion that is part of a non-traditional monetary transmission mechanism. This process inevitably translates into monetary over-issuance by the central bank in the long run for two main reasons: First, the economic growth model under this mechanism relies on high leverage of local governments, the lending sector and the household sector, and high leverage in the long run inevitably means fiscal deficits or a general debt crisis, which will lead to monetary policy having to issue more money to dilute debt through inflation, so the operation of this mechanism is The process of saving "future excess money", hence the name "monetary reservoir"; second, the complementary relationship between the monetary reservoir mechanism and financial intermediation leads to credit imbalances between the lending and non-lending sectors. Even if the monetary policy remains unchanged, financial intermediaries provide new financing for the lending sector through various financial instruments, while the non-lending sector has no access to it.
\section{The long term effect of monetary reservoir}
%在本小节本文扩展了Romer(1987，1990)构建的内生技术增长模型，假定劳动者可以选择将劳动供应当期产品部门或者投入研究部门Z_t获取未来收入的折现值，劳动在生产部门和研究部门的分配比例分别是α_t和1-α_t。生产部门是完全竞争的，意味着每一种投入要素的价格等于其边际产量，劳动力在贷款品和非贷款品两个部门的分配比例分别为w_(h,t)和1-w_(h,t)。贷款品和非贷款品两大部门的工资之和为:
\indent In this subsection the paper extends the endogenous technological growth model constructed by Romer(1987\citep{romer1987growth}, 1990\citep{romer1990endogenous}) by assuming that labourers have the choice of supplying labour to the current goods sector or investing it in the research sector $Z_t$ to obtain the discounted value of future income, and that labour is allocated in the production and research sectors in proportions $\alpha_t$ and $1 - \alpha_t$ respectively. The production sector is perfectly competitive, implying that each input factor The price of each input factor is equal to its marginal output, and labour is allocated in the loan and non-loan sectors in proportions $w_{h,t}$ and $1-w_{h,t}$ respectively. The sum of the wages of the two sectors, loan and non-loan goods, is :
\begin{equation}
\begin{aligned}
    W_{h,t}n_{h,t}+W_{f,t}n_{f,t}&\approx C^\sigma_t \omega_{h,t}(1+R_t)\frac{j_h}{R_t}(1-\psi_h-\phi_h)+j_f(1-\psi_f-\phi_f)
\end{aligned}
\end{equation}
\begin{equation}
    n_{h,t}+n_{f,t}=\alpha_tn_t
\end{equation}
%研究部门的增长率定义为公式53，其中$\eta$为未知的正实数：
\indent The growth rate of the study sector is defined by the equation 53, where $\eta$ is an unknown positive real number.
\begin{equation}
    \frac{A_{t+1}}{A_t}=1+\eta(1-\alpha_t)n_t
\end{equation}
%假设劳动者预期保持未来无限期分配给研发部门的比例不变，劳动总供给的增长率等于人口增长率，人口增长率的对数Log[n_t]服从均值为n_(g,t),方差为σ_(n,t)^2正态分布。研发部门的技术产品是垄断市场，因此全部利润均由劳动者获得。当期劳动者投入研发部门的(1-α_t ) n_t的预期折现值为：
\indent Assuming that workers expect to keep the proportion allocated to the R\&D sector constant for an infinite future period, the growth rate of total labour supply is equal to the population growth rate, and the logarithm of the population growth rate, $Log[n_t]$, follows a normal distribution with mean $n_{g,t}$ and variance $\sigma^2_{n,t}$. The R\&D sector has a monopoly market for its technology products, so all profits are earned by workers. The expected discounted value of $(1 - \alpha_t)n_t$ invested by labour in the R\&D sector in the current period is
\begin{equation}
\begin{aligned}
    &\mathbb{E}[\sum^{+\infty}_{k=t+1}{W_{h,k}n_{h,k}+W_{k,f,t}n_{f,k}}\frac{A_k}{(1+R_t)^{k-t}}]\\
    &\approx \eta(1-\alpha_t)n_t\mathbb{E}[\sum^{+\infty}_{k=t+1}\frac{exp[\sum^k_{i=t}{\epsilon_i}\{W_{h,k}n_{h,k}+W_{f,k}n_{f,k}\}]}{(1+R_k)^{k-t}}]
\end{aligned}
\end{equation}
%当达到平衡增长路径时，工人选择研究部门和生产部门是无差异的，研发技术的预期利润贴现流与从生产部门获取的劳动报酬等价。
\indent When the equilibrium growth path is reached, there is no difference in the choice of workers between the research and production sectors, and the expected discounted stream of profits from R\&D technology is equivalent to the labour remuneration received from the production sector.
$$
    \eta(1-\alpha_t)n_t\mathbb{E}[\sum^{+\infty}_{k=t+1}\frac{exp[\sum^k_{i=t}{\epsilon_i}\{W_{h,k}n_{h,k}+W_{f,k}n_{f,k}\}]}{(1+R_k)^{k-t}}]=1
$$
\begin{equation}
    \alpha_t=1-\frac{1}{\eta\mathbb{E}[\sum^{+\infty}_{k=t+1}\frac{exp[\sum^k_{i=t}{\epsilon_i}\{W_{h,k}n_{h,k}+W_{f,k}n_{f,k}\}]}{(1+R_k)^{k-t}}]}
\end{equation}
%由方程55，可以推测当人口增长率较快即n_(g,t)>0时，劳动者选择投入研发部门的劳动比例α_t→1，即几乎所有劳动都用于当期生产。在现实中人口增长率不可能维持长时间的正增长。因此上述结论说明的经济意义是在人口快速增长的时期，更少的资源投向研发部门，因为新增劳动供给替代了科技部门的进步，也就是生产率替代。值得注意的是，当预期未来贷款品杠杆ω_(h,k)增加时，劳动者从非研发部门所获得的预期工资将增加，因此α_t将增加，贷款品杠杆增加的过程也是排斥劳动力投向研发部门的过程，本文称之为杠杆率排斥。
\indent From equation 55, it can be surmised that when the population growth rate is fast i.e. $n_{g,t} > 0$, the proportion of labour that workers choose to put into the R\&D sector $\alpha_t \to 1$, i.e. almost all labour is used in current production. In reality the population growth rate is unlikely to remain positive for long. The economic implication illustrated by the above findings is therefore that in periods of rapid population growth, fewer resources are invested in the R\&D sector, as the additional labour supply substitutes for progress in the technology sector, i.e. productivity substitution. It is worth noting that when the expected future leverage of loan goods $\omega_{h,k}$ increases, the expected wage received by labour from the non-R\&D sector will increase and hence $\alpha_t$ will increase, and the process of increasing leverage of loan goods is also the process of excluding labour from investing in the R\&D sector, which is referred to in this paper as leverage exclusion.\newline
%为了维持固定的劳动力研发比例1-(α_t ) ̅，中央银行可以通过当期利率调控抵消生产率替代和杠杆率排斥的影响。当人口增长率或者贷款品杠杆率下降时，中央银需要行降低基准利率，反之同理。实际利率的调整还可以通过调整货币供应量或者央行沟通的方式影响通货膨胀率降低或提高实际利率。但是明显可见维持这一政策规则影响了货币政策的独立性。
\indent To maintain a fixed labour force R\&D ratio of $1-(\alpha_t)$, the central bank can offset the effects of productivity substitution and leverage exclusion through current interest rate regulation. The central bank needs to lower the benchmark interest rate when the population growth rate or the leverage ratio of loan products falls, and vice versa. Adjustments to the real interest rate can also affect inflation by adjusting the money supply or central bank communication to lower or raise the real interest rate. However, it is clear that maintaining this policy rule affects the independence of monetary policy.\newline
%由于A_t是关于时间t的单调增函数，更少的劳动投入将相对地形成更低的全要素生产率，最终长期增长受到货币蓄水池机制的抑制影响。因此在长期中央决策者面临短期增长和长期增长以及货币政策独立性的权衡。当中央决策者更重视短期增长时，贷款品部门杠杆提高，为了维持固定的研发投入比例，中央决策者通过各种方式提高当年实际利率，也通过增加应付本息的方式提升了当年各部门的债务，反之同理。如果中央决策者需要维护货币政策的独立性并且维持长期增长，高人口增长率和高杠杆率不可兼得。
\indent Since $A_t$ is a monotonically increasing function with respect to time t, less labour input will result in relatively lower total factor productivity and ultimately long-run growth is affected by the dampening effect of the monetary reservoir mechanism. In the long run, therefore, central policymakers face a trade-off between short-term and long-term growth and the independence of monetary policy. When central policymakers place more emphasis on short-run growth, the lending goods sector becomes more leveraged and, in order to maintain a fixed ratio of R\&D inputs, central policymakers raise the real interest rate for the year in various ways, also raising the debt of the sectors for the year by increasing principal and interest payable, and vice versa. If central policymakers need to preserve the independence of monetary policy and maintain long-term growth, high population growth and high leverage cannot be combined.
\section{Conclusion}
\indent In this paper we constructs a dynamic general equilibrium model including fiscal financing investment and financial accelerator based on the monetary reservoir phenomenon of "Chinese housing and US stocks" and China's unique fiscal and financial system to provide a explanation for the monetary transmission mechanism of monetary reservoir-fiscal financing and investment. Based on the theoretical model and numerical simulation results, the theory of monetary reservoir assets is proposed: an asset can become a monetary pool asset and be financed by a third party like local government if it satisfies three conditions: a leveraged trading system, a balance commitment payment and the existence of third-party utility. When it will lead to an increase in monetary over-issuance(bubble effect), forming a monetary transmission mechanism outside the central bank. Conversely, economic growth conceals the credit expansion of local governments, resulting in the monetary transmission mechanism being "invisible" to scholars. We also proposes a design for a de-bubble mechanism for monetary reservoir assets to address this problem.\newline
\indent Key findings, as well as policy implications, include: (i) Governments should pay attention to the accumulation of human capital, the introduction of innovative industries and the investment of public goods with strong positive externalities, so as to form a virtuous cycle of fiscal financing - tax return gain - refinancing of fiscal surplus;  (ii) The process of financial financing by local governments using monetary reservoir assets should be brought under the supervision of the central government to avoid the phenomenon that  local government officials merely focus on short-term local gains and neglect long-term overall gains and maintain the stability of the economic system and to maintain the consistency of monetary policy; (iii)Financial markets for assets that satisfy the nature of monetary reservoir, such as derivatives in equity markets, futures markets, securitised consumer loans, etc., should be constrained to limit the refinancing of collateral (pledges), leverage and market access in order to control the overall size of credit. (iv) Trading system in monetary reservoir market can be a noval and effective instrument for non-traditional monetary policy.\newline

\appendix

 \bibliographystyle{elsarticle-num} 
 \bibliography{cas-refs}
 %\nocite{*}





% \end{thebibliography}
\end{document}